\documentclass[journal]{IEEEtran}
\usepackage{epsfig,latexsym}
\usepackage{graphicx}
\usepackage{multirow}
\usepackage{float}
\usepackage{indentfirst}
\usepackage{amsmath}
\usepackage{amssymb}
\usepackage{times}
\usepackage{subfigure}
\usepackage{psfrag}
\usepackage{hyperref}
\usepackage{cite}
\usepackage{lastpage}
\usepackage{fancyhdr}
\usepackage{color}
 \usepackage{amsthm}
\usepackage{bigints}
\usepackage{setspace}
\usepackage{makecell}
\usepackage{chngpage}
\usepackage{booktabs}
\usepackage{array}
\usepackage{tabu}
\newcolumntype{"}{@{\hskip\tabcolsep\vrule width 0.5pt\hskip\tabcolsep}}

\begin{document}
\title{ \huge{ Safeguarding Next Generation Multiple Access Using Physical Layer Security Techniques: A Tutorial}}

\author{  Lu Lv,~\IEEEmembership{Member,~IEEE}, Dongyang Xu,~\IEEEmembership{Member,~IEEE}, Rose Qingyang Hu,~\IEEEmembership{Fellow,~IEEE},\\ Yinghui Ye,~\IEEEmembership{Member,~IEEE}, Long Yang,~\IEEEmembership{Senior Member,~IEEE}, Xianfu Lei,~\IEEEmembership{Member,~IEEE},\\ Xianbin Wang,~\IEEEmembership{Fellow,~IEEE}, Dong In Kim,~\IEEEmembership{Fellow,~IEEE}, and Arumugam Nallanathan,~\IEEEmembership{Fellow,~IEEE}
\thanks{Lu Lv and Long Yang are with the State Key Laboratory of Integrated Services Networks, Xidian University, Xi'an 710071, China (e-mail: \{lulv, lyang\}@xidian.edu.cn).}
\thanks{Dongyang Xu is with the School of Information and Communications Engineering, Xi'an Jiaotong University, Xi'an 710049, China (e-mail: xudongyang@xjtu.edu.cn).}
\thanks{Rose Qingyang Hu is with the Department of Electrical and Computer Engineering, Utah State University, Logan, UT 84322, USA (e-mail:
rose.hu@usu.edu).}
\thanks{Yinghui Ye is with the Shaanxi Key Laboratory of Information
Communication Network and Security, Xi'an University of Posts \&
Telecommunications, Xi'an 710121, China (e-mail: connectyyh@126.com).}
\thanks{Xianfu Lei is with the School of Information Science and Technology, Southwest Jiaotong University, Chengdu 610031, China (e-mail:
xflei81@gmail.com).}
\thanks{Xianbin Wang is with the Department of Electrical and Computer Engineering, Western University, London, ON N6A 5B9, Canada (e-mail: xianbin.wang@uwo.ca).}
\thanks{Dong In Kim is with the Department of Electrical and Computer Engineering, Sungkyunkwan University, Suwon 16419, South Korea (e-mail: dikim@skku.ac.kr).}
\thanks{Arumugam Nallanathan is with School of Electronic Engineering and Computer Science, Queen Mary University of London, E1 4NS, U.K. (e-mail: a.nallanathan@qmul.ac.uk).}
}
\maketitle

\begin{abstract}

Driven by the ever-increasing requirements of ultra-high spectral efficiency, ultra-low latency, and massive connectivity, the forefront of wireless research calls for the design of advanced next generation multiple access schemes to facilitate provisioning of these stringent demands. This inspires the embrace of non-orthogonal multiple access (NOMA) in future wireless communication networks. Nevertheless, the support of massive access via NOMA leads to additional security threats, due to the open nature of the air interface, the broadcast characteristic of radio propagation as well as intertwined relationship among paired NOMA users. To address this specific challenge, the superimposed transmission of NOMA can be explored as new opportunities for security aware design, for example, multiuser interference inherent in NOMA can be constructively engineered to benefit communication secrecy and privacy. The purpose of this tutorial is to provide a comprehensive overview on the state-of-the-art physical layer security techniques that guarantee wireless security and privacy for NOMA networks, along with the opportunities, technical challenges, and future research trends.

\end{abstract}

\IEEEpeerreviewmaketitle

\section{Introduction}

\subsection{Background}

Witnessed by the ever-increasing number of wireless devices and applications, such as holographic video, extended reality, intelligent transport, and Metaverse, the sixth generation (6G) mobile networks now face unprecedented challenges due to the extremely demanding and diverse requirements of these emerging devices and applications \cite{XiaohuYou_SCIS2021, Zhengquan_VTM2019,XiaohuYou_WCM2023}. Hence, this calls for the development of advanced multiple access schemes, namely the next generation multiple access (NGMA), to support massive connectivity and multiple functions of networks, such as communication, sensing, and computing, in a more resource- and complexity-efficient manner compared to existing multiple access schemes. Technically speaking, the aim of multiple access is to let multiple users share the same radio resource. Looking back at previous generations of mobile networks, orthogonal multiple access (OMA) schemes have been used where individually orthogonal resource blocks are allocated to different users, for example, orthogonal frequency/time/code/subcarrier/space resources in the multiple access schemes from 1G to 5G \cite{Vaezi_book2019}. However, conventional OMA schemes cannot meet the stringent communication requirements of the increased number of users, such as ultra-high spectral efficiency, ultra-low latency, and massive connectivity. This motivates the design principle of NGMA to shift from orthogonality to non-orthogonality, thus embracing non-orthogonal multiple access (NOMA) as a promising multiple access candidate in 6G networks \cite{Yuanwei_JSAC2022}. The underlying principle of NOMA is to encourage spectrum sharing among multiple users over the same resource block, and therefore help achieve high spectral and energy efficiency, extreme connectivity, and improved user fairness \cite{Lu_CM2018,Xianfu_JSAC2017, Zhiguo_CM2017,Yuanwei_PIEEE2017,DaiLinglong_COMST2018,Makki_OJCOMS2020}.

Given the aforementioned benefits, NOMA has received tremendous attention in both academia and industry during the past few years. Nevertheless, there are several pivotal security issues dedicated to NOMA due to the open nature of air interface and the broadcast characteristic of wireless communications \cite{Lu_Network2020, FengYouhong_Network2022,Vaezi_WCM2019,Lu_Proc2022,LiXing_IoT2023}. For example, although NOMA is able to utilize the available spectrum efficiently by allowing simultaneous transmissions of multiple users, the superimposed signal structure in NOMA makes an eavesdropper readily intercept more signals, leading to more severe leakage of confidential information \cite{HeBiao_JSAC2017,ZhangHaijun_JSAC2018}. Another fact is that strong users (i.e., with good channel conditions) can decode the signals of weak users (i.e., with poor channel conditions) by successive interference cancellation (SIC) in NOMA, therefore the security and privacy of weak users may be compromised \cite{ZhaoNan_TCOM2022,Zhiguo_TCOM2017,Mishra_TVT2020,Mishra_CL2022}. Furthermore, since NOMA is interference-limited, malicious users can insert harmful jamming signals to disrupt the ongoing NOMA communications between legitimate users \cite{Lu_WCL2019,Xingwang_TVT2020}. The above security and privacy issues pose great challenges in safeguarding NOMA communications, and typically, cryptographic techniques are employed to achieve NOMA communication confidentiality. Note that cryptographic techniques are implemented at upper layers and rely upon sophisticated computational problems, where the involved encryption and decryption procedures may increase the computational complexity and cause extra transmission delay, and hence offset the benefits of NOMA in terms of reducing access latency. As an alternative, physical layer security (PLS) becomes an efficient and cost-effective means to achieve secure NOMA communications \cite{WuYongpeng_JSAC2018,LiuYiliang_COMST2017,ChenXiaoming_COMST2017, ZouYulong_PIEEE2016,ZhaoNan_WCM2020,ZhaoNan_WCM2023,Mukherjee_COMST2014}. Specifically, by simply exploiting the physical characteristics of wireless channels such as fading, interference, and noise, PLS techniques can achieve different levels of security and privacy including information-theoretic security \cite{Hui-Ming_TSP2015,LiQiang_TSP2013,YangNan_CM2015}, post-quantum security \cite{Dongyang_TII2022,
Dongyang_TIFS2021}, as well as low probability of detection (a.k.a. covertness) \cite{ZhaoNan_COMST2022,Shihao_WCM2019,Bash_CM2015,
Tongxing_TWC2019}, without consuming excessive computation resources and/or signaling overheads.

Although the security and privacy issues in NOMA networks are more severe, the multiuser interference inherent in NOMA signaling, which is conventionally treated deleterious for communication, can be carefully managed to provide beneficial effects to security and privacy, i.e., stronger multiuser interference indicates greater potential to degrade eavesdroppers' capability \cite{ZhengGan_CM2014,Conti_ACSSC2013,Masouros_CM2013,Masouros_TSP2015,Conti_JSAC2018,
Masouros_TIFS2018,FanYe_TIFS2021,Ottersten_JSTSP2016,FanYe_TCOM2021,RenPinyi_TIFS2021,
LiMing_TIFS2020}. This opens up new opportunities and perspectives for the design and analysis of NOMA with intrinsic secrecy. Particularly, the superposition overt NOMA transmissions can be exploited as natural shields to enable covert wireless communication, where one can simply embed the covert messages in the existing broadcast channels to guarantee a low probability of detection \cite{Tan_TIFS2019,Bloch_TIFS2019,Lu_TWC2022,Zlatanov_CL2022}. Even if the existence of the NOMA signal transmission is detected by the eavesdroppers, the confidentiality of data carried by NOMA signal can be still guaranteed by specifically controlled multiuser interference in NOMA, such that it intentionally degrades the information reception quality at the eavesdroppers while having tolerable effects on the legitimate users \cite{ChenXiaoming_JSAC2020,FengYouhong_TWC2019,Lu_SCIS2021,Lu_SCIS2022,CaoYang_TWC2019}.

\subsection{Motivation and Contribution}

In view of the challenges and opportunities that arise in safeguarding NOMA communications using PLS techniques, as well as how the innovative 6G technologies, such as intelligent surfaces, cooperative relays, large antenna array, quantum computing, and ambient backscatter communication, can be integrated to achieve a new level of NOMA secrecy and covertness at the physical layer, more in-depth investigations are deserved. Therefore, in this paper, we are motivated to provide a comprehensive overview of state-of-the-art research results on the PLS techniques for NOMA in the envisioned 6G mobile networks. In particular, our main focus lies in the following aspects.

\begin{itemize}
  \item The fundamentals of NOMA and PLS techniques are reviewed first. Specifically, the analytical models of NOMA at the link- and system-level are discussed, and the several NOMA prototypes in real-world environment are also introduced. Subsequently, the family of PLS techniques, namely information-theoretic security, quantum-safe security, and low-detection covertness, are summarized.

  \item The emerging information-theoretic security techniques for NOMA communications are highlighted. The application of multiple antenna techniques is investigated, with an emphasis on unlocking the spatial degrees of freedom for security provisioning of NOMA. Then, enhancing the secrecy via cooperation techniques, e.g., active and passive cooperative mechanisms to smartly engineer the legitimate users' and/or eavesdropper's propagation environment, are surveyed. Furthermore, the use of machine learning (ML) and artificial intelligence (AI) approaches for secrecy resource allocation is presented, which provides a new and revolutionary paradigm to design next generation multiple access schemes with security and privacy guarantees.

  \item Quantum-safe technologies, including quantum encryption based on quantum physics and post-quantum cryptography based on novel NP-hard problems, are expected to play a key role in the design of secure and private NOMA communication. The quantum encryption case, i.e., quantum-key distribution assisted by physical layer security of NOMA over the air interface, is presented. The quantum-assisted multi-user detection technology in NOMA systems are also discussed by considering the iterative detection scheme based on quantum spatial coupling quantum noise channels.

  \item Besides protecting the information content, hiding the very existence of wireless communication to achieve a high-level privacy is required by the next generation multiple access design. The covert communication or low probability of detection communication technology can be applied in NOMA systems to tackle this issue, and the corresponding design principles are discussed in detail.

  \item Finally, we outline important research directions worthy of further investigation and promising topics pertaining to security and privacy provisioning for NOMA systems. In particular, the potentials and opportunities on exploiting the emerging technologies, including integrated sensing and communication (ISAC), satellite communication, new spectrum resource, net zero communication, and generative AI, to secure NOMA systems are presented for exciting future works.
\end{itemize}

\begin{table}[t]
\centering
\caption{Existing Survey Papers and Their Focused Issues} \label{mult}
\begin{tabu}{l|[0.6pt]l}
  \Xhline{0.6pt}
   Surveys&Focused issues related to security\\
  \Xhline{0.6pt}
\cite{Chorti_Mag2022}& Enhanced sensing, artificial intelligence \\
 \Xhline{0.6pt}
 \cite{LuXiaozhen_COMST2023}& Reinforcement learning\\
 \Xhline{0.6pt}
 \cite{WeiZhongxiang_CM2020}& Interference management\\
 \Xhline{0.6pt}
 \cite{Mitev_OJVT2023}& Wiretap coding, secret key generation\\
 \Xhline{0.6pt}
 \cite{Khalid_IoT2024}& Reconfigurable intelligent surface\\
 \Xhline{0.6pt}
 \cite{Arfaoui_COMST2020}& Visible light communication\\
 \Xhline{0.6pt}
 \cite{Angueira_COMST2022}& Industrial communication\\
 \Xhline{0.6pt}
 \cite{Illi_COMST2024}& Authentication, node detection\\
 \Xhline{0.6pt}
 \cite{Pirayesh_COMST2022}& Jamming attacks, anti-jamming techniques\\
 \Xhline{0.6pt}
 \end{tabu}
 \end{table}

\subsection{Related Surveys and Organization}

\begin{figure}[t]
  \centering
  \includegraphics[width=3.45in]{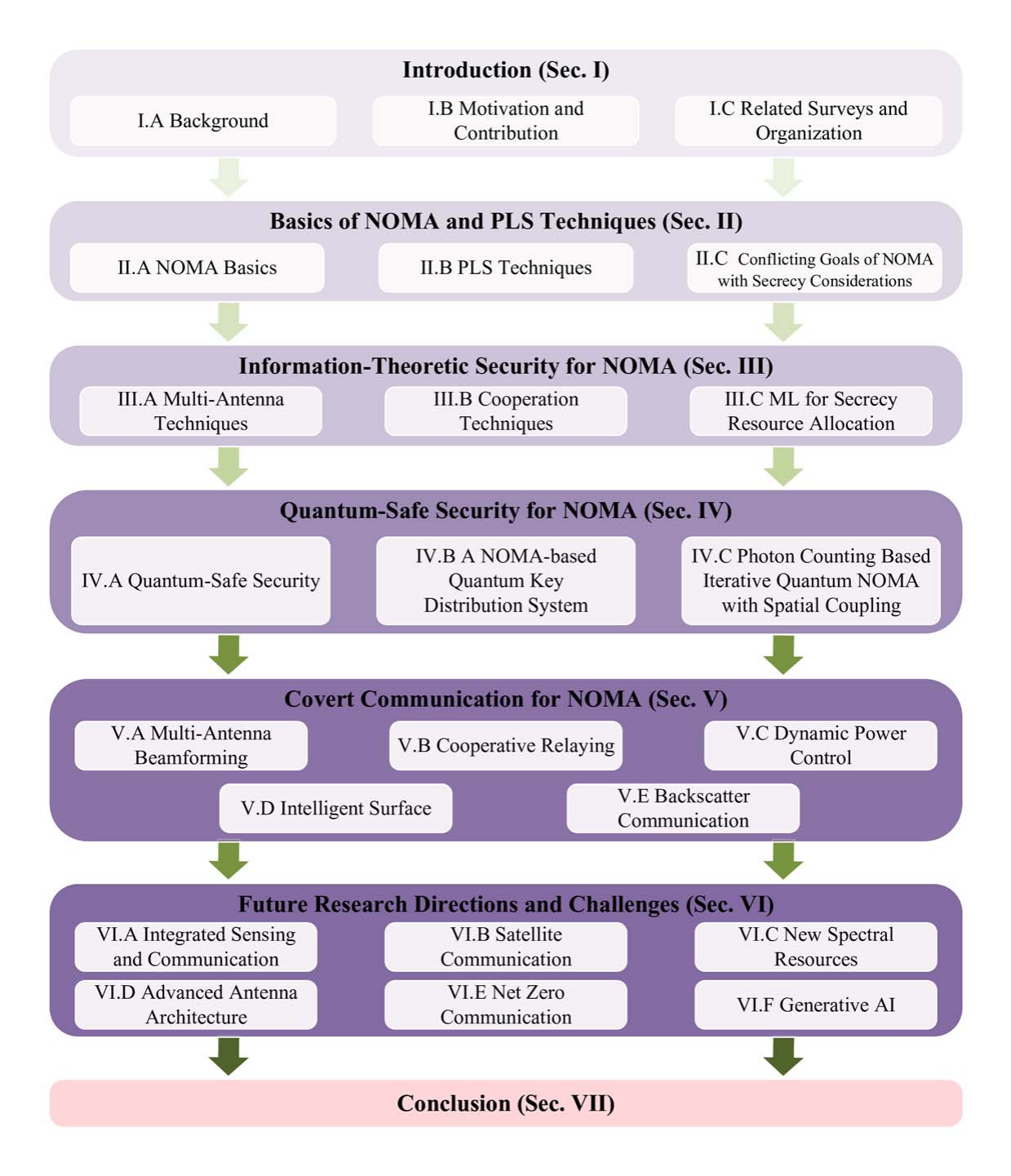}\\
  \caption{Structural diagram of the tutorial.}\label{fig:stuctural}
\end{figure}

Several survey papers \cite{Chorti_Mag2022,LuXiaozhen_COMST2023,WeiZhongxiang_CM2020,Mitev_OJVT2023,Khalid_IoT2024,Arfaoui_COMST2020,Angueira_COMST2022,Illi_COMST2024,Pirayesh_COMST2022} related to security issues have been reported involving different communication scenarios, emerging technologies, and enhanced management strategies, as shown in Table I. However, a comprehensive study of secrecy designs for the next generation multiple access is still missing. Moreover, existing studies have generally concluded that the use of NOMA poses a severe security risk to the system, while ignoring the constructive use of multi-user interference inherent in NOMA to improve security.

A structural diagram of this tutorial is provided in Fig. \ref{fig:stuctural}.
The remainder of this tutorial is organized as follows. Section \ref{Basics} reviews the basics of NOMA and PLS techniques, with a focus on both theoretical studies and practical implementations. Existing physical layer techniques for improving the performance of information-theoretic security, quantum-safe security, and low detection covertness in NOMA networks are surveyed in Sections \ref{ITS}, \ref{sec:Quantum}, and \ref{sec:covert}, respectively. Several future research directions and challenges for security provisioning of next generation multiple access are elaborated in Section \ref{sec:future}. Finally, the tutorial is briefly concluded in Section \ref{sec:conclusion}.

\section{Basics of NOMA and PLS Techniques}
\label{Basics}

In this section, we briefly introduce the basics and fundamentals of the considered communication techniques, namely NOMA and PLS. Afterwards, the conflicting goals of NOMA transmission towards increased spectral efficiency on the one hand and security provisioning of NOMA on the other hand are discussed. The latter is demanding because the combined signaling and propagation channel characteristics of NOMA transmission in wireless environments inherently induces vulnerability to intentional jamming.

\subsection{NOMA Basics}

\begin{figure*}[t]
  \centering
  \includegraphics[width=5.0in]{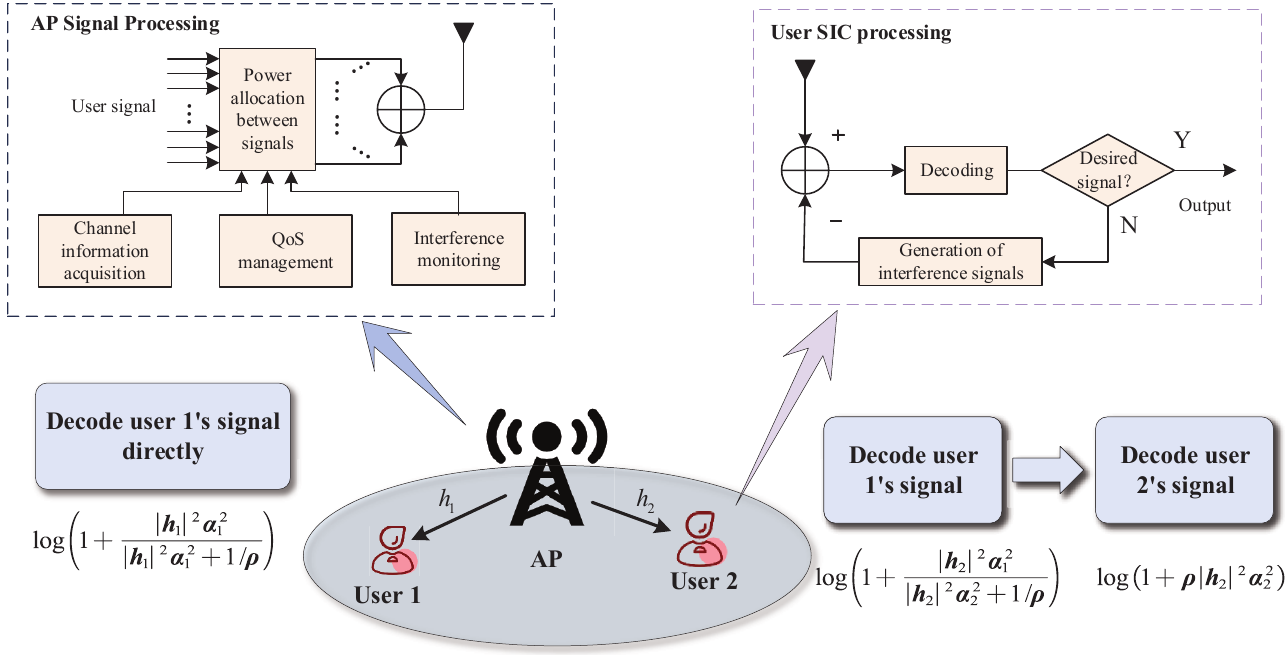}\\
  \caption{System model and signal processing of a two-user NOMA communication scenario.}\label{fig:NOMA_model}
\end{figure*}

\subsubsection{Theoretical Foundations}
The key principle of NOMA is to disregard the widely adopted resource orthogonality in multiple access, i.e., simultaneously serving multiple users in the same frequency/time/code/subcarrier/space and distinguishing them in the power domain \cite{Xianfu_JSAC2017}. Although there are many possible forms of NOMA that maintain non-orthogonality in resource allocation such as code-domain NOMA \cite{Hoshyar_TSP2018,Hoshyar_VTC2010,Nikopour_PIMRC2013, DaiXiaoming_WCM2018}, in this tutorial we focus only on the power-domain NOMA. Take a simple two-user NOMA scenario shown in Fig.~\ref{fig:NOMA_model} as an example, where each of the access point (AP) and the users is equipped with a single antenna. The AP exploits superposition coding (SC) to enable power-domain spectrum sharing and transmits a linear combination of the users' signals, i.e., $x=\sqrt{\alpha_1}s_1+\sqrt{\alpha_2}s_2$, where $s_i$ denotes the signal intended for user $i$ and $\alpha_i$ denotes the power allocation coefficient for user $i$'s signal satisfying $\alpha_1+\alpha_2\leq1$. At the user side, successive interference cancellation (SIC) technique is employed to mitigate the inter-user interference. Due to the sequential nature of SIC, the SIC decoding order is a key issue in the success of multiuser detection, which is in essence determined by the users' channel state information (CSI), power allocation coefficients, and quality of service (QoS) requirements \cite{Zhiguo_CL2020,Zhiguo2_CL2020}. In the literature, there are four main strategies that determine the SIC decoding order in NOMA, as follows:

\begin{itemize}
  \item \textbf{CSI-Based Strategy:} Selecting the SIC decoding order based on users' CSI conditions has long been viewed as a simple yet mature strategy and adopted since the invention of NOMA \cite{Zhiguo_SPL2014,Krikidis_SPL2015, YangZheng_TCOM2015,Rose_JSAC2017}. This strategy suggests that users are ordered according to their CSI conditions and more power is allocated to the user with a poor CSI condition. Here we assume that $|h_1|^2\geq|h_2|^2$, where $h_1$ and $h_2$ are the channel coefficients of user 1 and user 2, respectively. Then, user 1 applies SIC with the decoding order from user 2's signal first and then moving towards its own signal, and the achievable rates for decoding the two signals are given by $\log\big(1+\alpha_2|h_1|^2/(\alpha_1|h_1|^2+1/\rho)\big)$ and $\log\big(1+\alpha_1\rho|h_1|^2\big)$, respectively, where $\rho$ is the transmit signal-to-noise ratio (SNR). While user 2 directly decodes its signal by treating user 1's signal as noise, yielding the achievable rate given by $\log\big(1+\alpha_2\rho|h_2|^2\big)$.

  \item \textbf{Statistical CSI-Based Strategy:} Instead of the instantaneous CSI, the statistical CSI can also be used for the SIC decoding order design, especially for the scenarios when the instantaneous CSI changes very fast such as vehicular networks, ad-hoc networks, and cognitive radio networks. In particular, both the average channel gain \cite{Kim_CL2015,Gong_TVT2019,Lu_CR_TVT2018} and the location/distance information \cite{Yuanwei_JSAC2016,Xu_TWC2021, WangHong_WCL2021} can be utilized for the SIC ordering. Benefits of using such statistical CSI-based strategy include: no need of signaling overhead for the AP to obtain the instantaneous CSI of the users; no need of signaling overhead for the AP to inform the users of the SIC ordering in each communication slot. Furthermore, a surprising finding shows that under some specific system parameters, the use of the distance information for SIC ordering can even achieve a better outage performance than that of the instantaneous CSI-based strategy \cite{Ali_TCOM2019}.

  \item \textbf{QoS-Based Strategy:} Unlike the CSI-based strategy, the SIC decoding order of the QoS-based strategy is not decided by the users' channel conditions, but is determined by the users' QoS requirements. Here we assume that the users have heterogeneous QoS requirements, where user 1 is a delay-sensitive user with a low data rate $R_1$ but user 2 is a delay-tolerant user. The key idea of QoS-based strategy is to treat NOMA as a special case of a cognitive radio (CR) system, where user 1 is viewed as the primary user and user 2 is viewed as the secondary user. The CR ensures that the secondary user can be admitted into the spectrum that is solely occupied by the primary user, while satisfying the primary user's QoS requirement. This means that the user 1's signal is decoded first in the SIC stage, with the power allocation selected to meet its QoS requirement, i.e., $\log\big(1+|h_1|^2/(|h_2|^2+1/\rho)\big)\geq R_1$. If the first stage of SIC is successful, user 2's achievable rate is obtained as $\log\big(1+\rho|h_2|^2\big)$ \cite{Zhiguo_TVT2016}. This QoS-based strategy is particularly important for NOMA to offer significant performance gains over OMA when users are with similar channel conditions, and to reduce the access delay of massive Internet of Things (IoT) access \cite{Yinghui_TCOM2020}.

  \item \textbf{Hybrid CSI/QoS-Based Strategy:} Both the CSI- and QoS-based strategies face the same dilemma that the user in the first stage of SIC suffers from an error floor for the outage probability \cite{Zhiguo_CL2020}. Fortunately, it is reported in \cite{Zhiguo_TCOM2021} that the outage probability error floor can be avoided by using hybrid CSI/QoS-based strategy with an appropriate user scheduling, where the SIC decoding order is opportunistically determined. Assuming that there are one primary user and $N$ secondary users and defining a threshold as follows:
      \begin{equation}
        \tau=\max\bigg\{0,\frac{|h_0|^2}{2^{R_0}-1}-\frac{1}{\rho}\bigg\},
      \end{equation}
      where $h_0$ and $R_0$ are the channel coefficient and target rate of the primary user. Prior to user scheduling, the secondary users are divided into two groups based on $\tau$. One is the strong user group $\mathcal{G}_s$ consisted of the users whose channel conditions are better than $\tau$, i.e., $|h_n|^2>\tau$, and the other is the weak user group $\mathcal{G}_w$ contained the users whose channel conditions are worse than $\tau$, i.e., $|h_n|^2<\tau$. If the secondary user is scheduled from $\mathcal{G}_s$, the primary user's signal cannot be decoded in the first stage of SIC, since $|h_n|^2>\tau$ leads to $\log\big(1+|h_0|^2/(|h_n|^2+1/\rho)\big)<R_0$. This implies that the SIC decoding order should be from the secondary user's signal first and then the primary user's signal. On the other hand, if the secondary user is scheduled from $\mathcal{G}_w$, its signal can be decoded in either of the two SIC stages. Both the analytical and numerical results in \cite{Zhiguo_CL2020} show that by applying such hybrid CSI/QoS-based SIC strategy, the outage probability error floor can be effectively eliminated.
\end{itemize}

\begin{figure}[t]
  \centering
  \includegraphics[width=3.0in]{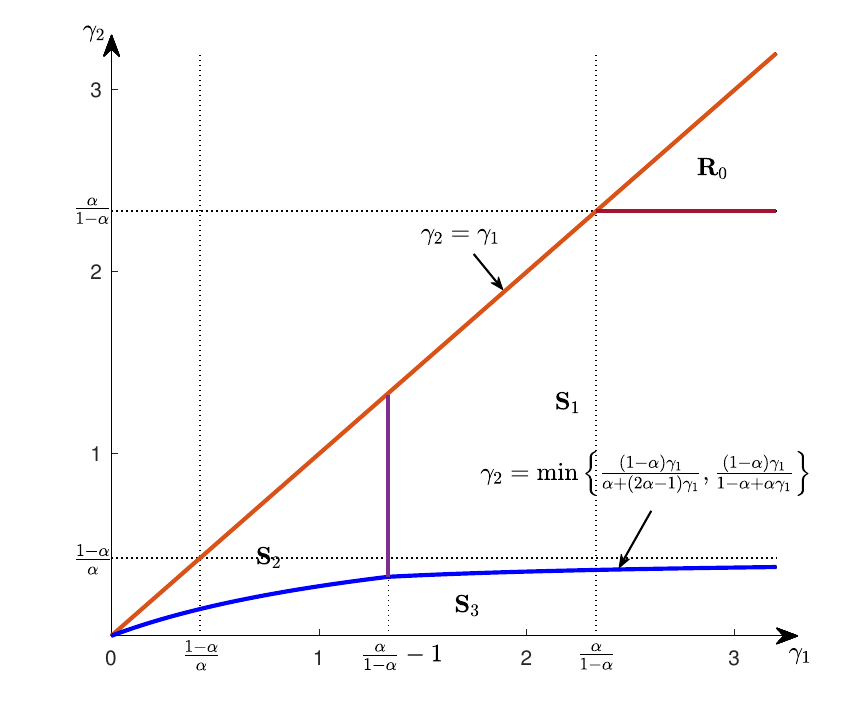}\\
  \caption{An illustration of threshold regions in terms of $(\gamma_1,\gamma_2)$.}\label{threshold_region}
\end{figure}

\begin{figure}[t]
  \centering
  \includegraphics[width=3.3in]{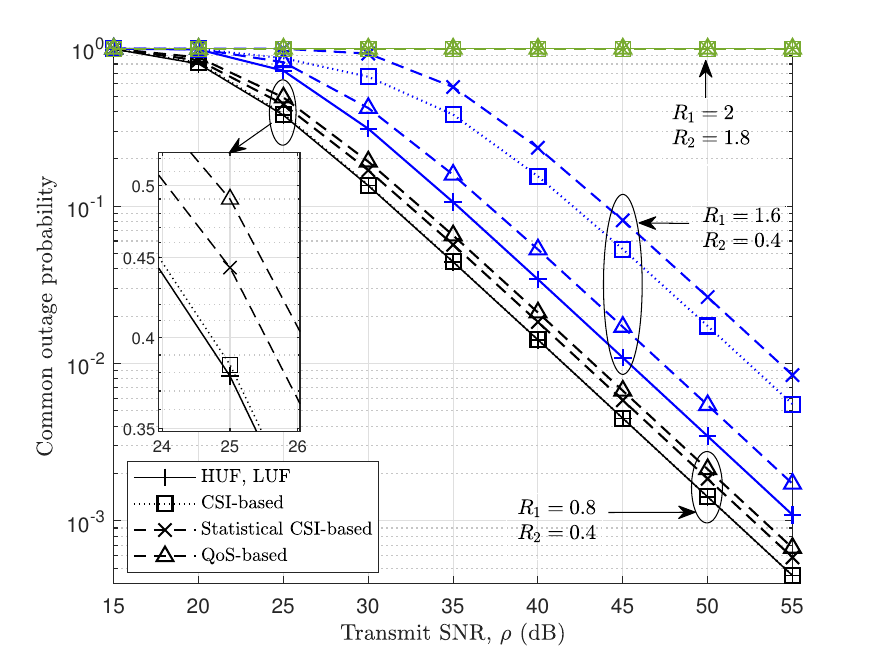}\\
  \caption{Common outage probability of different strategies. The distance between the AP and user 1 is $40\,m$, and the distance between the AP and user 2 is $30\,m$. Both small-scale Rayleigh fading and large-scale path loss are considered. The path loss is modeled by $(d_0/d)^\nu$, where $d_0=10\,m$ is the reference distance and $\nu=2.7$ is the path loss exponent. The power allocation coefficient is $0.7$.}\label{COP}
\end{figure}

It is important to point out that none of the above-mentioned strategies can guarantee the minimal outage probability of NOMA, since the SIC decoding order is indeed determined jointly by the users' CSI, power allocation coefficients, and QoS requirements \cite{Lu_TWC2023}. Theoretically, the outage occurs if the achievable channel capacity is lower than the target rate required by the NOMA transmission, where the channel capacity is a function of the CSI and power allocation coefficients, and the target rate is an indicator of the QoS requirement. As such, the gap between the channel capacity and target rate is also a metric to determine whether a channel is strong or weak, not merely depending on the channel coefficients as in the CSI-based strategy \cite{Zhiguo_SPL2014,Krikidis_SPL2015, YangZheng_TCOM2015,Rose_JSAC2017}. Moreover, this indicates that allocation of more power to the user whose channel gain is weak is not always optimal. To consider all these effects in the SIC decoding order of NOMA, a more recent work shows that the outage-optimal performance is achieved by using joint power allocation and decoding order selection (PA-DOS), where the SIC decoding order is independently done by each user, as explained in the following. We assume that $R_1$ and $R_2$ are the target rates for user 1 and user 2 with $R_1\geq R_2$, and the power allocation coefficient at the AP is $\alpha(>0.5)$. Define threshold regions in terms of $(\gamma_1,\gamma_2)$ for evaluating the users' target rates and power allocation, where $\gamma_i=2^{R_i}-1$, as shown in Fig.~\ref{threshold_region}. Some surprising findings are revealed: 1) If $(\gamma_1,\gamma_2)\in\mathbf{R}_0$, an outage inevitably occurs with arbitrary PA-DOS; 2) If $(\gamma_1,\gamma_2)\in\mathbf{S}_1$, the minimal outage probability of each user is achieved by allocating more power to the user whose target rate is low and decoding the low-rate user's signal first in SIC, without any CSI at the AP; 3) If $(\gamma_1,\gamma_2)\in\mathbf{S}_2\cup\mathbf{S}_3$, the minimal outage probability of each user can be achieved by the high-rate user first (HUF) and low-rate user
first (LUF) strategies. Interested readers may refer to \cite{Lu_TWC2023} for more details. In Fig.~\ref{COP}, the common outage probability is used as a metric for performance evaluation, where an outage event occurs for the common outage probability if any of the users are in outage. As can be observed from this figure, the use of the HUF and LUF strategies with joint PA-DOS provides a significant performance gain over the existing strategies, especially for the case with target rate pair $(R_1,R_2)=(1.6,0.4)$, i.e., $(\gamma_1,\gamma_2)\in\mathbf{S}_3$. As a further advance, by invoking the joint PA-DOS into opportunistic user scheduling in a multiuser NOMA system, the multiuser diversity gain can be exploited to significantly enhance the outage performance \cite{Mengqi_TCOM2023}.

\subsubsection{Implementation and Prototype}

From a theoretical point of view, the benefits of NOMA have been intensely studied based on either analytical models or simulations at link-level or system-level in the existing works. In order to transform NOMA into a mature technology in 6G networks, we believe it is of great importance to demonstrate how well does NOMA work in realistic deployments and operational conditions. Answering to this question motivates the research on the implementation and prototype designs of NOMA, where only very limited works can be found in the current literature (see, e.g., \cite{Benjebbour_WINCOM2015,Khorov_PIMRC2018, Khorov_Network2020,Khorov_InforCom2020,Khorov_JSAC2022} and references therein). In \cite{Benjebbour_WINCOM2015}, a typical NOMA downlink transmission testbed is developed, where both link-level and system-level computer simulations demonstrate the performance gains of NOMA over OMA. Later on, experimental downlink NOMA testbed is implemented over Wi-Fi, where the primary data stream is destined to a legacy device and the secondary data stream is sent to a new device \cite{Khorov_PIMRC2018,Khorov_Network2020,Khorov_InforCom2020,Khorov_JSAC2022}.

\begin{figure*}[ht]
  \centering
  \includegraphics[width=6.5in]{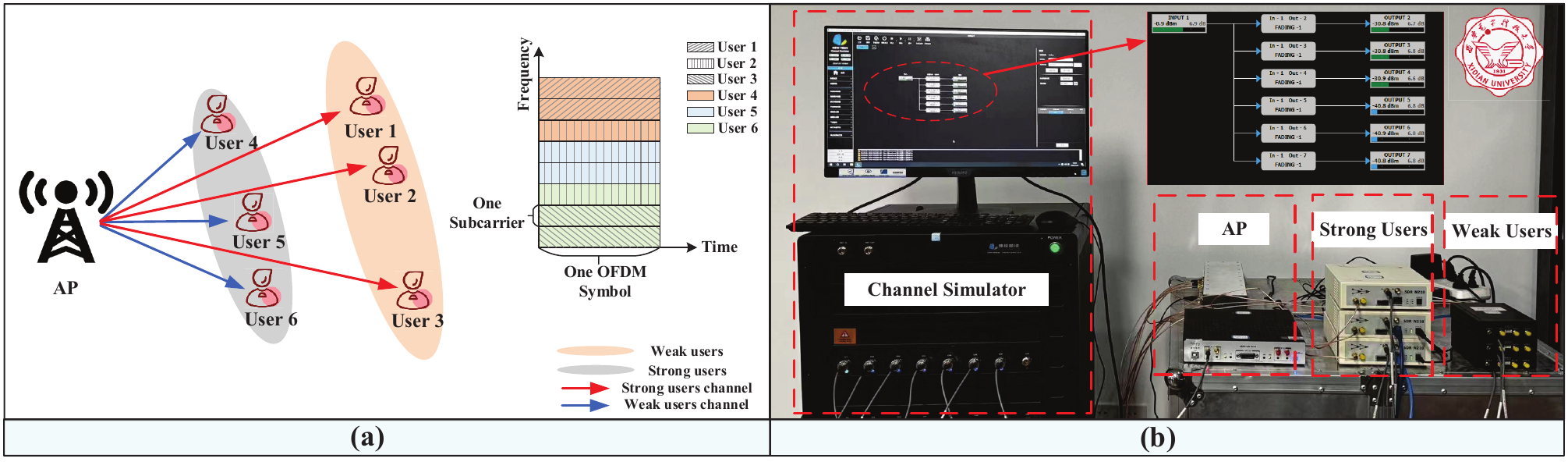}\\
  \caption{Scenario for a multi-carrier NOMA system: (a) Communication model; (b) Prototype.}\label{NOMA prototype}
\end{figure*}

\begin{figure}[t]
  \centering
  \includegraphics[width=3.3in]{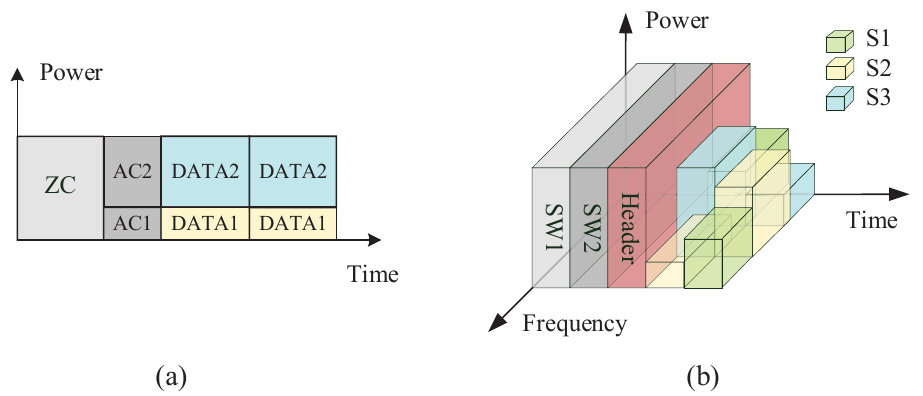}\\
  \caption{NOMA frame structures: (a) Single-carrier; (b) Multi-carrier.}\label{fig:NOMA_frame}
\end{figure}

\begin{figure}[t]
  \centering
  \includegraphics[width=3.3in]{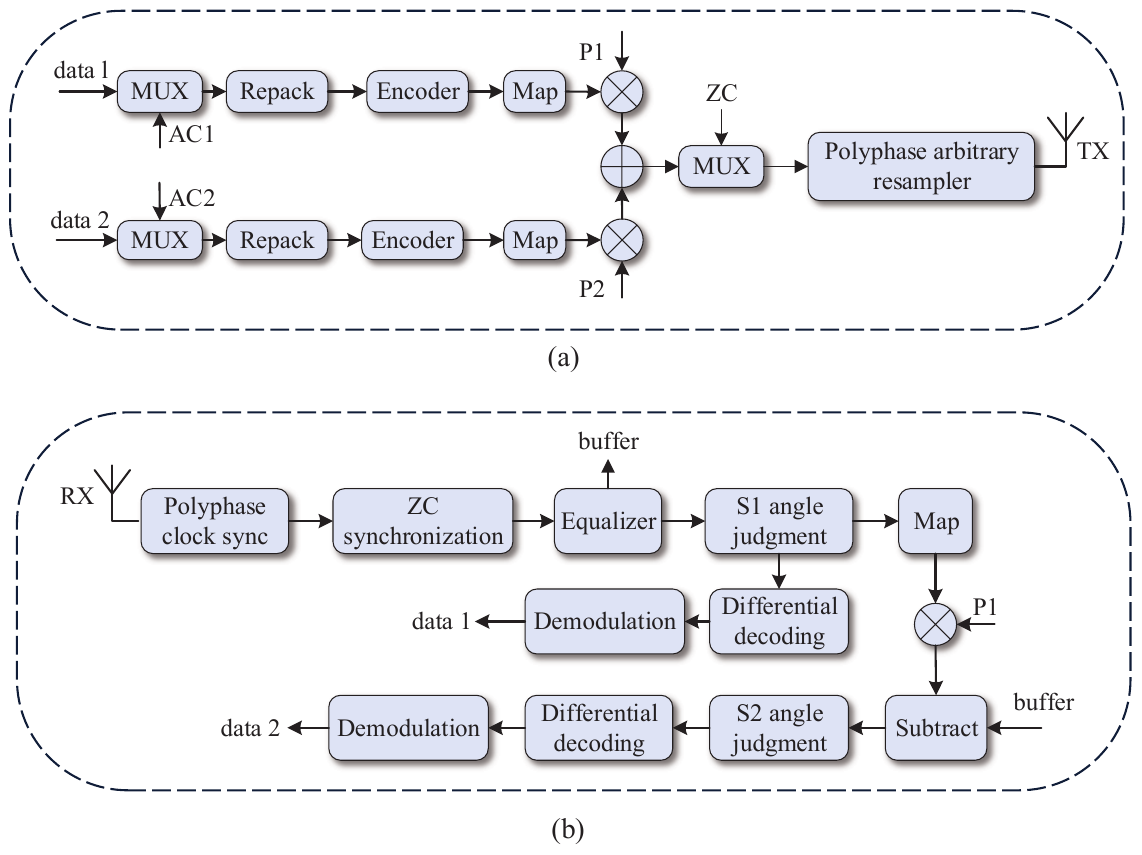}\\
  \caption{Transceiver architecture for single-carrier NOMA: (a) Transmitter with SC; (b) Receiver with SIC.}\label{fig:Singe_Carrier_NOMA}
\end{figure}

Most recently, our research group at Xidian University has presented pioneering single- and multi-carrier NOMA prototypes to support massive industrial IoT applications, as detailed in the following. In our prototype, we use universal software radio peripheral (USRP)-X310/N210/B210 that allows signal transmission and processing in a 40 MHz band with the potential increase up to 6 GHz. It has a large field-programmable gate array (FPGA) for applications demanding additional logic, memory and DSP resources. The FPGA also offers the potential to process up to 100 MS/s in both the transmit and receive directions. Fig.~\ref{NOMA prototype} shows a snapshot of our proposed NOMA prototype, composed of a host computer for performance monitoring, a channel emulator KSW-WNS02B to model the end-to-end real-world fading channels, and multiple USRPs for transceivers. While designing the NOMA frame structure, we focus on keeping it backward compatible, as shown in Fig.~\ref{fig:NOMA_frame}(a). Thus, we use the high-correlation value sequences, e.g., Zadoff Chu (ZC) sequence and access code (AC), to achieve the signal- and information-level synchronization. Fig.~\ref{fig:Singe_Carrier_NOMA} shows the single-carrier NOMA transmitter with SC, where the AP multiplies IQ samples of two data streams, denoted by data 1 and data 2, by coefficients $\sqrt{\alpha}$ and $\sqrt{1-\alpha}$, combines with the ZC and AC by MUX, and simply adds them together. The repack operation is to transform data bytes into bits to generate the corresponding quadrature phase shift keying (QPSK) constellation points. For the superposed signal reception, the correlation test is first performed to find the beginning of the NOMA frame with the ZC sequence and then remove the ZC sequence from the stream. The equalizer is used to mitigate the effect caused by the fading channels, and the results are saved in a buffer for the subsequent SIC after equalization. User 2 first decodes data 1 and then reconstructs the IQ samples of data 1 and saves them in a buffer. The bits stay in the buffer until the reconstructed IQ samples are ready for substraction. The residual frame is data 2 corrupted by noise. User 2 scales the residual frame to its original amplitude and then decodes it. Fig.~\ref{fig:Singe_Carrier_NOMA} shows the signal processing procedure of such SIC receiver.

\begin{figure}[t]
  \centering
  \includegraphics[width=3.3in]{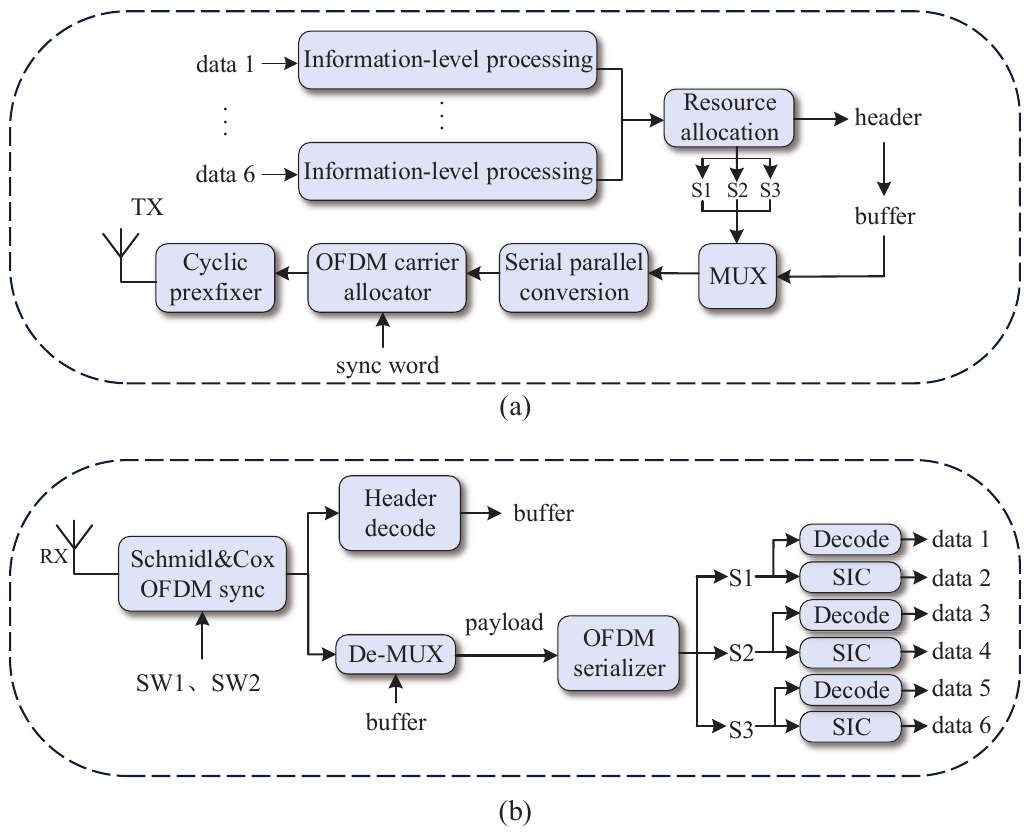}\\
  \caption{Transceiver architecture for multi-carrier NOMA based on OFDM: (a) Transmitter with SC; (b) Receiver with SIC.}\label{fig:Multicarrier_NOMA}
\end{figure}

\begin{figure*}[t]
  \centering
  \includegraphics[width=5.0in]{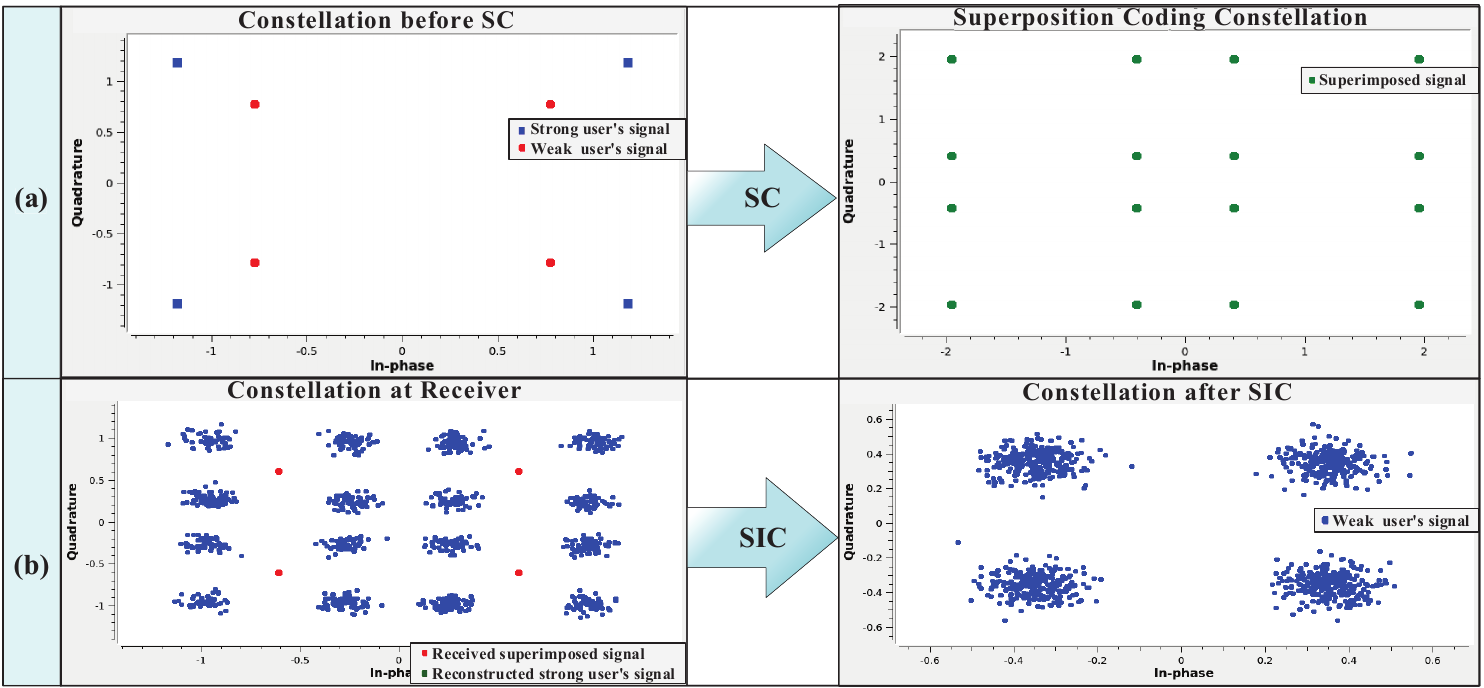}\\
  \caption{Performance of SC and SIC.}\label{fig:SC_and_SIC}
\end{figure*}

\begin{figure*}[t]
  \centering
  \includegraphics[width=5.0in]{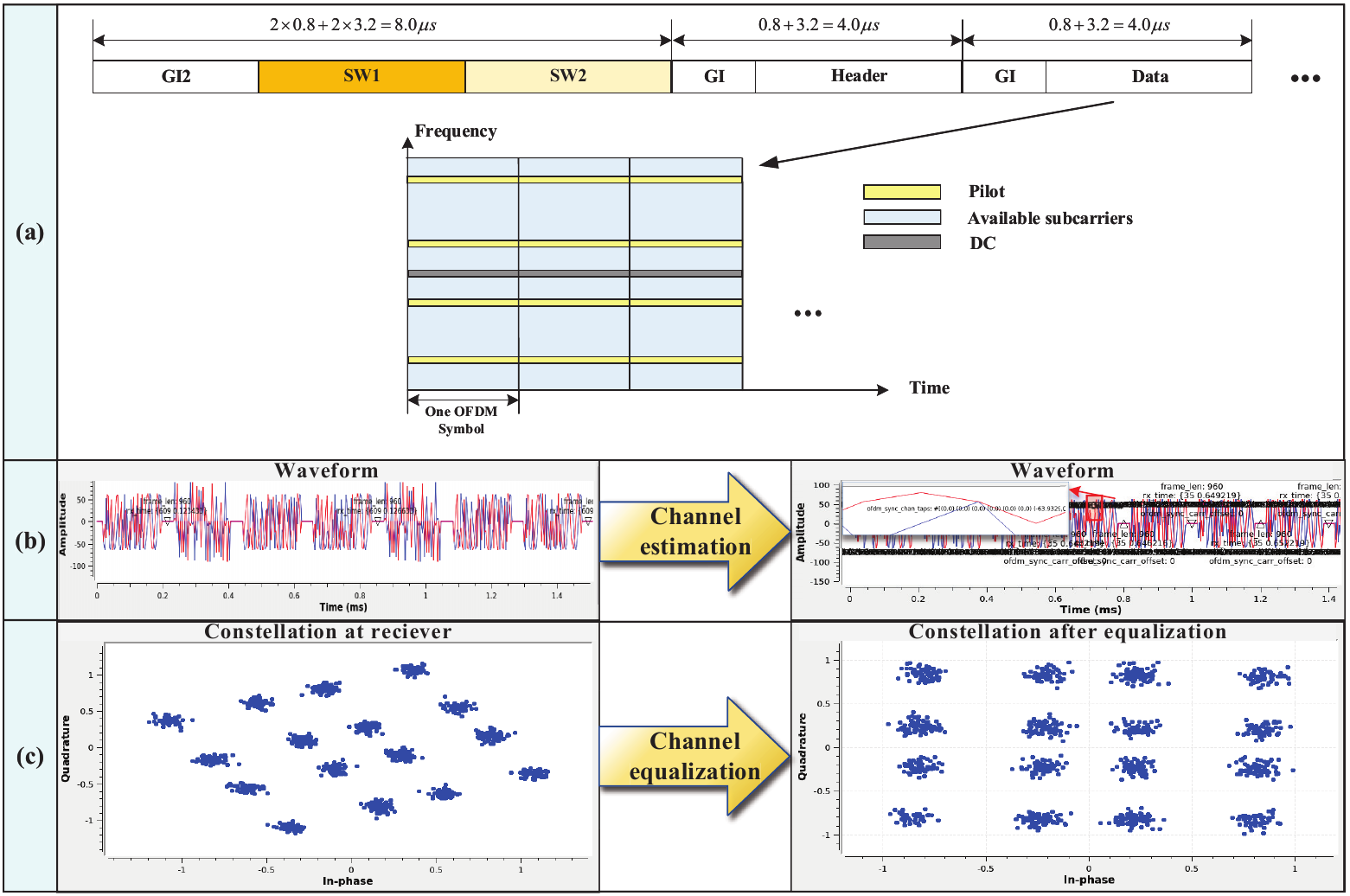}\\
  \caption{Performance of channel estimation and equalization.}\label{fig:Channel_estimation}
\end{figure*}

\begin{figure*}[t]
  \centering
  \includegraphics[width=5.0in]{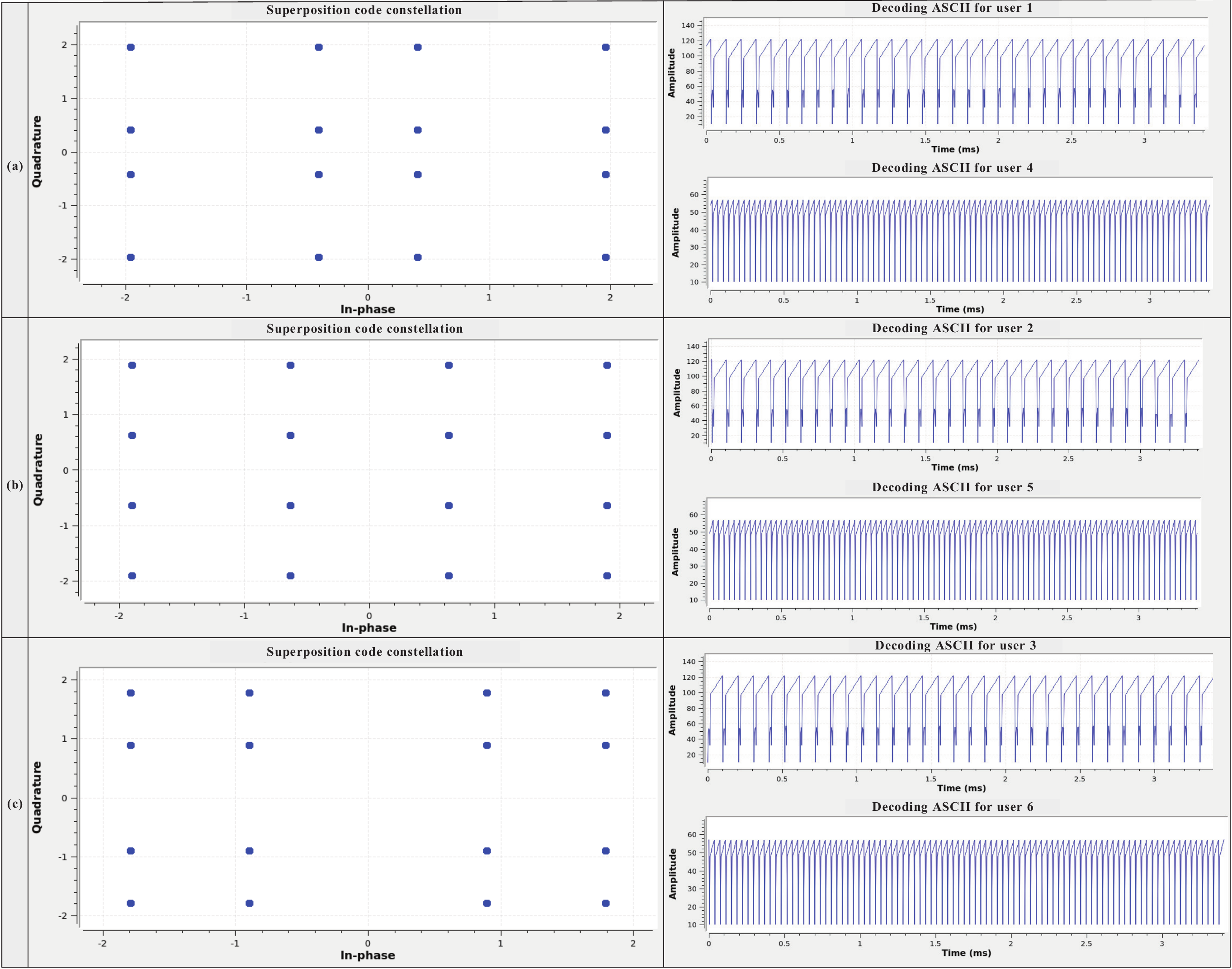}\\
  \caption{Performance of signal constellation and ASCII decoding.}\label{fig:ASCII_code}
\end{figure*}

An example of transceiver structures for our multi-carrier NOMA prototype based on orthogonal frequency division multiplexing (OFDM) is illustrated in Fig.~\ref{fig:Multicarrier_NOMA}. To reduce the complexity of the algorithm and promise low communication latency, we consider each time-frequency block is allocated to one NOMA signal that contains two data streams. All the parallel data streams are processed by an information-level progressing block consisted of synchronization, difference encoding and constellation mapping. At the receiver side, the Schmidl cox OFDM sync block is adopted for synchronization and coarse-fine frequency offset correction. Fig.~\ref{fig:NOMA_frame}(b) shows the multi-carrier NOMA frame structure. Unlike the single-carrier NOMA frame, here we use two sync words, namely SW1 and SW2, at the beginning of the frame to realize synchronization and frequency offset. The length of the sync words is equal to the number of subcarriers of one OFDM symbol, in order to prevent sync information from being mixed with data bits. After that, a header is produced to convey the information of resource allocation. The experiment results of our multi-carrier NOMA prototype are provided in Figs.~\ref{fig:SC_and_SIC}--\ref{fig:ASCII_code}. The performance of SC and SIC is shown in Figs.~\ref{fig:SC_and_SIC}(a) and (b). Recall that the success of SIC relies heavily on the CSI accuracy, Fig.~\ref{fig:Channel_estimation} provides the channel estimation and equalization results of our NOMA prototype. In the proposed design, we use both the synchronization word and pilot information in OFDM symbols to perform channel estimation based on the received signals, as shown in Fig.~\ref{fig:Channel_estimation}(a). The channel estimation results are added to the time-domain signal waveform in the form of labels, and then the amplitude and phase offset are corrected by the equalizer, shown in Figs.~\ref{fig:Channel_estimation}(b) and (c).
Furthermore, it is seen from Fig.~\ref{fig:ASCII_code} that each user has a clear signal constellation for its QPSK modulation and can decode its own signal correctly.

\subsection{PLS Techniques}

\subsubsection{Information-Theoretic Security}

The basics of information-theoretic security can be understood using the discrete
memoryless wiretap channel model introduced by Wyner \cite{Wyner_BSTJ1975}, where a legitimate transmitter Alice wants to reliably and securely communicate with a legitimate receiver Bob via the main channel, while an eavesdropper Willie tries to intercept this communication via the wiretap channel. The work in \cite{Wyner_BSTJ1975} proved that if the main channel capacity is larger than the wiretap channel capacity, perfect secure transmission from Alice to Bob can be achieved. The result was further extended to an additive white Gaussian noise (AWGN) channel in \cite{Leung-Yan-Cheong_TIT1978}, where the notion of secrecy capacity was introduced as follows
\begin{equation}
  C_s=C_m-C_w,
\end{equation}
where $C_m=\log\big(1+\frac{P}{n_m}\big)$ denotes the main channel capacity and $C_w=\log\big(1+\frac{P}{n_w}\big)$ denotes the wiretap channel capacity. $P$, $n_m$, $n_w$ are the transmit power, noise power of the main channel, and noise power of the wiretap channel, respectively. If the secrecy capacity falls below zero, i.e., $\frac{1}{n_m}<\frac{1}{n_w}$, it is impossible to achieve perfect secure transmission from Alice to Bob and Willie is capable of intercepting the confidential messages. Furthermore, by considering channel fading \cite{Gopala_TIT2008}, the secrecy capacity is defined as
\begin{equation}
  C_s=\bigg[\log\Big(1+\frac{P|h_m|^2}{n_m}\Big) -\log\Big(1+\frac{P|h_w|^2}{n_w}\Big)\bigg]^+,
\end{equation}
where $h_m$ and $h_w$ are the channel coefficients of the main and wiretap channels, and $[x]^+=\max(x,0)$. The definition of the secrecy capacity with multiple eavesdroppers can be found in \cite{Khisti_TIT2010}. Based on the definitions of secrecy capacity under different wireless channel models, it is clear that the intrinsic elements of physical channels, such as fading, noise, and interference, can be exploited to increase the secrecy capacity and guarantee secure communication. This motivates the investigations of sophisticated signal processing techniques, such as multi-antenna beamforming and cooperative communication for wireless information-theoretic security, as will be discussed in detail in Section \ref{ITS}.

\begin{figure*}[t]
  \centering
  \includegraphics[width=5.0in]{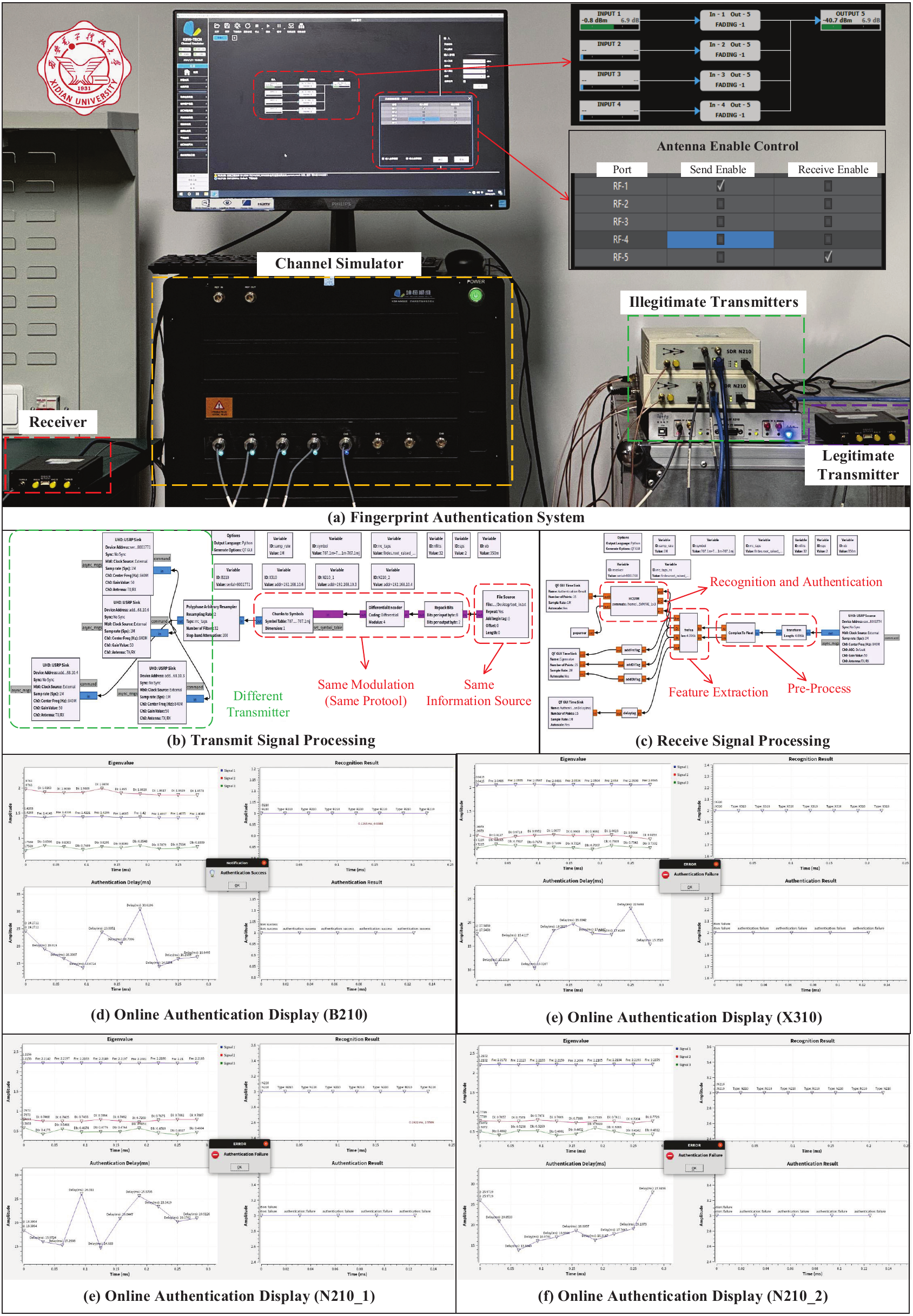}\\
  \caption{An RF fingerprint authentication prototype system.}\label{Fingerprinting prototype}
\end{figure*}

\begin{figure*}[t]
  \centering
  \includegraphics[width=6.5in]{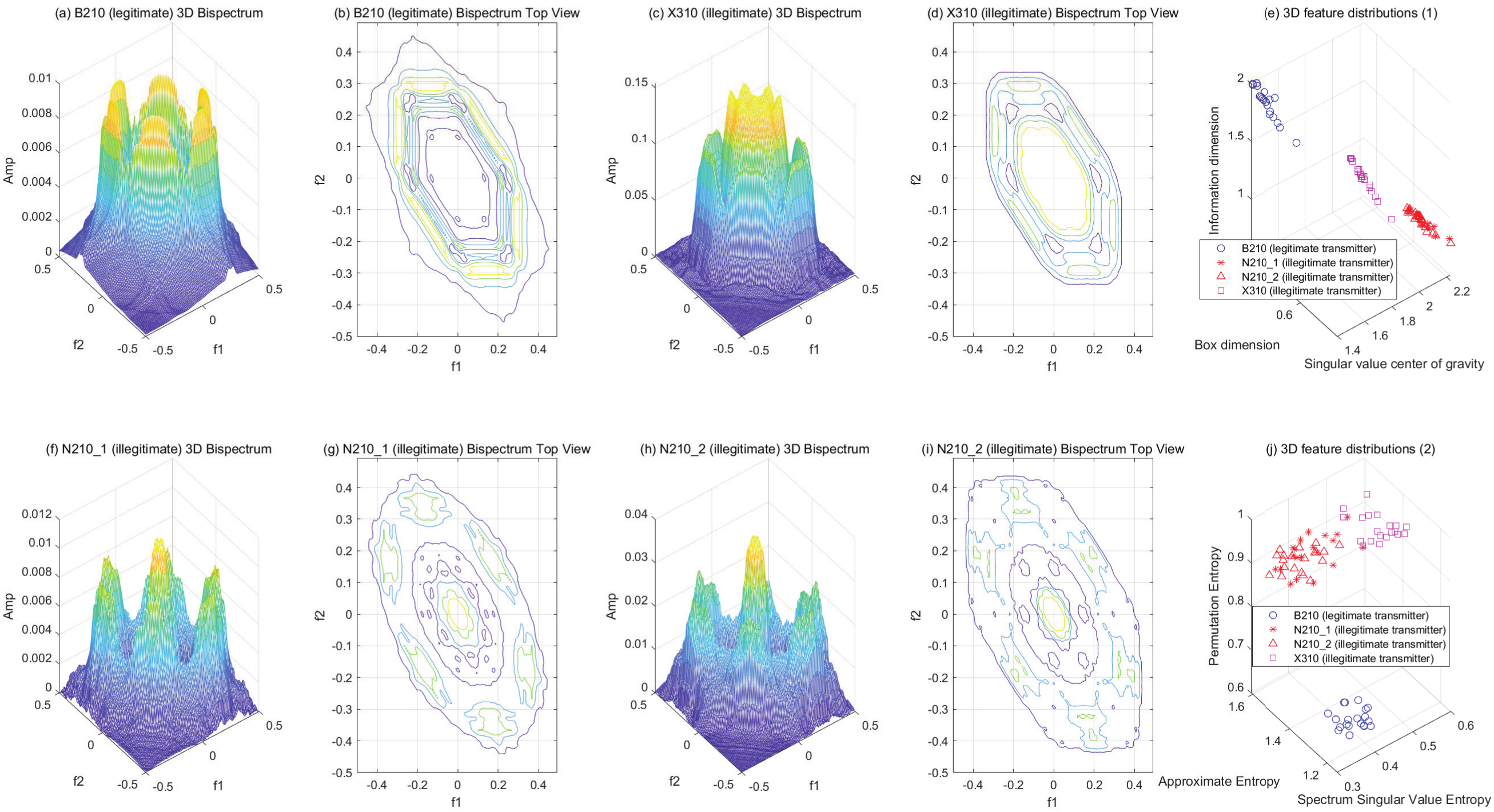}\\
  \caption{Offline experiment results of the RF fingerprint authentication prototype system.}\label{result_fingerprint}
\end{figure*}

Another critical aspect of information-theoretic security is to guarantee that the confidential signals can be decoded only by their intended receivers. This introduces physical layer authentication schemes to enable
the legitimate receiver to detect whether its received signal is forged or illegitimately modified by some malicious users other than its desired transmitter \cite{ZouYulong_PIEEE2016}. Existing efforts that use physical layer dynamics as authentication keys for the legitimate transmitter follow various approaches. One possible approach to first adopt a pre-shared secret key hidden in the modulation scheme and then detect the secret key at the receiver side \cite{Yu_TIFS2008}. On the other hand, wireless fingerprinting is a low complexity and keyless-transmitter based approach, where device specific non-ideal transmission parameters are extracted from
the received signal. They are identified as characteristics of the
legitimate transmitter and then compared with those from previous
authenticated signals \cite{Daniels_2005,Faria_ACM2006}. More recently, our research group at Xidian University has developed an RF fingerprint authentication prototype system as shown in Fig.~\ref{Fingerprinting prototype}(a), where we use two USRP-B210s to serve the legitimate transmitter and receiver, respectively, and two USRP-N210s and one USRP-X310 to serve as three illegitimate transmitters.
The received signal is fingerprint authenticated on the GNU Radio platform to identify the camouflaged signal of any illegitimate transmitter at the electromagnetic signal level. Fig.~\ref{Fingerprinting prototype}(b) shows the transmit signal processing, where both the legitimate and illegitimate transmitters adopt the same signal modulation and transmission protocol. Fig.~\ref{Fingerprinting prototype}(c) shows the legitimate receiver architecture, where the USRP-B210 performs feature extraction and identification authentication based on its received signals, and then obtains the final authentication results. The online authentication results are illustrated in Fig.~\ref{Fingerprinting prototype}(d) to Fig.~\ref{Fingerprinting prototype}(e), which respectively represent the performance of the eigenvalue curve, the device type, the authentication delay, and the final authentication result. Moreover, the offline authentication results are provided in Fig.~\ref{result_fingerprint}, where Figs.~\ref{result_fingerprint}(a), (c), (f), and (h) are the 3D bispectrum diagrams for USRP-B210, -X310, -N210$\_$1, and -N210$\_$2, respectively, and Figs.~\ref{result_fingerprint}(b), (d), (g), and (i) are the corresponding top views of their 3D bispectrum diagrams. From these figures, it is clear that there exists a significant difference in the 3D bispectrum diagrams between the legitimate and illegitimate transmitters, i.e., the 3D bispetrum is a concave shape for the legitimate transmitter whilst convex shapes for the illegitimate transmitters. More specifically, although the 3D bispectrum shapes of USRP-N210 and -X310 are similar, their top views of the 3D bispectrum are quite different. Figs.~\ref{result_fingerprint}(e) and (j) illustrate the 3D feature distributions. It can be seen that the feature points of the same type of transmitters show obvious aggregation characteristics, while the feature points of different types of transmitters exhibit evident clustering characteristics.

\subsubsection{Quantum-Safe Security}

As quantum computing power grows exponentially, theoretical mathematics-based encryption algorithms could no longer be secure and could be cracked within seconds, which would pose a serious threat to network security. These existing public key cryptosystems have the weakness of being attacked by quantum computing, and will become a recognized fact in the cryptography industry, information security industry and quantum computing experts. At present, the most affected by quantum computing attacks is the public key cryptosystem built on computational complexity such as integer decomposition and discrete logarithm, including RSA, DSA, DH, ECDH, ECDSA and other variants based on these passwords. Both security protocols and systems based on the above public key passwords will be threatened by quantum computing. However, at present, most cryptographic systems need to use these types of cryptographic algorithms, so it is very important to study the quantum security technology to replace these cryptographic algorithms to resist quantum attacks, which has become an urgent need.

{\bf Post-Quantum Cryptography:}
The ways to realize quantum security are mainly divided into two categories: one is the classical cryptographic algorithm that can resist known quantum computing attacks. The security of this cryptographic algorithm also depends on the computational complexity. This kind of algorithm or protocol is usually called anti-quantum computing cryptographic (Quantum Resist Cryptography, QRC) or post-quantum cryptographic (Post Quantum Cryptography, PQC).

Post-quantum cryptography is a type of cryptography that assumes an attacker has a large quantum computer. This field of research has had some success in identifying mathematical operations for which quantum algorithms do not have a speed advantage and building cryptographic systems around them. The central challenge facing post-quantum cryptography is to meet cryptographic usability and flexibility requirements without sacrificing trustworthiness. Based on the classification of the underlying mathematical difficulty problem, there are five main technical routes to study post-quantum cryptographic algorithms, including lattice-based ciphers, coding-based ciphers, multivariate-based ciphers, hash-function-based signatures, and curve homology-based ciphers.

(1) Lattice-based post-quantum cryptography

The lattice is a discrete subgroup in a linear space over the real number field that can be expressed as a linear combination of integer coefficients of a set of vectors. Given a set of linearly independent vectors  ${{\mathbf{b}}_{1}},{{\mathbf{b}}_{2}},\cdots ,{{\mathbf{b}}_{n}}$, the lattice they generate is the set  $\left\{ {{z}_{1}}{{\mathbf{b}}_{1}}+{{z}_{2}}{{\mathbf{b}}_{2}}+\cdots +{{z}_{n}}{{\mathbf{b}}_{n}}\mid {{z}_{1}},{{z}_{2}},\cdots ,{{z}_{n}}\in \mathbb{Z} \right\}$, where n is called the dimension of the lattice, ${{\mathbf{b}}_{1}},{{\mathbf{b}}_{2}},\cdots ,{{\mathbf{b}}_{n}}$ are called a set of bases of this lattice. An n-dimensional lattice has many lattice bases, and any two sets of its lattice bases differ from each other by an nth-order youngest mode matrix. Computing the shortest nonzero vectors in a lattice is an important and difficult problem , and is the security cornerstone of contemporary lattice ciphers. The Short Integer Solution (SIS) problem proposed by Ajtai \cite{X1} and the Learning with Errors (LWE) problem proposed by Regev \cite{X2} opened the way for practical provably secure lattice cipher research.

The current practical and secure encryption and decryption lattice system \cite{X3} and digital signature lattice cryptosystem are designed based on the above two problems. On the one hand, lattice cryptography has the above difficult problems as the security basis of the theoretical specification. On the other hand, the space and time resource consumption of lattice cryptography is moderate. Furthermore, lattice-based cryptography can be used to design advanced cryptographic application algorithms such as attribute encryption \cite{X5} and homomorphic encryption.

(2) Coding-based post-quantum cryptography

The encoded information is transmitted on the channel, and error is generated due to noise, which is recovered by decoding algorithm at the receiving end. The theoretical basis of encoding-based cryptography is that decoding random linear codes is a difficult problem. In 1978, McEliece used Goppa code to design a public-key encryption scheme \cite{X9}. The generation matrix $G$ of [n,k,t]-Goppa code was used as the core, and invertible matrix $S$ and random permutation matrix $P$ were applied on the left and right sides respectively to cover $G$. The private key of the scheme consisted of matrices $S$, $G$, and $P$. The public key contains the matrix ${{G}^{\prime }}=SGP$ and the error correction capability $t$ of the Goppa code. When encrypting, the plaintext is encoded into a vector $m$, and the ciphertext $c=m{{G}^{\prime }}+e$ is calculated, where $e$ is a random vector weighing no more than $t$. During decryption, $c{{P}^{-1}}=\mathbf{m}{{G}^{\prime }}{{P}^{-1}}+e{{P}^{-1}}$ is calculated and the decoding operation is applied. In recent years, code-based cryptography has considered using rank distance codes and polarization codes \cite{X10}.

Encoding-based cryptography has its own advantages, but also faces challenges. On the one hand, compared with the existing public key cryptography algorithms, the encoding based cryptography has the characteristics of faster encryption. On the other hand, the public key size of the cryptographic algorithm based on encoding is very large, which affects the application field of the algorithm. In addition, the situation of NIST post-quantum standardization shows that \cite{X11}, the development of codes-based encryption algorithms is good, but the design of a secure and efficient practical signature system based on coding is still a challenging research work.

\begin{figure*}[t]
	\centering
	\includegraphics[width=5.0in]{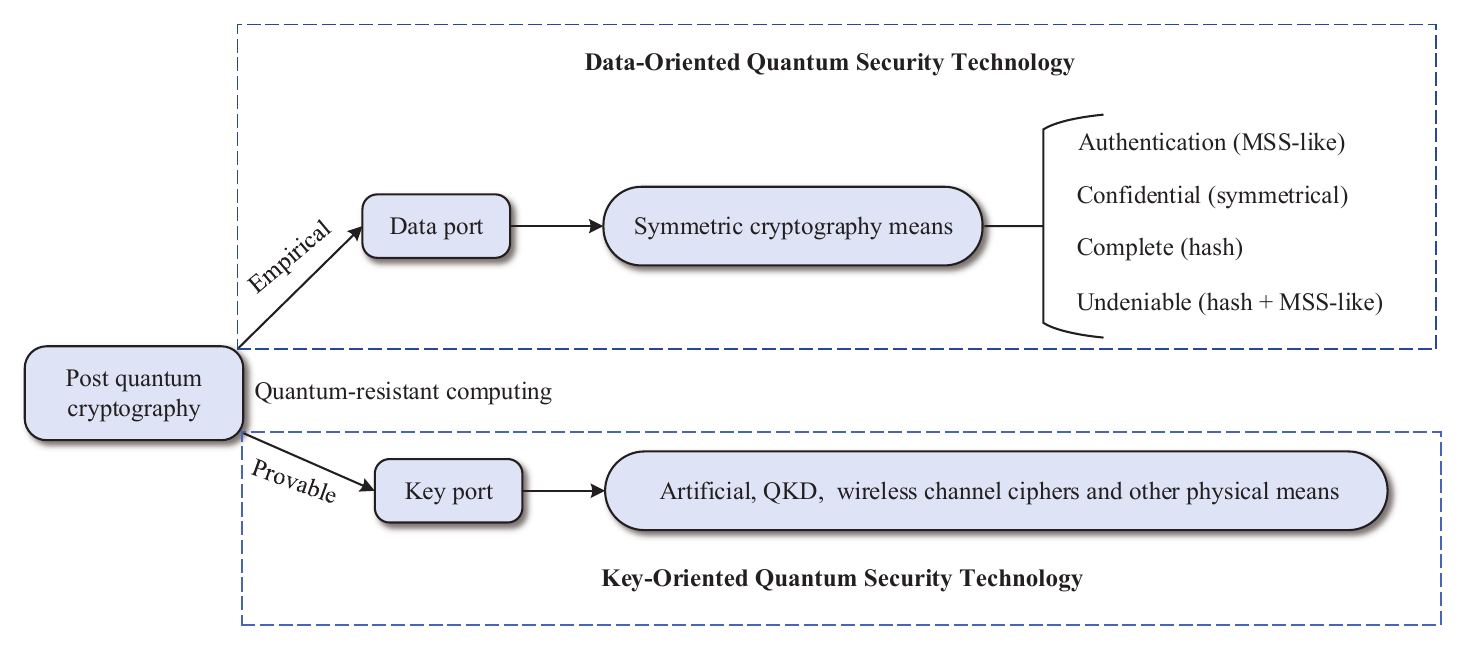}
	\caption{Post quantum security technology system.}
	\label{fig1}
\end{figure*}

(3) Multivariate-based post-quantum encryption

The security basis of multivariable based cryptography comes from the fact that solving higher order multivariable equations is a NP problem. The mapping $\text{F}$ of higher dimensional space over finite field is constructed, so that its inverse operation is easy. Introduce two linear (or affine) mappings $\text{S}$, $\text{T}$ on the left and right ends to mask $\text{F}$. The private key consists of mappings $\text{S}$, $\text{F}$, $\text{T}$, and the public key is their composite mapping $P=S{}^\circ F{}^\circ T$, which behaves like a random mapping. Representative cryptographic algorithms based on multiple variables include GeMSS signature scheme of HFEV-type \cite{X12} and Rainbow signature algorithm of UOV type \cite{X13}, both of which have entered the third round of NIST post-quantum standardization selection. However, the minimal rank distance attack proposed by Tao et al. \cite{X14,X15} in 2020 greatly reduces the security of GeMSS signature algorithm, followed by the attack algorithm of Rainbow signature algorithm proposed by Beullens \cite{X16} in 2022, which greatly improves the efficiency of key recovery attack. Therefore, both GeMSS and Rainbow algorithms failed to enter the fourth round of NIST post-quantum standardization selection.

Multivariable based cryptography has resource advantages, but its security needs to be further studied. On the one hand, multivariable based cryptography has higher computing speed and smaller signature size. On the other hand, similar to coded cryptography, the public key size of the multivariable based cryptography algorithm is also large, and the practicality of the algorithm is limited. In addition, the situation of NIST post-quantum standardization shows that \cite{X3,X11}, it is a challenging research work to design a secure and efficient encryption scheme based on multivariables.

(4) Hash-based post-quantum encryption

Block ciphers and hash functions are cryptographic components with well-studied and well-standardized objects \cite{X18,X19}. The design of quantum cryptography algorithm based on symmetric cryptography has a good tool foundation. In the signature system based on hash function, the one-time signature proposed by Lamport \cite{X20} is used as the leaf node, and the digital signature scheme is constructed by Merkle tree structure . Representative algorithms include XMSS \cite{X21} and SPHINCS+\cite{X22}. The former is a stateful signature and the latter is a stateless signature. Both algorithms have made important progress in post-quantum standardization.

In the post-quantum standardization of NIST, Picnic, a signature scheme based on block cryptography and non-interactive zero-knowledge proof, has the characteristics of high efficiency of symmetric components used, strong modular design, small public key size, large signature size, and fast signature speed.

(5) Isogeny-based cryptography

Isogenies are mappings between two elliptic curves that maintain group homomorphism. A homologous expression is often as follows: Given two elliptic curves ${{E}_{1}}$ and ${{E}_{2}}$ defined in ${{\mathbb{F}}_{q}}$, denoted by the mapping $\varphi :\varphi \left( {{P}_{1}} \right)={{P}_{2}},{{P}_{1}}\in {{E}_{1}},{{P}_{2}}\in {{E}_{2}}$ (${{P}_{1}}$, ${{P}_{2}}$ are rational points on ${{E}_{1}}$ and ${{E}_{2}}$, respectively). In 2011, Jao and others \cite{X23} for the first time put forward the hypersingular homologous Diffie-Hellman problem (Supersingular Isogeny Diffie - Hellman, SIDH), and design a public-key cryptosystem based on hypersingular homologous SIKE, This algorithm was also submitted to NIST as the only homologous class candidate algorithm. The biggest difference with the other 4 candidate algorithms is that the parameter size of SIDH based algorithm is very small and the calculation speed is slow. However, since 2022, SIDH and its encryption algorithm SIKE have been subjected to the key recovery attack proposed by Castryck et al. \cite{X24}, and were finally fully deciphered.

{\bf Quantum Key Distribution:}
Another kind of quantum security password is based on the principle of quantum physics to achieve classical cryptography goal quantum password (Quantum Cryptography), one of the most representative and practical is the quantum key distribution (QKD). QKD is an implementation method of Shannon's information security theory. Quantum key distribution uses the entangled state of quantum to distribute keys. Both parties to the communication hold entangled quanta, respectively, and the quantum state can create a certain connection so that they can still change with the changes of the other party no matter how far apart they are. By randomly changing the quantum state, the communicating parties generate and share a random key by measuring the quantum state. In addition, based on the uncertainty principle or non-cloning property of quantum, it is impossible for any unknown quantum state to carry out the exact same replication process. Because the premise of replication is measurement, and measurement generally changes the state of the quantum, and the existence of the third party (and the amount he intercepts) can be detected, this ensures the privacy of the key distribution between the communicating parties. Quantum key distribution uses the entangled state and uncertainty characteristics of quantum to effectively realize a private key channel between the communicating parties, theoretically providing the possibility of one-time padding.

QKD has information theory security, which means that QKD is still safe even when the attacker has infinitely strong computing resources, which naturally includes the security of classical and quantum computing. The function of QKD is to realize the negotiation and generation of symmetric key, and the combination with symmetric cryptographic algorithm. Combined with one secret (OTP), it can realize the information theory security of information encryption, while the symmetric cryptographic algorithm combined with quantum security realizes quantum security.

{\bf Quantum-Secure Transmission at the Physical Layer:}
For improving the security of optical transmission, many physical layer encryption approaches have been proposed, such as basing on quantum noise stream cipher (QNSC), chaos, optical stealthy, quantum key distribution (QKD), optical code division multiple access (OCDMA), and etc. QNSC coming from Y-00 protocol realizes security of transmission with quantum noise masking signal, which uses both physical encryption and mathematical one without bandwidth expansion. In recent years, a lot of research work has been conducted, which focus on the security of the QNSC including the security analysis on transmission distance and security proof for high rate. Synchronous chaotic spectral phase encryption and decryption are utilized to secure high-rate optical communication \cite{X25}. The 40 Gb/s 16-QAM coherent optical chaotic security communication is demonstrated and evaluated \cite{X26}. Chaotic-based secure communication over two kinds of multi-mode fiber has been experimentally proved \cite{X28}. These have greatly promoted the development of high rate optical chaotic communication. Furthermore, the remaining gap between two adjacent communicating channels is used as a secure channel for stealthy transmission in the WDM network \cite{X29}. Security of the QKD in information theory has been proven and potentiality in practice has been excavated. In particular, the analysis of finite key security has been studied extensively. OCDMA technique has been a research hotpot all long due to one of its advantages is increasing channel secrecy. It is also compatible with existing WDM networks while protecting data transmission from eavesdropping.

There are deficiencies in the security of a single key in the face of big data and cloud processing, so limiting
attackers from intercepting ciphertexts or altering keys becomes increasingly important in secure transmission
at the physical layer. In asymmetric key encryption, it is especially critical to ensure the secure transmission
of the key. Some scholars have proposed to utilize quantum key distribution algorithms and enhance security
by masking the ciphertext with quantum noise. This noise masking phenomenon is generated by a combination
of quantum noise and amplified spontaneous emission noise, making it impossible for an attacker to intercept
the ciphertext. Therefore, the use of QKD method allows encryption of signals at the physical layer. Through
the interplay of QKD and physical layer encryption algorithms, optical physical layer encryption and quantum
key distribution are utilized to protect high-speed data transmission and achieve physical layer security for data
transmission.

The physical layer of a data transmission system usually includes two channels: the classical channel and the
quantum channel. The classical channel is used to transmit encrypted data, while the quantum channel is used to
transmit the parameters of the encrypted data. The data at the transmitting end is encrypted and encoded through
the physical layer, amplified by an amplifier, and then fed into the classical channel. QKD can be combined with
the Y-00 protocol, thus enhancing the security of high-speed data transmission at the physical layer.

Here, we use the QKD system to transmit the necessary parameters ensuring the security of data transmission in the classical channel, which is a symmetric-key data encryption system basing on quantum noise Y-00 cipher.
The encryption principle of the IM Y-00 cipher is that 2$M$ intensity levels are divided into $M$ bases, $(a_0 , a_M); (a_1 , a_{M+1});...; (a_{M-1}, a_{2M-1})$.
The two levels in each base are encoded as $(0, 1)$. Setting the base in this way, the difference between the two intensities is large enough to distinguish 0 and 1 correctly.
When the intensity difference is sufficiently small, for two adjacent bases, e.g. $a_M$ and $a_{M+1}$, the correct signal cannot be tapped by the eavesdropper due to the noise masking the signal level.
The user with the shared key knows the mapping relationship in every base and restores the original binary data by accurately judging the encrypted signal. Therefore, the higher security is achieved by this cipher rather then only mathematical encryption.

In the encryptor, a running key is generated by the PRNG from a seed key sharing in the encryptor and decryptor. The polarity of the input binary data is scrambled in the XOR operation, and a multiary digital signal is formed by the scrambled data and mapped data. An IM-based electrical signal is output by the DAC. Finally, a multilevel optical signal is produced. In the decryptor, after direct detection, the original binary signal is demodulated and restored by the reverse process using the shared seed key.

\begin{figure}[t]
  \centering
  \includegraphics[width=3.0in]{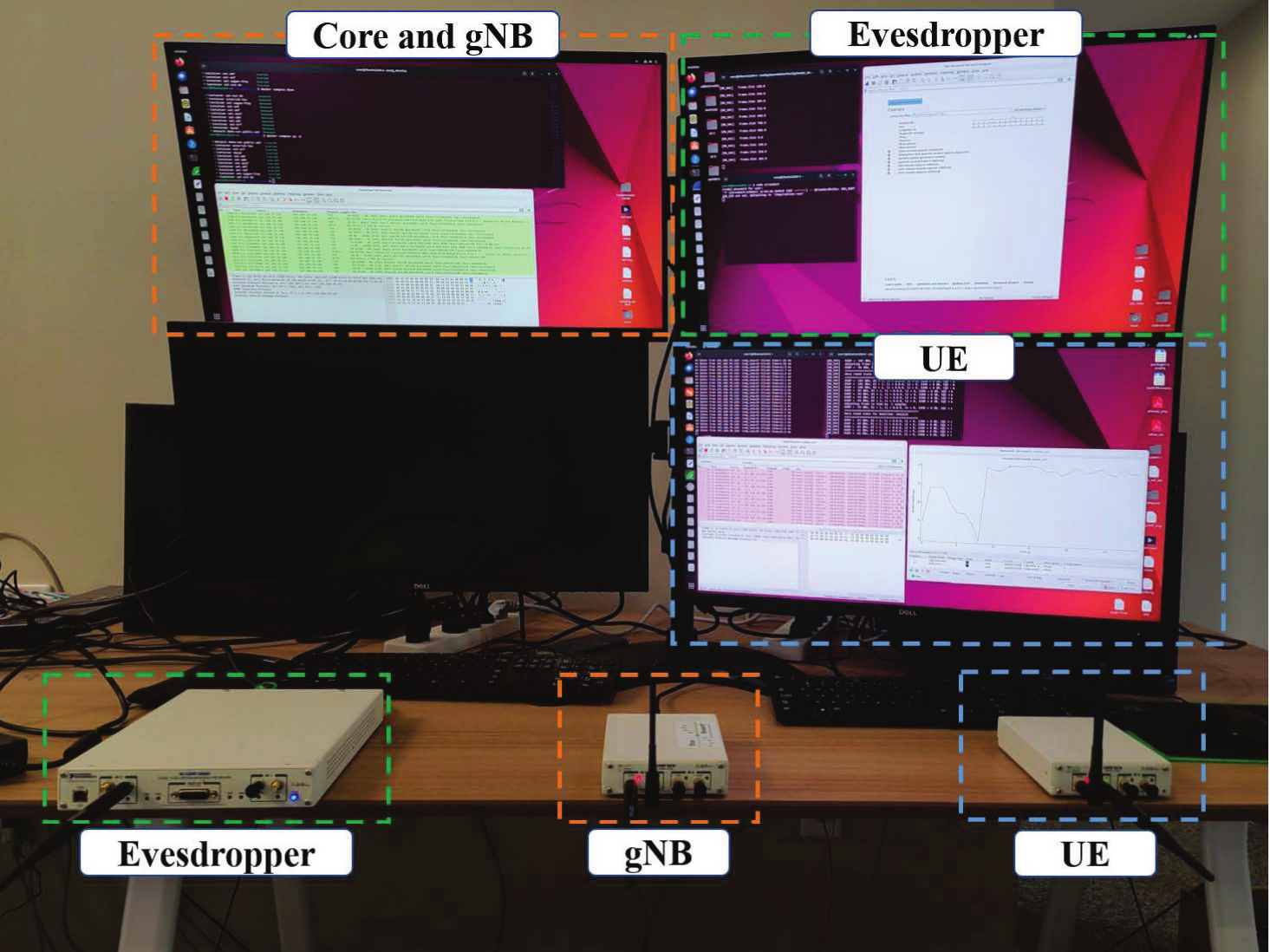}\\
  \caption{A post quantum security system.}\label{fig:Pls_demo}
\end{figure}

The semi-realistic simulation system with post quantum secure communication algorithms in a 5G local area network based on the USRP platform, developed by our Xi'an Jiaotong University group, is illustrated in Fig.~\ref{fig:Pls_demo}. In this system, a host computer constructs a mirror image of the core network elements and connects to a USRP B210, serving as the base station. Another host computer is connected to a USRP B210, functioning as the user terminal. Additionally, a third host computer is linked to a USRP 2954R, which is more powerful than USRP B210, acting as an eavesdropper. The core network elements can communicate with the base station via the host computer, while the connection between the base station and the user is established wirelessly through the USRP on the radio side. The eavesdropper operates in a passive listening mode, attempting only to receive and decipher messages transmitted by legitimate users, without engaging in feedback or active attacks. Within this system, legitimate users implement all physical layer data processing of the LTE Release 15 protocols. This system serves as a platform for experimenting and validating various physical layer security algorithms, including those related to NOMA.

\subsubsection{Low-Detection Covert Communication}

\begin{figure}[t]
	\centering
	\includegraphics[width=2.8in]{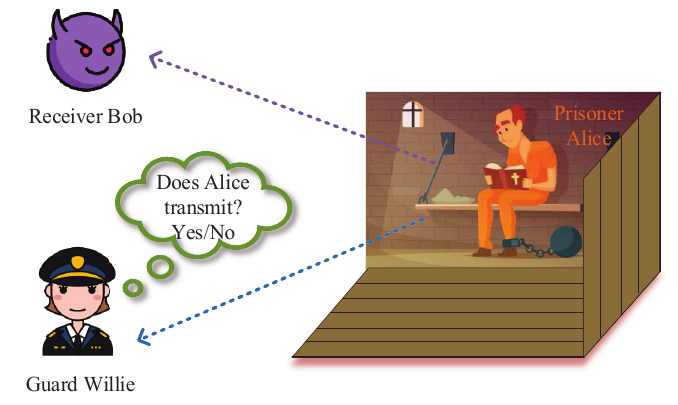}
	\caption{A typical model of prisoner's covert communication.}
	\label{fig:Covert_example}\vspace{-3mm}
\end{figure}

In wireless networks, low-detection covert communication can be vividly depicted using a prisoner problem as shown in Fig.~\ref{fig:Covert_example}, where a prisoner Alice would like to talk to her friend Bob about her escape while a guard Willie keeps monitoring the jail and tries to detect any abnormal behavior of Alice that may occur. The goal of Alice is to hide her wireless transmission to Bob or even her existence, since any reasonable suspicion about her transmission at Willie will lead to a failure of the escaping plan. This indicates that there is a competition relationship between Alice and Willie. From Willie's perspective, he is going to detect whether Alice transmits or not, regardless of the content of the transmission. While from Alice's perspective, she wants to covertly communicate with Bob subject to an extremely low probability of being detected by Willie \cite{Shihao_WCM2019}.

The notion of covert communication can be traced back to the 20th century, when the world's first international workshop on ``Information Hiding'' was held at Cambridge, UK in 1996 \cite{Anderson_1996}, and the potential of using spread spectrum for covert communication was widely discussed. In theory, spread spectrum is a technique that increases signal bandwidth far beyond the minimum necessary for data transmission, such that the power spectral density of the signal is below the noise floor, making it difficult to detect. Therefore, it is ideal for information hiding and achieves widespread use in military applications. However, the fundamental performance limit of spread spectrum cannot be explicitly analyzed. Without such performance limit analysis, it is rather difficult to know the maximum data bits that can be covertly transmitted by Alice, as well as the exact detection probability at Willie about the ongoing covert transmission. Until recently, covert communication with proven performance has received increasing attention in \cite{Bash_JSAC2013,Bloch_TIT2016,LiFeng_TIT2016, Mukherjee_CCNC2016}, where an important square root law (i.e., at most $\mathcal{O}(\sqrt{n})$ bits can be reliably transmitted from Alice to Bob subject to a detection error probability at Willie being higher than $\varepsilon$) was demonstrated rigorously under covert wireless communication.

A typical model of covert wireless communication is shown in Fig.~\ref{fig:Covert_example}, where Alice transmits her covert signal $s(i)$ to Bob with a probability of $\pi_1$, otherwise she remain silent with a probability of $\pi_0$. Here we emphasis that signaling design is of significant importance for covert communication. The Gaussian signaling is known to be optimal from a legitimate transmitter to its receiver, but is not the best choice from the aspect of covert communication since the Gaussian signaling cannot optimally hide the very existence of the secret signal from the warden \cite{Shihao_TWC2019}.
Furthermore, the transmission probability $\pi_1$ can be optimized to guarantee the low probability detection of the Willie, where the result shows that $\pi_1=\pi_0=0.5$ is not always optimal for Alice to transmit the covert signal \cite{Justin_WCL2021,Guoxin_CL2022}.
At the same time, Willie tries his best to detect whether Alice is transmitting or not. The observations at Willie can be described by two hypotheses as follows
\begin{equation}
\label{binary_test}
  \begin{cases}
    \mathcal{H}_0: y_w(i)=I(i)+n(i), \\
    \mathcal{H}_1: y_w(i)=s(i)+I(i)+n(i),
  \end{cases}
\end{equation}
where $I(i)$ is the aggregate received interference, $n(i)$ is the AWGN. $\mathcal{H}_0$ is the null hypothesis indicating that Alice is keeping silent, and $\mathcal{H}_1$ is the alternative hypothesis indicating that Alice is transmitting to Bob. In particular, Willie adopts Neyman-Pearson test based on a summation of his received signal sequence $y_w(i)$ for $i=1,\dots,n$, where $n$ is the number of channel uses, and his decision rule is described as
\begin{equation}
\label{test-rule}
  \bar{P}_w=\frac1n\sum_{i=1}^n|y_w(i)|^2
  \underset{\mathcal{D}_0}{\overset{\mathcal{D}_1}{\gtrless}}\gamma,
\end{equation}
where $\gamma>0$ is the preset power threshold of Willie's detector, $\mathcal{D}_1$ and $\mathcal{D}_0$ are the binary decisions in favor of $\mathcal{H}_1$ and $\mathcal{H}_0$, respectively. From a worst-case perspective of covert communication, Willie is able to optimize $\gamma$ to improve its detection performance. Owning to the unawareness between Alice and Willie, it is reasonable to assume that Willie has the optimal detection probability when designing the covert transmission strategies at Alice.
There are two types of errors for Willie's hypothesis tests, namely false alarm with probability $P_\text{FA}=\mathbb{P}(\mathcal{D}_1|\mathcal{H}_0)$ and miss detection with probability $P_\text{MD}=\mathbb{P}(\mathcal{D}_0|\mathcal{H}_1)$. Hence, the performance of Willie's hypothesis test can be measured by the detection error probability $\xi$, defined as
\begin{align}
\label{DEP}
  \xi&=P_\text{FA}+P_\text{MD}=1-\mathcal{V}_T(\mathbb{P}_0,\mathbb{P}_1)\nonumber\\
  &\geq1-\sqrt{\frac12\mathcal{D}(\mathbb{P}_0\|\mathbb{P}_1)},
\end{align}
where $\mathbb{P}_1$ denotes the probability of Willie's observation when Alice is transmitting, and $\mathbb{P}_0$ represents the probability of Willie's observation when Alice remains silent. $\mathcal{V}_T(\mathbb{P}_0,\mathbb{P}_1)$ is the total variation distance of $\mathbb{P}_0$ and $\mathbb{P}_1$. By applying the Pinsker's inequality, the last equality in \eqref{DEP} is obtained. $\mathcal{D}(\mathbb{P}_0\|\mathbb{P}_1)$ is the Kullback-Leibler divergence, also known as relative entropy, from $\mathbb{P}_0$ to $\mathbb{P}_1$. Here $\xi=0$ implies that Willie can perfectly detect the covert signal without error, while $\xi=1$ implies that Willie cannot detect the covert signal and his behavior is equivalent to a random guess. Based on \eqref{test-rule} and \eqref{DEP}, it is know that one need to make the received signal at Willie as random as possible to degrade the detection performance of Willie, which can be achieved by using, for example, the noise uncertainty at Willie \cite{Goeckel_CL2016,BiaoHe_CL2017, Cho_TIFS2021}, the aggregate random interference \cite{BiaoHe_TWC2018,Sobers_TWC2017}, and opportunistic transmission of Alice \cite{Tan_TIFS2019,Bash_TWC2016}, etc.

\begin{figure}[t]
	\centering
	\includegraphics[width=3.0in]{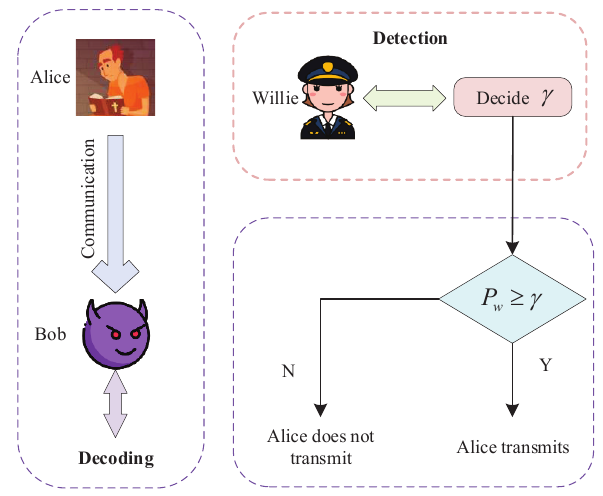}
	\caption{System model and signal processing of covert wireless communication.}
	\label{fig:Covert_model}
\end{figure}

\begin{figure*}[t]
  \centering
  \includegraphics[width=4.5in]{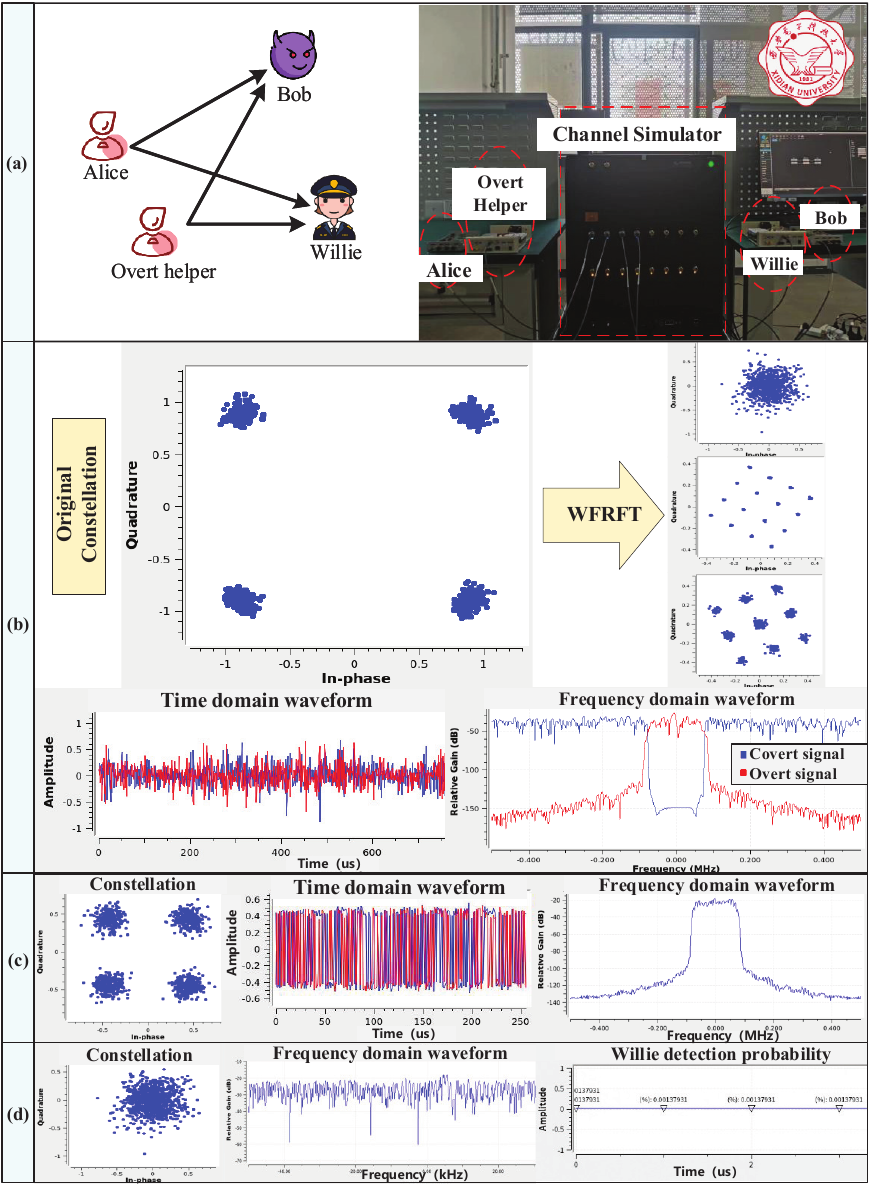}\\
  \caption{Scenario for a covert wireless communication system: (a) Model and prototype; (b)--(d) Experiment results.}\label{covert prototype}
\end{figure*}

Although many fruitful research outcomes on covert wireless communication have been discussed above, the effectiveness of the developed schemes is verified only via theoretical analysis and numerical simulations. The implementation and validation of covert transmission schemes in real-world environment are rarely reported in the literature. Recently, our research group at Xidian University has developed a covert wireless communication prototype based on the joint time-frequency, power and modulation domains scheme. The system setup is shown in Fig.~\ref{covert prototype}(a). To exploit the time- and modulation-domain resource, we propose to use the weighted fractional Fourier transform (WFRFT) technique to hide the IQ samples. The WFRFT can calculate by
\begin{align}
  F_{4\mathbb{W}}^{\alpha,\mathbf{V}}[X_0(n)]&=\omega_0(\alpha,\mathbf{V})X_0(n)
  +\omega_1(\alpha,\mathbf{V})X_1(n)\nonumber\\
  &\quad+\omega_2(\alpha,\mathbf{V})X_2(n)
  +\omega_3(\alpha,\mathbf{V})X_3(n),
\end{align}
where $X_i(n)$ is the $i$th discrete Fourier transform (DFT) and $\omega_i(\alpha,\mathbf{V})$ is the weight factor. Here, $\omega_i(\alpha,\mathbf{V})$ can be expressed as
\begin{align}
  \omega_i(\alpha,\mathbf{V})&=\frac1M\sum_{k=0}^{M-1}\mathrm{exp}\bigg\{\pm\frac{2\pi j}{M} \Big[(Mm_k+1)\nonumber\\
  &\quad\times\alpha(k+Mn_k)-ik\Big]\bigg\},
\end{align}
where $\alpha$ is the order of transform, $\mathbf{V}=[\mathbf{MV},\mathbf{NV}]$ is the scale factor, and $m_k$ and $n_k$ are the $k$th elements of $\mathbf{MV}=[m_0,m_1,\dots,m_{M-1}]$ and $\mathbf{NV}=[n_0,n_1,\dots,n_{M-1}]$, respectively. One can control $\alpha$ and $\mathbf{V}$ to achieve the constellation rotation and splitting of QPSK modulation.

Moreover, we consider to use the frequency-hopping technique based on the frequency domain resource, in which a pseudo-random sequence is input to the frequency synthesizer to generate the hopping carrier wide-band signal, as follows
\begin{equation}
  s(t)=\mathrm{cos}(2\pi f_it+\phi_i).
\end{equation}

\begin{figure*}[t]
  \centering
  \includegraphics[width=5.0in]{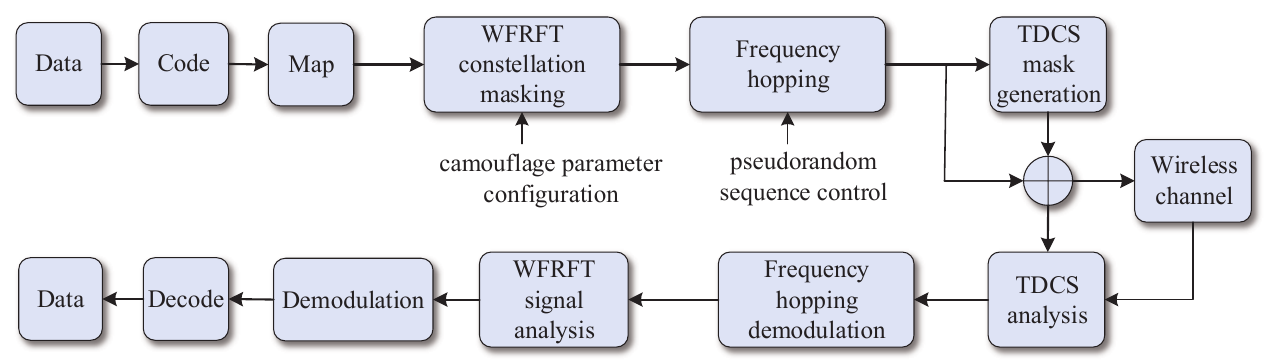}\\
  \caption{Transceiver signal processing for the developed covert communication prototype.}\label{fig:Covert_TxRx}
\end{figure*}

Finally, we apply the transform domain communication system (TDCS) technique to adapt to a fast changing electromagnetic environment and achieve the power-domain covert communication. Fig.~\ref{fig:Covert_TxRx} describes the detailed transceiver signal processing of our developed prototype. The experiment results are provided in Figs.~\ref{covert prototype}(b)--(d). The red line represents the spectrum of the covert messages that can be easily discovered by spectrum and energy detection. The blue line is the power mask signal through our proposed scheme. After signal superposition of TDCS, we obtain a wide-band signal with an even distribution power that can be hardly detected by Willie.

\subsection{Conflicting Goals of NOMA with Secrecy Considerations}

As stated in the paper, NOMA transmission requires sophisticated SIC at the receiving end because of superposed transmission. The latter is the prerequisite for increasing the critical spectral efficiency in wireless communications with limited spectrum resources. This approach was first proposed for the early-stage code division multiple access (CDMA) system especially when a short data message is sent with low duty cycle, i.e., intermittently \cite{DongIn_TCOM2001,DongIn_TCOM2003}. In this case, the power control overhead can be avoided for fast access, unlike a constant speech frame which requires the initial warm-up (i.e., power control) time for connection. The former does not require the power control to normalize the received signal power at the AP with wide coverage for optimal multi-user operation. This is because the power-domain user differentiation in fact exploits the varying received power levels that result from the random location of contending users in a serving area around the AP. Based on these observations, it is clear that the NOMA approach induces multi-user interference on top of the broadcast nature of wireless transmission. Consequently, the security issue becomes more prominent because of the signaling characteristics that also exploit the propagation channel characteristics, such as the distance dependent path loss with shadowing and multipath fading, leading to fluctuating power levels to facilitate the SIC.

It should be noted that the early-stage NOMA (i.e., CDMA) was designed to rely on the processing gain via wideband NOMA signaling where the pseudo-noise (PN) sequences are used for user-specific encoding. In particular, each user spreads its signal over the entire bandwidth, such that when demodulating any specific user's signal, other users' signals appear as pseudo white noise.
Hence, the processing gain provides the advantages of low probability of interception and anti-jamming capability while suppressing multi-user interference. However, the narrowband NOMA, namely with no encoding via PN sequences, appears to be more vulnerable to intentional jamming (e.g., cyberattacks). The latter requires more protection against cyberattacks to ensure robust narrowband NOMA transmission, compared to the early-stage wideband NOMA transmission.

In this context, the design philosophy of secure NOMA systems can be broken down into two conflicting goals: 1) the first goal is to increase the spectral efficiency (i.e., data rate) via narrowband NOMA signaling, and 2) the second goal is to add more protection against cyberattacks for robust information transmission. The two goals are conflicting because the first goal reduces the required transmission bandwidth, while the second goal increases the bandwidth to strengthen information encryption.

For clear understanding, the overall performance of secure NOMA transmission will be evaluated in various wireless environments, in which the large-scale antenna arrays, such as active and/or passive, are employed in resource-constrained wireless networks (e.g., AI-native wireless networks) for offering the increased signal dimension, either on-site AP/relays with backhaul connection, or in distributed way (e.g., user cooperation), in subsequent sections.

\section{Information-Theoretic Security for NOMA}
\label{ITS}

As we discussed in Section \ref{Basics}.B, the information-theoretic secrecy performance is mainly determined by the capacity difference between the main channel and the wiretap channel. In order to enhance secrecy, one should improve the strength of the main channel while simultaneously degrading the strength of the wiretap channel. However, this cannot always be guaranteed in wireless communications, especially in NOMA networks. Due to the geographical distribution of different NOMA users, it is possible that interception distances between the transmitter and the eavesdropper are shorter than those between the transmitter and some NOMA users. For these NOMA users, the secrecy capacities are typically zero and secure transmissions are impossible. In such a context, multi-antenna techniques and cooperative communication are regarded as two promising methods to enhance the wireless security of NOMA. Furthermore, dynamic resource management is crucial for realizing secrecy performance gains of NOMA, which can be handled by utilizing advanced mathematical optimization tools, e.g., ML.
In this section, we review state-of-the-art research results on multi-antenna and cooperation techniques for secure NOMA communications, as well as the ML tools for resource allocation with secrecy considerations.

\begin{table*}[t]
\caption{Summary of Multi-Antenna Techniques for NOMA Secrecy}
\centering
\small
\begin{tabular}{|l"c"c"c"c"c|}
\Xhline{1.2pt}
 \makecell[c]{\bf Technique}&{\bf Characteristics}&\makecell[c]{\bf Number of \\ \bf Antennas}&\makecell[c]{\bf Power \\ \bf Consumption}&\makecell[c]{\bf Hardware  \\ \bf Cost}&{\bf Complexity}\\
\Xhline{1.0pt}
\makecell[c]{Secure\\ Precoding}&\makecell[c]{Adaptively change the\\ signal direction and\\ strength via beamforming}&Medium&Medium&Medium&Medium\\
\Xhline{0.5pt}
\makecell[c]{Jamming-Aided \\Beamforming}&\makecell[c]{Coordinately send jamming \\ signal to interfere with\\ eavesdropper deliberately}&Medium&High&Medium&Medium\\
\Xhline{0.5pt}
\makecell[c]{Near-Field \\Beamfousing}&\makecell[c]{Generate pencil beams based\\ on spherical-wavefront \\to reduce signal leakage}&\makecell[c]{Extremely\\large}&Higher&Higher&Difficult\\
\Xhline{0.5pt}
\makecell[c]{Hybrid \\Beamforming}&\makecell[c]{Integration of digital\\ and analog beamforming\\ to create narrow beams}&\makecell[c]{Very large}&Higher&High&Difficult\\
\Xhline{0.5pt}
\makecell[c]{Directional \\Modulation}&\makecell[c]{Modulate phase/frequency \\ to generate constellation \\ distortion at eavesdropper}&Large&High&High&Difficult\\
\Xhline{0.5pt}
\makecell[c]{Antenna \\Selection}&\makecell[c]{Select the best antenna\\ among all antennas \\for signal transmission}&Large&Low&Low&Simple\\
\hline
\end{tabular}
\label{table-multiantenna}
\end{table*}

\subsection{Multi-Antenna Techniques}

With spatial degrees of freedom and space-time signal processing, multi-antenna techniques are capable of improving the secrecy capacity by simultaneously increasing the main channel capacity while decreasing the wiretap channel capacity. Hence, an appropriate design of transmit beamformer or signal processing is necessary, which is categorized as secure precoding, jamming-aided beamforming, near-field beamfocusing, hybrid beamforming, directional modulation, and antenna selection, as summarized in Table~\ref{table-multiantenna}.

\subsubsection{Secure Precoding}

Consider a special scenario where the channels of the legitimate NOMA users are orthogonal to that of the eavesdropper, one can simply direct the confidential signals towards the legitimate users without information leakage to the eavesdropper. However, this is an ideal scenario and serves as a performance upper bound for secure precoding. In practical NOMA networks, not all the legitimate users' channels are orthogonal to the eavesdropper's channel. In this case, zero-forcing (ZF) precoding can be applied at the transmitter, which sends the confidential signals in the null space of the eavesdropper's channel such that the eavesdropper cannot overhear any information \cite{Quanzhong_TVT2017,Alouini_ICC2011}. Note that the ZF precoding works well when the interference dominates the AWGN, while at low and medium signal-to-noise (SNR) regimes, the noise will degrade the performance significantly. This requires the regularized channel inversion precoding to balance the interference and noise power \cite{Schober_TWC2016}. Furthermore,
as can be observed from Fig.~\ref{fig:beamforming}(a), the ZF precoding is not optimal for maximizing the secrecy capacity, because purely nulling out the signal in the eavesdropper's channel may lead to direction mismatches in the legitimate channels, and hence sacrifices spatial degrees of freedom for improving the quality of legitimate channels. In addition, the ZF precoding requires the number of antennas at the transmitter larger than that at the eavesdropper. Alternatively, interference alignment (IA)-based precoding can relax the strict restriction on the number of antennas at the transmitter \cite{Nandan_CL2018}. The use of IA can ensure that the aligned interference channel matrix is reshaped, such that the secret signal cancellation at the legitimate users can be avoided \cite{Hu_TIFS2023}, which balances the tradeoff between improving the quality of legitimate channels and deteriorating the quality of the wiretap channel. As a further step, the secrecy capacity of NOMA can be maximized by optimizing the signaling design, such as power splitting, linear precoding, and power allocation, at the transmitter \cite{Vaezi_TSP2022}. Unfortunately, the optimal covariance matrix is challenging to obtain due to the non-convex objective function of the secrecy capacity, where advanced machine learning (ML) algorithms can be used to efficiently find the suboptimal precoding and power allocation matrices \cite{Vaezi_WCL2022}.

\subsubsection{Jamming-Aided Beamforming}

\begin{figure*}[t]
  \centering
  \includegraphics[width=6.6in]{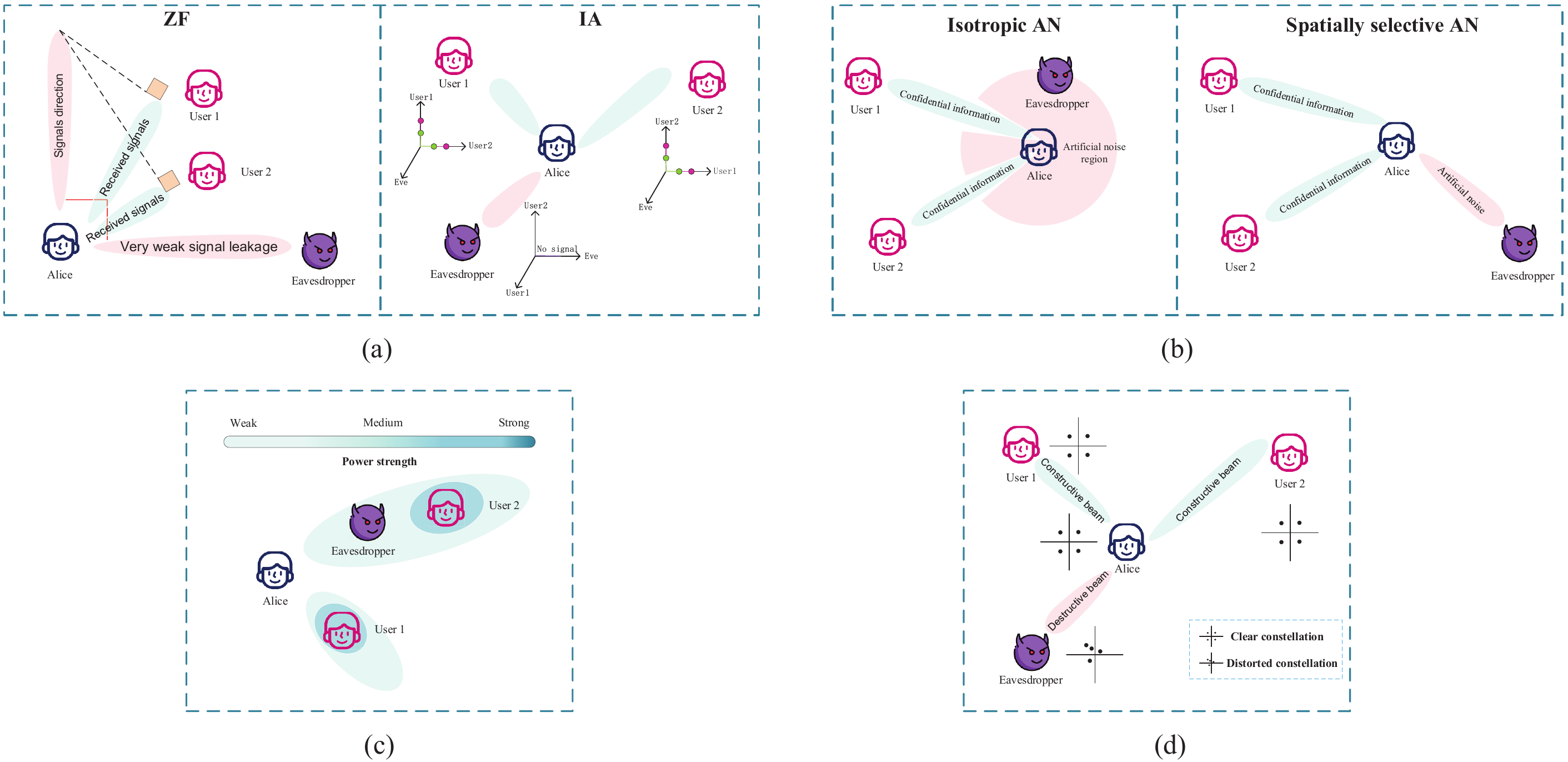}\\
  \caption{Illustrative diagrams of multi-antenna techniques for secure NOMA transmissions: (a) Secure precoding; (b) Jamming-aided beamforming; (c) Near-field beamfocusing; (d) Directional modulation.}\label{fig:beamforming}
\end{figure*}

The above secure precoding aims to transmit the signal in the direction of the legitimate users while reducing power leakage to the eavesdropper, which, however, is not sufficient in some cases with a low spatial resolution between the users and the eavesdropper and/or a close eavesdropper's location to the transmitter. In this challenging scenario, by coordinately sending artificial jamming to interfere with the eavesdroppers deliberately, i.e., jamming-aided beamforming, is a more active approach. Among jamming signal designs, artificial noise (AN) has shown a significant potential in security enhancement \cite{Negi_TWC2008}. The availability of the eavesdropper's CSI plays an important role in the design of AN-aided beamforming, as explained next. If the eavesdropper's CSI is not available, it is optimal to uniformly spread the AN on the legitimate channels' null space and transmit the desired signals toward the NOMA users to facilitate SIC \cite{Rose_JSAC2018,Jia_CL2021,Yuanwei_TWC2017, LuLv_TVT2018}. By doing so, no interference will be received by the legitimate NOMA users and only the reception of the eavesdropper will be seriously degraded by the AN. Fig.~\ref{fig:beamforming}(b) gives an example of how this isotropic AN-aided beamforming works. When the eavesdropper's CSI is available, more spatial degrees of freedom can be exploited for AN-aided beamforming design, such that spatially selective AN, rather than keeping AN isotropic, is generated to block the eavesdropper much more effectively \cite{Wang_TCOM2020,ZhaoNan_TCOM2019}, as illustrated in Fig.~\ref{fig:beamforming}(b). The case with the known eavesdropper's CSI corresponds to scenarios where the eavesdropper is also a legitimate user of the network but is not scheduled for transmission in the current communication slot \cite{Huang_TSP2011,Ding_JSAC2012}, or an estimation of the CSI from the active
eavesdropper's transmission \cite{Mukherjee_ICASSP2012}. Nevertheless, it is usually challenging to acquire the perfect CSI of the eavesdropper, and a more general and important issue is about robust and secure AN-aided beamforming designs in the presence of imperfect eavesdropper's CSI \cite{Rose_JSAC2019,Sun_TCOM2018,Fu_TWC2020}. More specifically, if the transmitter is capable of the more advanced sensing capability, the knowledge of the eavesdropper's CSI can be obtained more easily by arranging the transmitter to perform wireless sensing first for channel estimation \cite{Yang_TCOM2022,Luo_CL2022}. It is worthy to note that the AN-aided beamforming works ineffectively if the eavesdropper is equipped with nearly twice as many antennas as that of the transmitter, because in this case interception of the eavesdropper is always successful no matter how much power is allocated to the AN \cite{Khisti_TIT2010, Huiming_TWC2015}. To overcome this challenge, the AN-aided randomized beamforming scheme was presented to corrupt the received signal at the eavesdropper by an
unknown time-varying and multiplicative noise, which prevents blind channel estimation and permits only non-coherent detection at the eavesdropper \cite{Li_JC2007,Fan_TVT2019}.
Apart from using AN, the inter-user interference inherently existing in NOMA networks can be utilized as a useful jamming source to disrupt the eavesdropper. For example, consider a task offloading scenario in edge computing, a two-slot hybrid secrecy cooperation scheme for NOMA was proposed, where the interference to the eavesdropper comes from the task signals of users at the first slot as well as the jamming signal from the user who has completed the task offloading at the second slot \cite{Li_TVT2021,Rose_TCOM2020,Xianfu_TCOM2022}.

\subsubsection{Near-Field Beamfousing}

\begin{figure}[t]
	\centering
	\includegraphics[width=3.5in]{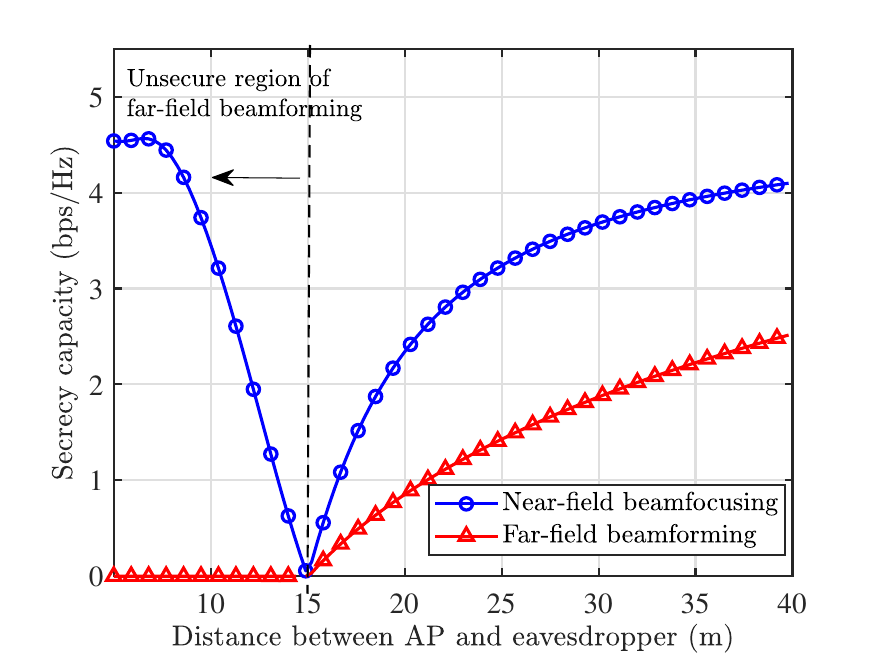}
	\caption{Secrecy capacity achieved by near-field beamfocusing and far-field beamforming. The distances from the AP to the legitimate user and the eavesdropper are 15 m and 5 m, respectively. The LoS signal propagation is considered for the wireless channels. The free space path loss model is given by $\frac{c}{4\pi f d}$, where $c=3\times10^{8} m/s$, $f = 28$ GHz, and $d$ denote the velocity of light, the carrier frequency, and the corresponding distance between the user and the AP, respectively. The transmit power is set as $P_{\mathrm{max}}=-15$ dBm.}
	\label{fig:Numerical_beamfocusing}\vspace{-3mm}
\end{figure}

Extremely large antenna array (ELAA) and high-frequency bands are two key features of the next generation mobile networks, and a combination of the two often results in the communication transceivers working in the near-field (Fresnel) region, accompanied by the spherical-wavefront transmission model rather than the plane-wavefront in the conventional far-field region \cite{Zeng_TWC2022,Zhang_TWC2022}. In the radiative near-field with ELAA, beamforming not only creates narrow beams but  also generates beams with finite depth. As such, ELAA can focus its large array gain in a particular zone, yielding a beamfocusing effect \cite{Dai_CM2023}. This helps reduce the information leakage in both the angle and location domains and diminish the eavesdropping zone, thus exhibiting a large potential to enhance the wireless security of NOMA \cite{Yuanwei_2023,Anaya_VTC2022}. Fig.~\ref{fig:beamforming}(c) shows a picture to illustrate the basic idea of near-field beamfocusing for secrecy, and the corresponding performance advantage is shown in Fig.~\ref{fig:Numerical_beamfocusing} \cite{ZhangZheng_2023}. It is observed from Fig.~\ref{fig:Numerical_beamfocusing} that a non-zero secrecy rate can be achieved even if the eavesdropper is located closer to the transmitter than the legitimate user, whereas the perfect secure communication is indeed impossible for the traditional far-field beamforming. The use of beamfocusing with ELAA lays a heavy burden on computational resources and accurate CSI acquisition, which may be prohibitively expensive in such a large-scale scenario. In this regard, low-complexity precoding schemes such as mean-angle based ZF and tensor ZF, can be applied to realize cost-effective beamfocusing and guarantee secrecy \cite{Ribeiro_ESPC2021}.

\subsubsection{Hybrid Beamforming}

It is important to point out that as the number of antennas becomes extremely large, e.g., the ELAA system, the computational complexity of fully digital beamforming (FDB) is increased significantly, and the circuit costs of the FDB structure would be too expensive for commercial applications. This motivates the emergence of the hybrid digital and analog beamforming structure, which is less expensive and with low complexity but comparable performance \cite{Yu_JSAC2017,Molisch_CM2017}. In typical hybrid beamforming architecture with secrecy considerations, a large number of economical and energy-saving analog phase shifters (PSs) are employed to implement high-dimensional analog beamformers to achieve orthogonality between legitimate and eavesdropping links, while only a few radio frequency (RF) components are utilized to realize low-dimensional digital beamformers for multiplexing or multi-streaming \cite{Tian_SJ2020}. Furthermore, for secrecy multi-user MIMO-NOMA systems, the hybrid beamforming algorithm also needs to be designed with the proper user grouping strategies \cite{CaoYang_TWC2022,Zhu_TWC2019,Dai_JSAC2019}. The major challenge of the hybrid beamformer design lies in the magnitude/phase constraints of the analog beamformer. Thus, it is necessary to incorporate hybrid precoders with low-resolution PSs and algorithms for both codebook-free and codebook-based analog beamformer designs \cite{Chang_TWC2024,LiMing_JSTSP2018}.

\subsubsection{Directional Modulation}

Directional modulation (DM) is a promising technology to guarantee information security under unknown eavesdroppers' locations. Specifically, by performing the modulation at the antenna level, a clear constellation with a low bit-error rate (BER) is generated in the desired direction, while the constellations in other undesired directions can be distorted significantly \cite{Daly_TAP2009,DingYuan_2016}, as shown in Fig.~\ref{fig:beamforming}(d). Contrary to the conventional beamforming, the modification rate of antenna array weights depends on the transmission rate instead of the dynamics of the wireless channel \cite{Alrabadi_ICC2012}. The synthesis for DM can be implemented by phased array (PA) or frequency diversity array (FDA). PA-based DM only uses the angle information of legitimate receivers. The work in \cite{Christopher_CL2020} introduced PA-based DM to a downlink MISO NOMA network, where the symbols of weak users can be distorted at the strong user and thus the secrecy rate was optimized. Theoretical results show that the DM-aided NOMA achieves a non-zero secrecy rate while achieving the sum rate close to the conventional NOMA. However, when the eavesdropper moves in the same direction as legitimate receivers, the confidential message can be intercepted. To tackle this issue, FDA-based DM is proposed, which can achieve location-specific wireless secure transmission by jointly exploiting angle and range dimensions \cite{Sammartino_TAES2013}. In practice, direction and range are typically coupled, there exists a certain number of locations for eavesdroppers to receive the same message as legitimate users. To decouple the direction and range, the random frequency direction array (RFDA) was proposed, where random frequency increments are adopted for all antenna elements \cite{Liu_JSTSP2017}. By applying the RFDA-DM in the NOMA network, the constellation of the confidential message at the external or internal eavesdropper (i.e., strong users) can be disturbed or eliminated by jointly optimizing power allocation and beamforming design. Moreover, DM can be combined with traditional baseband-level modulation to increase the sum rate with perfect security guarantee \cite{Hong_AWPL2011}.

Another constellation-shaping technique is WFRFT, which can simultaneously change the
time domain and frequency domain characteristics of signals for wireless communication security. Specifically, the signal modulated by WFRFT can achieve constellation splitting, rotation, uniform distribution of bit energy, and excellent resistance to parameter scanning by eavesdropper, hence effectively realizing secure communication without relying on various resources such as antenna and
power \cite{Mei_WFRFT2013}. The WFRFT can be categorized into single parameter (SP)-WFRFT and multiple parameter (MP)-WFRFT. The SP-WFRFT solely incorporates transform order parameters, whereas the MP-WFRFT contains both transform order and scaling vector parameters. In MP-WFRFT, the processing of time domain and frequency domain components corresponds to distinct characteristics. The time domain item corresponds to the splitting characteristics of constellations, while the frequency domain term corresponds to the characteristics of constellation noise \cite{Zhang_IoT2020}. As a consequence, the MP-WFRFT offers a more flexible camouflage design capability and is more preferable with security considerations.

\subsubsection{Antenna Selection}

In addition to the above-mentioned beamforming designs, the transmitter can also exploit the time-varying channel characteristics by selecting one of its antennas, thereby improving the secrecy performance with reduced RF chains and signal processing overheads \cite{Assad_JSAC2018}. Intuitively, if the transmitter has the global CSI of the eavesdropper, the optimal antenna selection (OAS) scheme can be designed to maximize the secrecy capacity. On the other hand, if the eavesdropper's CSI is unknown to the transmitter, a suboptimal antenna selection (SAS) scheme can be used, which only maximizes the capacity of the main channel \cite{ZhuJia_TVT2016,Lei_TVT2018}. Theoretical and numerical results in \cite{ZhuJia_TVT2016} and \cite{Lei_TVT2018} demonstrate that both the OAS and SAS schemes can achieve a higher security level than the traditional space-time transmission scheme. The best transmit antenna for security enhancement can also be selected by using an ML-based approach proposed in \cite{He_WCL2018}. In a more challenging eavesdropping scenario where multiple eavesdroppers can form collusion to enhance their interception capability via cooperation, the work in \cite{Pei_TVT2020} developed a max-min transmit antenna selection strategy to combat concluding eavesdropping. Furthermore, when the transmitter has the full-duplex capability, a portion of the antennas can be selected to transmit confidential messages, whilst the remaining antennas are employed to receive information in the same time-frequency domain. In this way, the concurrent uplink and downlink information can be regarded as beneficial jamming signals to enhance secrecy \cite{Gao_TVT2021}.

\subsection{Cooperation Techniques}

Fortunately, although the wireless channels cannot be controlled, with proper signal design and optimization, one can construct equivalent channels to increase the capacity difference between the main channels and the wiretap channel. Hence, this motivates the use of cooperation techniques to enhance wireless secrecy. In recent years, various secrecy cooperation strategies have been proposed, which are divided into two categories: active cooperation and passive cooperation. Table~\ref{table-cooperation} summarizes these methodologies.

\begin{table*}[t]
	\caption{Summary of Cooperation Techniques for NOMA Secrecy}
	\centering
	\small
\resizebox{0.9\linewidth}{!}{
\begin{tabular}{|c|c"c"c"c"c|}
	\Xhline{1.2pt} Category  & Technique & Characteristics & \begin{tabular}{c}
		CSI \\
		Requirement
	\end{tabular} & \begin{tabular}{c}
		Power \\
		Consumption
	\end{tabular} & Complexity \\
	\Xhline{1.0pt}  \multirow{10}{*}{\begin{tabular}{c}
			Active \\
			Cooperation
	\end{tabular}} & \begin{tabular}{c}
		User \\
		Cooperation
	\end{tabular} & \begin{tabular}{c}
		Exploit multi-user\\ diversity
		and/or\\ friendly jamming
	\end{tabular} & Local CSI & Low & Low \\
	\Xcline { 2 - 6 } {0.5pt}& \begin{tabular}{c}
		Single-Relay \\
		Cooperation
	\end{tabular} & \begin{tabular}{c}
		Select the best\\ relay to
		utilize\\ cooperative diversity
	\end{tabular} & Local CSI & Low & Low \\
	\Xcline { 2 - 6 } {0.5pt} & \begin{tabular}{c}
		Multi-Relay \\
		Cooperation
	\end{tabular} & \begin{tabular}{c}
		Utilize multi-path\\ gain for relaying\\
		 and/or jamming
	\end{tabular} & Full CSI & High & High \\
	\Xcline { 2 - 6 } {0.5pt} & \begin{tabular}{c}
		Untrusted Relay \\
		Cooperation
	\end{tabular} & \begin{tabular}{c}
		Employ untrusted\\ relay with
		minimal\\ information leakage
	\end{tabular} & Local CSI & Low & Low \\
	\Xcline { 2 - 6 } {0.5pt} & \begin{tabular}{c}
		Movable Relay \\
		Cooperation
	\end{tabular} & \begin{tabular}{c}
		Enable movement\\ of relays
		for favorable\\ channel conditions
	\end{tabular} & \begin{tabular}{c}
		Constantly \\
		changing CSI
	\end{tabular} & Medium & High \\
	\Xhline{1.0pt} \multirow{3}{*}{\begin{tabular}{c}
			Passive \\
			Cooperation
	\end{tabular}} & \begin{tabular}{c}
		Backscatter \\
		Communication
	\end{tabular} & \begin{tabular}{c}
		Use backscattering\\ modulation
		for low-\\power transmission
	\end{tabular} & Cascaded CSI & Very low & Low \\
	\Xcline { 2 - 6 } {0.5pt} & \begin{tabular}{c} Intelligent\\ Surface\end{tabular} & \begin{tabular}{c}
		Exploit reflection\\/refraction
		beamforming\\ and/or jamming
	\end{tabular} & Cascaded CSI & Very low & Low \\
	\hline
\end{tabular}\label{table-cooperation}
}
\end{table*}

\subsubsection{Active Cooperation Techniques}

Generally, the use of cooperative nodes or relays with active components can improve the quality of the equivalent main channels and/or degrade that of the equivalent wiretap channel, thus exhibiting great potential to secure NOMA transmissions. Significant research efforts have been made on this topic, which are detailed in the following.

\textbf{User Cooperation:}
The multi-user cooperation is an effective solution to enhance wireless security via the spatial multi-user diversity. Consider an uplink NOMA scenario with one AP, one primary user, multiple secondary users, one eavesdropper. Assume that the primary user is delay-sensitive and is allowed to solely occupy a resource block. Each secondary user competes with other secondary users and only one of them is granted to access the resource block occupied by the primary user. To ensure the QoS of the primary user, the AP sets a threshold $\tau$ and the secondary user whose resultant SNR is below $\tau$ constitutes a candidate secondary user group. Without the CSI of the eavesdropper, scheduling the secondary user with the highest SNR to the AP in the candidate group can provide diversity gains for secure NOMA transmissions \cite{Kunrui_IoT2022}. If the CSI of the eavesdropper is available, it is intuitive to schedule the user with the highest SNR to the AP and the lowest SNR to the eavesdropper for the secrecy capacity maximization \cite{Lei_TWC2024}. However, when the eavesdropper's channel is stronger than all the NOMA users' channels, relying only on user scheduling cannot guarantee a non-zero secrecy capacity. In this context, it is meaningful to schedule another user serving as a jammer to transmit AN to protect the NOMA transmissions. In \cite{Kunrui_TCOM2020}, a joint design of user and jammer scheduling in NOMA was proposed, where the user that minimizes the received signal quality at the eavesdropper is selected from the idle users as a friendly jammer. Specifically, the idle users compete with each other to serve as a friendly jammer for the sake of obtaining a high priority of transmission in subsequent time slots. In fact, it is a non-trivial task to encourage idle uses to participate in jamming as a friendly jammer, since
the transmission of AN leads to additional energy consumption at the jammer, which may dampen users' enthusiasm for the jamming services. Leveraging wireless power transfer to feed the jamming power provides a good incentive to NOMA users to act as friendly jammers \cite{Kunrui_TIFS2021}. However, this will increase the possibility of creating extra interference to legitimate communication users due to the simultaneous transmission of information and energy signals, which deserves an elaborated design to balance the energy efficiency, reliability, and security performance in wireless powered NOMA systems \cite{Rose_JSAC2018}.

\begin{figure*}[t]
  \centering
  \includegraphics[width=5.6in]{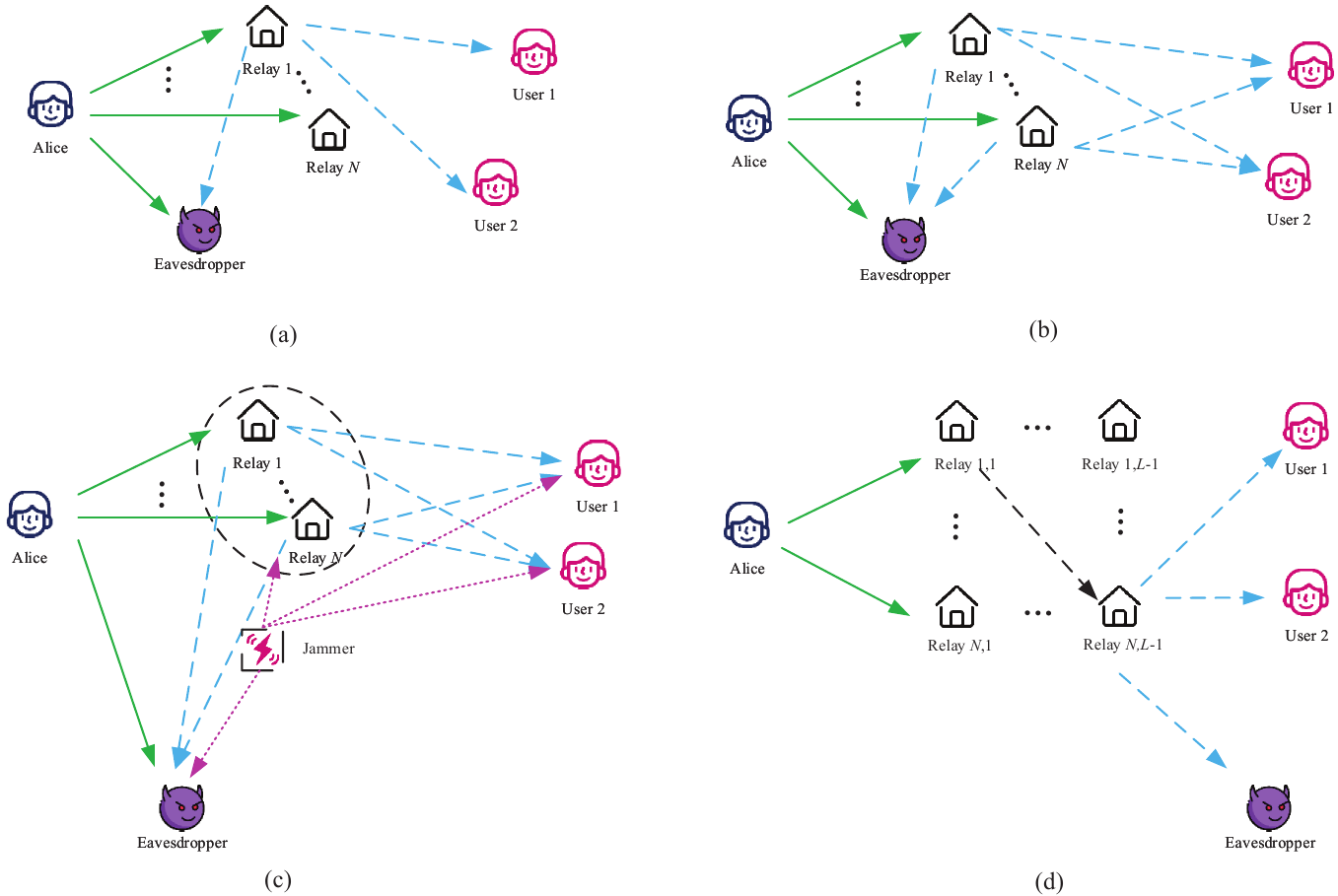}\\
  \caption{Illustrative diagrams of single- and multi-relay cooperation techniques for secure NOMA transmissions: (a) Relay selection; (b) Distributed beamforming; (c) Coordinate relaying and jamming; (d) Multi-hop relaying.}\label{fig:relay_cooperation}
\end{figure*}

\textbf{Single-Relay Cooperation:}
Apart from user cooperation, the cooperative relaying has been considered as an effective way to wireless communication security. By choosing an appropriate relay with strong channels to the NOMA users but a weak channel to the eavesdropper, as illustrated in Fig.~\ref{fig:relay_cooperation}(a), secrecy capacity can be increased via cooperative diversity \cite{LiBin_TCOM2019,Feng_TIFS2019, Haiyang_TVT2020}. The cooperative diversity can be also combined with the multi-user diversity to further improve the secrecy performance. This performance enhancement can be demonstrated by conceiving a multi-user and multi-relay cooperative network, where the best relay and the best user pair are selected to assist the two-hop secure transmission, while the rest relays and users keep silent \cite{Fan_TCOM2014}. It is known that the performance of single-relay cooperation depends strongly on the number of available links for selection. When the number of available links is limited, the secrecy performance cannot be effectively improved. Fortunately, one can introduce data buffers at the relays to overcome this limitation. By selecting the best transmission link among all available transmitter-relay and relay-receiver links, a nearly two times diversity order compared to traditional relays without buffers is achieved and thus a lower secrecy outage probability is obtained \cite{Gaojie_TIFS2014,Gaojie_IoT2019, Xianfu_TCOM2021}. Furthermore,
if the relay is equipped with multiple antennas, relay beamforming gain can be utilized for secrecy. In \cite{Huiming_TIFS2014}, a joint beamforming design of source and relay precoding was proposed, where the transmitter adopts a generalized singular value decomposition (SVD)-based precoding to transmit the signal in the first phase, and in the second phase, the relay forwards the received signal based on the SVD precoding in the null-space of the wiretap channel. We highlight that guaranteeing secure communication with active eavesdropping \cite{Bai_IoT2023,Long_TWC2017} is a challenging task and requires more elaborate designs. Take the following active eavesdropping where the eavesdropper can adaptively choose to perform eavesdropping or jamming as an example. With the full CSI of the eavesdropper, the transmitter is able to utilize the variable rate for codeword transmission and choose the best relay to maximize the secrecy capacity. While with only the statistical eavesdropper's CSI, the transmission outage will happen if the main channel capacity is below the code transmission rate. Thus, relay selection should be designed to minimize the secrecy outage probability while at the same time ensuring that the transmission outage probability is bounded within an acceptable level \cite{Long_TWC2017}.

\textbf{Multi-Relay Cooperation:}
By extending single-relay cooperation to multi-relay cooperation, additional multi-path gain can be exploited to benefit secrecy. In general, there are three ways to improve the secrecy capacity via multi-relay cooperation: 1) {\it Distributed beamforming:} The secrecy capacity can be maximized by optimizing the distributed beamformers at the relays. As illustrated in Fig.~\ref{fig:relay_cooperation}(b), the forward signals of the relays can be designed to superimpose constructively at the legitimate users and/or null out at the eavesdropper, which increases the secrecy capacity \cite{Huiming_TSP2012,MultiRelay1}. 2) {\it Coordinate relaying and jamming:} Relays can also serve as jammers to inject AN to cover the confidential signal transmissions. In particular, the relays can be divided into two sets, i.e., relay set and jammer set. The nodes in the relay set perform distributed beamforming to enhance the legitimate signal reception quality while the nodes in the jammer set use cooperative jamming to impede the capability of the eavesdropper \cite{WangChao_TWC2015,CaoYang_TCOM2019,LiDongdong_TVT2021}, shown in Fig.~\ref{fig:relay_cooperation}(c). Such coordinate transmission requires global CSI, high synchronization (i.e., precise AN coordination), and central optimization among the relays, resulting in high complexity in implementation. As a low-complexity alternative, the relays can transmit independent AN in an uncoordinated way. The independent AN is received by both users and the eavesdropper, and thus, it is important to optimize the AN power for striking a balance between impeding the capability of the eavesdropper and affecting the reception quality of the legitimate users.
Another low-complexity approach is joint relay and jammer selection, where only the best relay in the relay set and the best jammer in the jammer set are opportunistically selected to optimize the security \cite{Hui_SPL2015,Yu_CL2019,JointRJ1,JointRJ3}.
3) {\it Multi-hop relaying:} As depicted in Fig.~\ref{fig:relay_cooperation}(d), by jointly assigning the available relays and choosing the best one for signal forwarding in each hop, the secrecy degrees of freedom are improved significantly \cite{MultiHop1,MultiHop2,MultiHop3,MultiHop4,Chen_Multihop2021}.

\textbf{Untrusted Relay Cooperation:}
In heterogeneous and tactical intelligent networks, the cooperative relay may be untrusted and try to decode the confidential information for its own concern \cite{Yener_TIT2010,Huang_TSP2013,Wang_WCL2014}. Wireless secrecy may be compromised with such an untrusted relay, and it is not clear whether the use of an untrusted relay is beneficial for NOMA security. The answer to this question can be found in \cite{Poor_TIFS2020}, where it is demonstrated that rather than treating the untrusted relay as a pure eavesdropper, utilizing the untrusted relay for cooperation is helpful to improve secrecy while cooperative jamming from NOMA users is indispensable. In cooperative jamming designs in OMA networks, the jamming signal sent by the receiver can be easily canceled by itself since the jamming signal is a copy of what the receiver previously transmits \cite{Wang_WCL2014}. However, NOMA needs extra efforts in dealing with the jamming signal design, due to the fact that the jamming signal sent by a user may be a non-trivial task to be canceled by other NOMA users, thus affecting their signal reception quality significantly \cite{Xiang_WCL2019}. To this end, it is wise to select the user who has the minimal impact on other NOMA users' signal reception to emit jamming \cite{He_CL2020}. Another viable approach is to enable jamming signal cancellation by using SIC \cite{Lu_TCOM2020}. Specifically, with a careful design of the transmission rate and jamming power allocation, the jamming signal can be only canceled by the legitimate NOMA users but not the untrusted relay. As such, a non-zero secrecy capacity is guaranteed and an improved secrecy capacity scaling law is obtained \cite{Lu_TCOM2020}. A common feature of the above works is that the direct communication link from the transmitter to NOMA users is not utilized for legitimate signal transmission. This may cause a reduction of the secrecy capacity. Fortunately, such secrecy capacity reduction can be circumvented by arranging the transmitter to send a superimposed confidential signal and jamming signal with NOMA, and designing the jamming signal to intentionally impede the untrusted relay's interception capability. Therefore, a considerable security enhancement is achieved \cite{Lu_TIFS2019}.
We refer interested readers to \cite{Lu_TCOM2020,Lu_TIFS2019} for more details. In addition, multi-user secrecy diversity gains can be explored under untrusted relay cooperation in a multi-user cooperative network \cite{He_TVT2019}.

\textbf{Movable Relay Cooperation:}
Different from the fixed relay, the movable relay, such as an unmanned aerial vehicle (UAV) that has high mobility and flexibility in positioning, can be close to the legitimate users and away from the eavesdropper for security guarantee \cite{WuQQ_WCM2019}. Compared to terrestrial networks that in general suffer from severe path loss, shadowing, and multi-path fading, the high altitude of UAVs leads to more dominant line-of-sight
(LoS) channels with the ground users, which can increase the main channel capacity and enhance secrecy capacity. Specifically, by jointly optimizing the UAV trajectory and time schedule, the secrecy capacity can be improved significantly \cite{ZhaoNan_TVT2020,UAV_Cach_TCOM2019}. However, UAVs are generally considered as energy-constrained nodes. This motivates the integration of the energy harvesting capability to UAVs. For example, in \cite{UAV_tra_TCOM2020}, the UAV employs the power-splitting scheme to simultaneously receive information and harvest energy from the source, and then exploits the time-switching protocol for information relaying. Secure transmission is promised with the help of a full-duplex receiver that simultaneously receives confidential signals and cooperatively transmits AN to confuse the eavesdropper. Besides, the strong LoS links of UAVs can also cause severe interference to the terrestrial eavesdropper. Some UAVs may act as aerial jammers to perform cooperative jamming to degrade the signal reception of the eavesdropper, thus achieving better secrecy communication performance \cite{UAV_3,UAV_2,UAV_1}.

\subsubsection{Passive Cooperation Techniques}

The active cooperation techniques require costly radio frequency (RF) chains (e.g.,
oscillators, mixers, and digital-to-analog converters)  and complex signal processing to guarantee high performance gains, which are highly spectrum and energy inefficient and become the main roadblocks to achieve green and sustainable wireless communications. This is indeed the motivation of applying passive cooperation techniques, such as backscatter communication (BC) and intelligent surfaces, to enhance the security of NOMA transmissions, as discussed next.

\begin{figure}[t]
  \centering
  \includegraphics[width=3.2in]{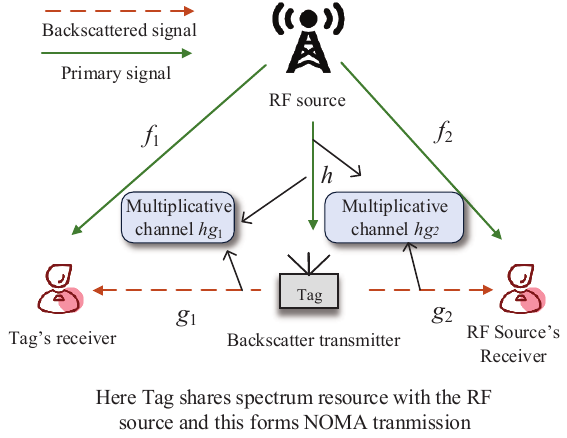}\\
  \caption{Signal processing of a BC-enhanced NOMA communication system.}\label{fig:Backscatter_basic}\vspace{-3mm}
\end{figure}

\textbf{Backscatter Communication:}
In the conventional  NOMA, it is assumed that all NOMA users are powered-sufficient, ignoring the heterogeneity of NOMA users' battery life. Particularly, when the power-limited  nodes share the same resource allocation with  powered-sufficient nodes, it is necessary to consider the extra-low power communication technique for the power-limited  nodes and this motivates us introducing BC to NOMA, yielding a BC-NOMA. In BC-NOMA, the energy-limited  node modulates and reflects  signals emitted by the  powered-sufficient node just via  adjusting reflection coefficients, rather than generating carriers by itself,  thus  its circuit design is simple and enjoys low-power consumption \cite{8454398,9051982,9328518}. Fig.~\ref{fig:Backscatter_basic} shows the detailed signal processing procedure of the BC. However, the  simple design also makes the power-limited node susceptible to eavesdropping. Accordingly, one of most important tasks in BC-NOMA is to improve the security of power-limited  nodes via using AN \cite{6836141,9745773,7556997}, beamforming \cite{9848786,10102794}, etc. Although the way to improve the wireless secrecy in BC-NOMA is similar to that of conventional NOMA, there are still two different points. One is that the multiplicative channel is present  in BC-NOMA but does not exist in conventional NOMA \cite{9051982}. This indicates that the secure rate of BC-NOMA partly depends on the
CSI from the powered-sufficient node to the powered-limited one, while in conventional NOMA, such CSI has no impact on its secure rate. Besides, when the imperfect CSI is considered, the multiplicative channel results in different robust optimization frameworks compared to conventional NOMA \cite{10102794,10109835}. This is because the distribution of the multiplicative channel's estimation error is quite different from that of the conventional NOMA. Another is that BC-NOMA introduces a new variable called as the  reflection coefficient. In BC-NOMA,  the rates for both legacy channel and eavesdropping channel increase with the  reflection coefficient \cite{9864647}. Accordingly, the reflection coefficient has a significant impact on the secrecy rate of the power-limited node, and should be carefully considered in designing security-enhanced schemes.

\begin{figure*}[t]
  \centering
  \includegraphics[width=6.3in]{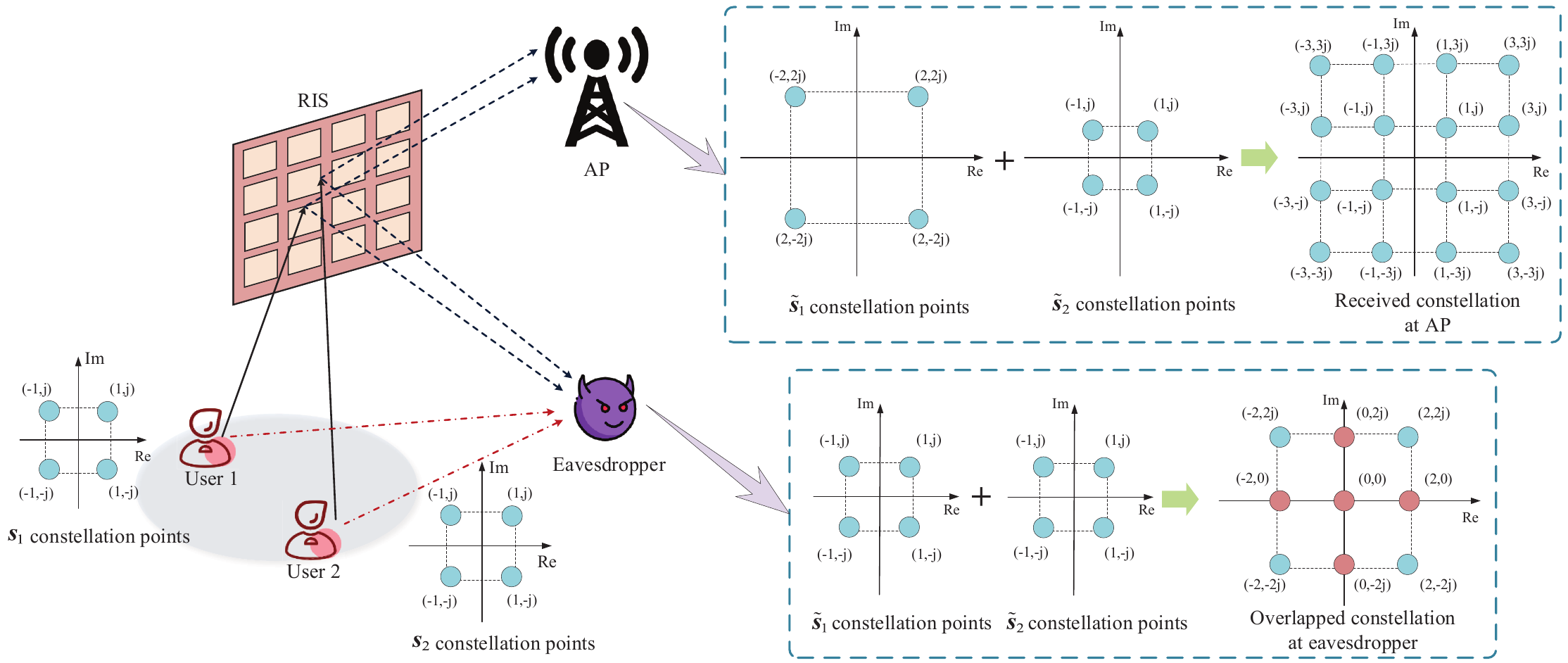}\\
  \caption{Illustration of the RIS-NOMA enabled constellation overlapping at the eavesdropper.}\label{fig:RIS_overlapped_constellation}
\end{figure*}

\textbf{Intelligent Surface:}
Due to their strong capability of constructing favorable wireless channels and radio propagation environments, intelligent surfaces have shown great potential to enhance the security performance \cite{Wu_arXiv2024,Cao_IOTJ2023,Zheng_TVT2023}. Reconfigurable intelligent surface (RIS) is the most common type of intelligent surface, which is a two-dimensional planar array with the electromagnetic functionality of passive reflection \cite{QingqingWu_CM2020,Y.Liu_ArXiv2020}. Unlike the above BC technique, RIS aims to enhance the communication link quality instead of delivering its own information by reflection. In particular, by appropriately optimizing the phase shifts and amplitudes of the RIS, the signals received from the direct transmission links and the RIS reflection links can be added coherently at the legitimate users and/or non-coherently at the eavesdropper, thereby benefiting the wireless secrecy \cite{YangLong_TWC2023,Lu_CL2021,Luo_TWC2021}. With the help of the AN, more sources of interference can be introduced to degrade the reception quality of the eavesdropper \cite{Rose_TWC2021,Feng_CL2022,Zheng_CL2021}.
Another outstanding merit of RIS is its flexibility of deployment, such that it can be easily deployed to enable various secure NOMA transmission scenarios. It is demonstrated in \cite{Zheng_TCOM2022} that by spreading the total reflection elements over distributed RISs to form higher multi-path diversity and collaborative reflection beamforming, a larger sum secrecy rate can be achieved. Besides utilizing RIS as a reflector, it is also possible to use RIS for signal modulation, i.e., spatial modulation, index modulation, and etc \cite{Basar_TCOM2020, Basar_Access2019,Lin_TWC2021}. Considering this merit, the use of RIS can generate artificial jamming based on the incident signals to degrade the eavesdropper's capability \cite{Wu_CL2023,XuSai_IOTJ2022,Ji_TCOM2023}. More specifically, by optimally dividing the RIS elements into two groups, where elements in one group perform signal reflection to enhance the legitimate signals' quality while elements in the other group generate artificial jamming to interfere with the eavesdropper \cite{Ji_TCOM2023}, a balanced reliability and security tradeoff can be achieved. In addition, exciting results in \cite{Lu_Network2020} and \cite{Lu_Proc2022} show that the operating principle of RISs opens up a new design opportunity for improving security, that is, adjusting the phase shifts and amplitudes of the RIS to produce a constellation overlapping at the eavesdropper deliberately and achieve over-the-air encryption. We use the RIS-aided uplink NOMA example shown in Fig.~\ref{fig:RIS_overlapped_constellation} to clarify this idea. Assume both users apply QPSK modulation, i.e., their signals $s_1$ and $s_2$ take values from the alphabet $\{1+j, -1+j,1-j,-1-j\}$. After RIS reflection, these two users' signals are added up at the eavesdropper to produce a 9-point quadrature amplitude modulation (QAM) overlapped constellation set shown in Fig.~\ref{fig:RIS_overlapped_constellation}. As such, two different received signal pairs $(\tilde{s}_1,\tilde{s}_2)=(1-j,-1-j)$ and $(\tilde{s}_1,\tilde{s}_2)=(-1-j,1-j)$ yield the same constellation point $(0,-2)$ for $\tilde{s}_1+\tilde{s}_2$, which makes the eavesdropper difficult to identify the individual signals $\tilde{s}_1$ and $\tilde{s}_2$ and leads to a poor BER performance at the eavesdropper. While the legitimate AP observes a clear one-to-one mapping QAM constellation between $(\tilde{s}_1,\tilde{s}_2)$ and $\tilde{s}_1+\tilde{s}_2$ and hence ensures a detection with almost no performance degradation.
Motivated by the above secrecy advantages offered by the RISs, it is expected that several new intelligent surface architectures that expand their electromagnetic functionalities from passive reflection to active amplification \cite{Ji_ICCC2023,Dong_TVT2022,GuoYuan_TCOM2023,Hao_WCL2023}, simultaneous transmitting and reflecting (STAR) \cite{Lu_TWC2024,LiXingwang_TVT2023,Zheng_TWC2022,LuoHao_WCL2023}, as well as intelligent computing \cite{YangBo_WCM2023,An_JSAC2023}, can be applied to obtain further security enhancement.

\subsection{ML for Secrecy Resource Allocation}

\subsubsection{Fundamentals of ML}

ML, a key technology supporting AI, has the ability to solve complex problems without explicit programming.  It includes the ability to extract knowledge from data and subsequently adapt the learner's behavior based on the knowledge gained.  Machine learning has distinct advantages over traditional algorithms.  First, it can absorb valuable information from input data, thereby improving system performance.  Second, machine learning-based algorithms for resource management, network optimization, and mobility management can be effectively tailored to dynamic environments.  Finally, machine learning helps to achieve the goal of system adaptivity.

The fundamental concept of information-theoretic security lies in leveraging the characteristics and impairments of the wireless channel, encompassing noise, fading, interference, dispersion, and diversity, to ensure successful data decoding by the intended recipient while thwarting eavesdroppers from deciphering the target user's information.        Consequently, the primary objective in designing physical layer security is to enhance the performance disparity between legitimate receiver links and eavesdropper links through meticulously devised transmission schemes.

The physical layer security framework enables the redesign and application of conventional wireless communication technologies, originally lacking security considerations, for confidential information transmission.     From a system design perspective, research topics on physical layer security primarily focus on secure resource allocation, beamforming/coding with enhanced security measures, secure antenna/node selection and collaboration, as well as comprehensive considerations based on the aforementioned strategies.

Resource allocation has been widely used in traditional communication and is an effective way to improve the security of the physical layer.   For resource management problems, knowledge base cooperation may be utilized to enhance the PLS on top of secure communication based NOMA system in resource-constrained networks (e.g., AI-native wireless networks).
Model flaws are when there is not enough information to know or the mathematical model does not exist, while algorithm flaws are when there are well-developed mathematical models, but optimizing these models using existing algorithms is very complex.   In such cases, low-complexity solutions using machine learning are preferred.   Similarly, when contextual information is important to the decision-making process, machine learning techniques are most appropriate.   There are two types of machine learning commonly used.   One type is reinforcement learning (RL), which learns optimal controls by monitoring unknown parameters and system behavior in repeated experiments.   The other type is deep learning (DL), which uses hierarchical artificial neural networks for the learning process.   These artificial neural networks work in a similar way to the human brain, with neuron nodes connected like a network.   While traditional machine learning analyzes data in a linear fashion, hierarchical deep learning can take a non-linear approach to data.

In recent years, optimization methods based on ML have been rapidly developed in wireless communication \cite{X30}, unsupervised DL strategies have been used to solve resource allocation problems \cite{X31}. In \cite{X32}, a machine learning-based antenna selection scheme for massive MIMO systems is proposed. It utilizes $K$-nearest neighbor (KNN) learning algorithm and support vector machine (SVM) to select the best data-driven antenna. In \cite{X33}, KNN supervised learning is used to allocate $N$ beams among $K$ users. In \cite{X34}, a multi-parameter Q-learning method is proposed to maximize the total signal-to-noise ratio of the system. In \cite{X35}, Q-learning is used to cooperate with power allocation, indicating that Q-learning algorithm is the most deeply studied reinforcement learning algorithm. In \cite{X36}, deep reinforcement learning (DRL) methods have been used for dynamic power allocation in single-user cellular networks. In \cite{X37}, deep learning is applied to resource allocation in massive MIMO systems.

{\bf Deep Learning:}
In the field of machine learning, deep learning has recently shown encouraging results in solving complex wireless network problems with large amounts of data. DL belongs to the supervised learning category of machine learning, which uses a priori known models and labels to estimate and predict unknown parameters.

DL is based on artificial neural networks.
The input layer is used to obtain data given to deep learning by the external environment and the output layer is used to send data features made by deep learning after training. Each neuron weights and biases the data transmitted by the previous layer of neurons and outputs it to the next layer of neurons through an activation function, with no direct connection between the two discontinuous layers. After training, the deep neural network can approximate the functional relationship present in the dataset. It is expected that this functional relationship will generalize to data beyond the training dataset. Deep learning can be used to solve a variety of problems such as prediction, classification, resource allocation, spectral state estimation, target detection, speech recognition and language translation. Deep learning learns features in multiple abstraction layers and allows the system to learn complex functions and map inputs directly from data to outputs. In deep learning, the first hidden layer learns raw features from the data by looking for combinations of frequently occurring numbers. These identified features are then fed into the next layer, which trains itself to recognize more complex (i.e., identified combinations) features.

{\bf Reinforcement Learning:}
RL is an approach to machine learning where the basic idea is to enable the learner to select the best action or strategy and achieve the desired outcome by constantly interacting with the dynamic environment. In other words, it enables the agent to perceive its surroundings and translate states into actions to maximize a specific metric (usually defined as reward). Being ignorant of action selection, the learner must gain experience through the training phase to discover which actions bring the greatest rewards. Furthermore, an important feature of reinforcement learning is that taking a certain action affects not only the immediate reward, but also the next state and all subsequent rewards. Trial-and-error search and delayed reward are then two key features of reinforcement learning. In summary, reinforcement learning enables learners to gain knowledge from their environment during the interactive process of receiving feedback on their own experiences.

The interaction between the learner and the environment can be modeled as a Markov decision process with infinitely discounted rewards. Q-learning is considered to be one of the most popular algorithms in reinforcement learning. This kind of learning always tries to perform the action that obtains the most reward in the state action space according to some strategy (e.g., greedy strategy).

\subsubsection{Apply ML for Secrecy Resource Allocation}

The benefits of using ML tools for realizing long-term secrecy performance gains in NOMA networks are discussed in this section below.

{\bf Intelligent Power Control and Resource Allocation:}
Power allocation is an important study of resource allocation in multicarrier NOMA systems, which generally includes power allocation among subcarriers and power allocation among users on subcarriers. The optimal power allocation scheme aims to improve the physical layer safety performance of the system by optimally allocating power resources to maximize the safety and rate of the system when the system transmit power is fixed.
Different from traditional algorithms based on iterative search, deep learning can be used for inter-subcarrier power allocation schemes to obtain efficient power allocation results without extensive computation.

\begin{figure}[t]
	\centering
	\includegraphics[width=3.2in]{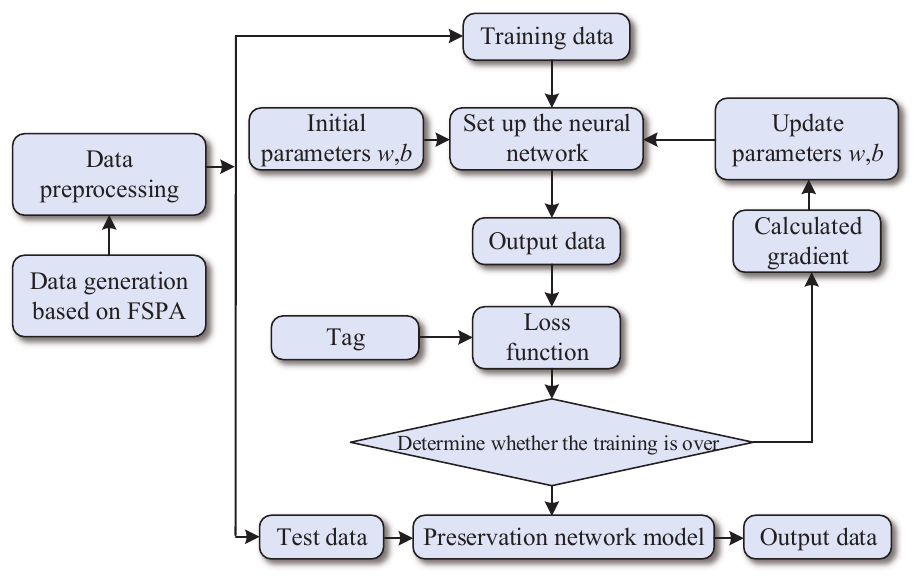}
	\caption{The power allocation scheme based on DL.}
	\label{fig6}
\end{figure}

Consider a case study, we introduce a power allocation scheme based on DL. The basic unit of neural network is neuron, and the basic elements of neuron include input $\overrightarrow{x}$, output label $\overrightarrow{y}$, weight $\overrightarrow{w}$, bias $b$, excitation function ${{f}_{a}}(\centerdot )$ and loss function ${{f}_{l}}(\centerdot )$. The input feature $\overrightarrow{x}$ and the output label $\overrightarrow{y}$ together constitute the training data and the test data of the neural network. Weight $\overrightarrow{w}$ and bias $b$ are the parameters to be optimized by neural network. Excitation function ${{f}_{a}}(\centerdot )$ and loss function ${{f}_{l}}(\centerdot )$ are known important parameters. The power allocation scheme based on DL consists of three parts: data generation and processing, neural network training stage and neural network testing stage. The power distribution scheme based on DL is shown in Fig.~\ref{fig6}.

{\it Data Generation and Processing Stage:}
The power allocation scheme of a NOMA system is a non-convex optimization problem that is difficult to solve by optimization methods. The optimal scheme for power allocation is known to be the global search power allocation, i.e., the optimal allocation scheme is found by traversing all possible allocations. The disadvantage of this scheme is that the complexity of the algorithm grows exponentially as the number of elements involved in the allocation increases.

The presented study of the DL algorithm is verified to be able to approach the original algorithm by continuously learning from the data of the original algorithm.
The scheme presented in this part selects some scenarios with limited allocation elements and finds the optimal allocation scheme by traversing all allocation possibilities through a global search of power allocation, with optional parameters being the channel fading coefficients and power allocation coefficients between the central base station and the legitimate user nodes.
The optimal allocation coefficients corresponding to the channel fading coefficients are found by globally searching the power allocation. After preprocessing, they are used as input features and output labels, respectively. They are used as training data and test data for neural network.

{\it Neural Network Training Stage:}
A neural network model is built, the weights and biases are initialized, and the excitation and loss functions are set up. Using the training data generated by global search weight assignment, after the initial weight and bias, forward training is carried out according to the back propagation algorithm of the gradient descent formula, and the parameters are optimized. After several rounds of training, the trained neural network model is used for data testing.

{\it Neural Network Test Stage:}
The test data is fed into the trained neural network model, and the accuracy of the neural network is calculated by comparing the labels of the test data with the estimated output data of the neural network. Once the trained neural network can guarantee a certain accuracy, it proves the learning ability of the neural network. The input data composed of the communication parameters generated in real time can be fed into the neural network, and the output results can be sent back to the communication system to complete the strategy optimization problem of the communication system.

{\bf User Grouping in Multi-Carrier NOMA:}
When the number of users in the system increases and the number of users assigned to each subcarrier increases, the influence of user grouping scheme on system performance becomes more prominent. With the increase in the number of users in the system and the increase in the number of subcarriers, the grouping situation in the system becomes more complicated. It is very difficult to go through all the group cases with the exhaustive method, and it is even more difficult to achieve in practice. Therefore, how to achieve the optimal user grouping scheme is an important problem in the multi-carrier NOMA system.

Consider a multiuser group NOMA system, meaning that in a multicarrier NOMA system, multiple users share the same subcarrier resource. In this system, the complexity of the user grouping scheme increases with the number of users. In this section, a user grouping scheme is introduced to improve the security performance of the physical layer of the system. The q-learning based user grouping scheme models the optimization problem as a Markov decision-making process to find the optimal behavior during continuous interaction with the environment to obtain an approximately optimal allocation scheme.

Q-learning is an algorithm based on the idea of optimal value, which focuses on the value function, learns the value model through the interactive sequence of information, and updates the strategy through the value model. The idea of Q-learning is very similar to the value iteration method. Q-learning is a typical model-free learning.

Reinforcement learning methods require the definition of a set of states $a\in A$, a set of actions $s\in S$, and a reward function that represents the effect of the selected action on the environment, and this reward function will influence the user to choose the next action from $A$. Assume that all users have the same set of actions and states. Each user $k$ selects a finite discrete space of candidate actions to find the best solution. Each user $k$ observes the current state $s\in S$ and takes an action $\pi (s)\in A$ based on a certain strategy $\pi $, generating an immediate reward $R_{t}^{k}$. The problem here is to find a strategy to maximize the discount reward $V$ received.

The system action $A$ of a user in this system is represented by some columns ${{x}_{n,k}}$, ${{x}_{n,k}}=0$ indicates that the user is assigned to a subresource, and ${{x}_{n,k}}=1$ user is not assigned to a subresource. The state space $S$ can be represented by a series of $\vec{S}_{K}^{T}$, $\vec{S}_{K}^{T}={{\overrightarrow{s}}_{0}}$ means that the user has not made a choice and $\vec{S}_{K}^{T}={{\overrightarrow{x}}_{k}}$ means that the user has made a choice.
The policy ${{\pi }^{\varepsilon }}$ is a nondeterministic policy and adopts the $\varepsilon $ greedy strategy. Reward $R_{S}^{{{A}_{k}}}$ represents the return reward value after making an action ${{A}_{k}}$ in the state $S$, when $R_{{{S}^{*}}}^{{{A}_{k}}}>R_{S}^{{{A}_{k}}}$, $R_{{{S}^{*}}}^{{{A}_{k}}}={{R}_{ssum}}$ is satisfied, otherwise, $R_{{{S}^{*}}}^{{{A}_{k}}}=0$. This indicates that after each action selection, the upper limit of the number of users on the subresource cannot be breached. If the number of users on all sub-resources did not exceed the limit, the system security and rate after this iteration were used as rewards. Otherwise, a penalty will be imposed. The above definition is introduced into the system action state value function for training to obtain the user grouping scheme.

\section{Quantum-Safe Security for NOMA}
\label{sec:Quantum}

Quantum secure communication has gained increasing attention for its incomparable advantages of secure key transmission. QKD ensures two individuals to establish a secret key by exchanging photon quantum states, which ensures security and  assists future wireless communications. As a promising candidate for 6G wireless, NOMA has great potential application in QKD networks for supporting more secure transmission simultaneously. In this context, secure data transmission in networks combining NOMA and quantum security techniques is worth studying. In this section, we explore NOMA-based secure transmission schemes for quantum-assisted wireless communications and review NOMA-based quantum key distribution systems and quantum-coupled NOMA iterative detection schemes.

\subsection{Quantum-Safe Security}
Security has become an essential requirement in modern communication systems. Due to the broadcasting feature of wireless channels, wireless communication systems are inherently vulnerable to eavesdropping and security cannot be absolutely guaranteed \cite{Dongyang_IoTJ2019,Dongyang_TIFS2019,Dongyang_TIFS2018}. However, in quantum communication, the security can be theoretically guaranteed based on the quantum no-cloning theorem and the fundamental postulate of quantum physics that every measurement perturbs a system. Quantum cryptography allows two legal parties, namely Alice and Bob, to achieve the information-theoretic security over public channels. In this context, wireless QKD has become a research hotpot recently for building future highly secure free-space communication networks \cite{X38,X40,X41}.

In \cite{X42}, a wavelength division multiplexing (WDM) and orthogonal angular momentum (OAM) based multiple-input-multiple-output (MIMO) QKD system is studied for achieving the high-speed secure key transmission over atmospheric turbulent channels. In \cite{X43}, a network-coded cooperative QKD system is conceived over free space optical (FSO) channels, in which network coding is adopted for improving the key exchange in multi-user QKD systems. In \cite{X44}, a subcarrier intensity modulated (SIM) QKD scheme with a dual-threshold/direct detection (DT/DD) receiver is designed. It proves that the QKD protocol can be implemented in standard pulse-based optical signal systems. Another key requirement of QKD networks is expansion from point-to-point to multiple-access QKD communications \cite{X45,X46,X47}. In \cite{X45}, time/code division multiple access (TDMA/CDMA) approaches are considered in the star-coupler quantum networks, where multiple QKD users can exchange secret keys simultaneously. In \cite{X46}, a wavelength division multiplexing access QKD system (WDMA-QKD) is designed and analyzed. Compared to a point-to-point link, this scheme achieves a higher secure key rate for each user. In \cite{X47}, an orthogonal frequency division multiplexing QKD (OFDM-QKD) scheme is proposed. Through numerical simulations, the scheme can achieve a key rate of 20 Kbps when the link distance is 50 km, thus validating the suitability of the proposed scheme for long-haul secure communications.

\subsection{ A NOMA-based Quantum Key Distribution System}

\begin{figure}[t]
	\centering
	\includegraphics[width=3.2in]{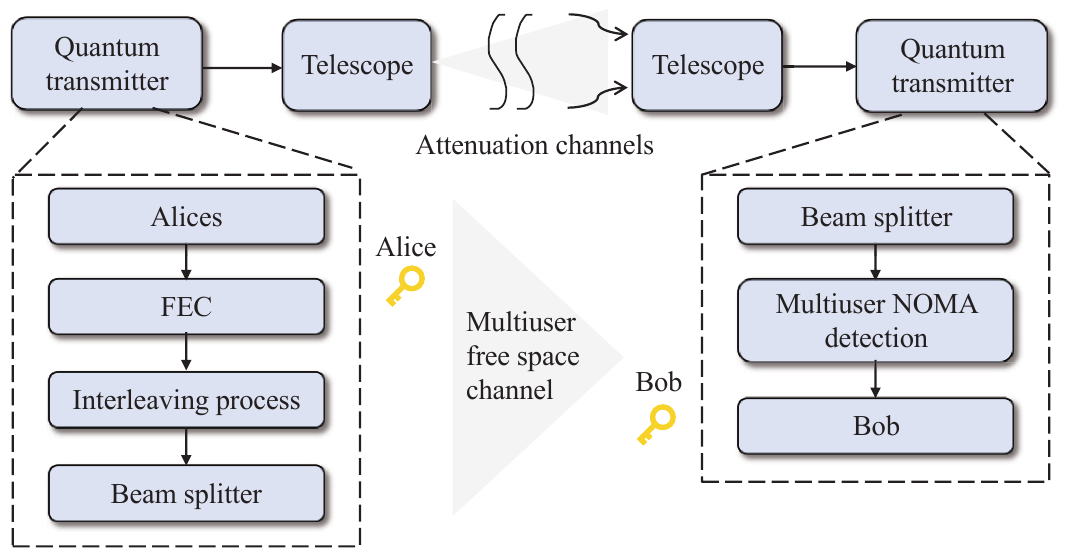}
	\caption{The schematic of the proposed NOMA-MQKD QKD system.}
	\label{fig8}
\end{figure}

It is worth noting that, as a candidate for 6G communications, NOMA can support massive secure connections in QKD networks with enhanced throughput. As illustrated in Fig.~\ref{fig8}, the NOMA-based QKD system consists of a quantum transmitter, a quantum receiver, and a multi-user NOMA iterative detection module.

\subsubsection{Transmitter Design}

At the $k$th Alice, the information bits sequence
is encoded by the repetition coder,
and then, each encoded sequence of Alice is interleaved with different interleaving to generate the transmitted sequence.
Thereafter, the quantum transmitter randomly chooses one of the polarization base group to emit the permuted sequence.
In a two-dimensional Hilbert space, the quantum information bit can be described by the polarization quantum state.
Each Alice controls the laser diode (LD) to emit an optic beam to the beam splitter (BS). Then, the BS maps the information `0' and `1' to a random polarization state in the $l$th slot. After that, the bit sequences are transmitted via atmospheric attenuation channels.

\subsubsection{Receiver Design}
At the receiver, the detection consists of two parts: the quantum measurement and the multi-user iterative detection.

{\bf Quantum Measurement:} During the quantum transmitting process, the state of the quantum combination system is the maximum space of each system. As a result, the received quantum signal state in the $k$th slot can be described as the direct product state.
Then, the polarizing beam splitter (PBS) randomly divides the received photons into two different branches. Each branch includes a BS for positive-operator valued measuring (POVM) and two APDs for photon counting.
Then, the BS can split and transmit every photon into the right APD for photon counting. However, if the receiver selects an incompatible polarization basis, the BS can not split and transmit every photon into the
right APD.

{\bf Multiuser Iterative Detection:} The received sequences are transmitted to the NOMA multiuser iterative detection module after the quantum measurement. For the NOMA receiving factor graph, there are only two lines that pass in and out of the received sequence variable node of the $k$th Alice's $i$th bit, so the information will not be updated when it passes through the variable nodes, and the update of information only occurs in factor nodes: multiple accessing nodes and decoding nodes.

Additionally, the safety of the physical-layer transmission becomes a research problem. Some mathematical encryption algorithms have been used to prevent cracking ciphertext, but it would be disastrous if they are broken. In addition, many encryption algorithms are applied at the upper layer of the physical layer, and it is necessary to transform the optical signal of the physical layer into the electrical signal, and then convert it back to the optical signal after encryption.

\subsection{Photon Counting Based Iterative Quantum NOMA with Spatial Coupling}

As we all know, the SIC, superposition coding, and power control are important issues in NOMA systems. Inspired by iterative SIC and user-specific interleaving, interleave division multiple access (IDMA) has become an attractive method of NOMA. It is worth noting that in recent NOMA studies, spatially coupling is a representative technology for achieving high spectrum efficiency and reducing computational complexity.

Quantum-assisted multi-user detector (QMUD) has been widely concerned and developed because of its capability of parallel processing. However, in most existing studies of quantum-assisted multi-user detectors, signals are transmitted over classical RF channels. In this part, a quantum-spatially coupled NOMA (q-SC-NOMA) iterative detection scheme over quantum-spatially coupled quantum noise channels is introduced.

In this system, the encoded information from the user is modulated in a coherent state. Then the coherent state signals are spatially coupled and combined into quantum noise channels. In an iterative quantum communication receiver, the quantum noise can be reduced after several iterations. The quantum measurement part of the receiver mainly includes a single photon detector (SPD) unit, an electro-optical modulator (EOM), and a displacement operation unit. The receiver uses quantum adaptive frequency division interval feedback measurement and inter-SIC process to measure the NOMA superposition coherent state signal. A multi-user detector (MUD) and $K$ decoder (DEC) make up the inter-SIC module.

\begin{figure}[t]
	\centering
	\includegraphics[width=3.2in]{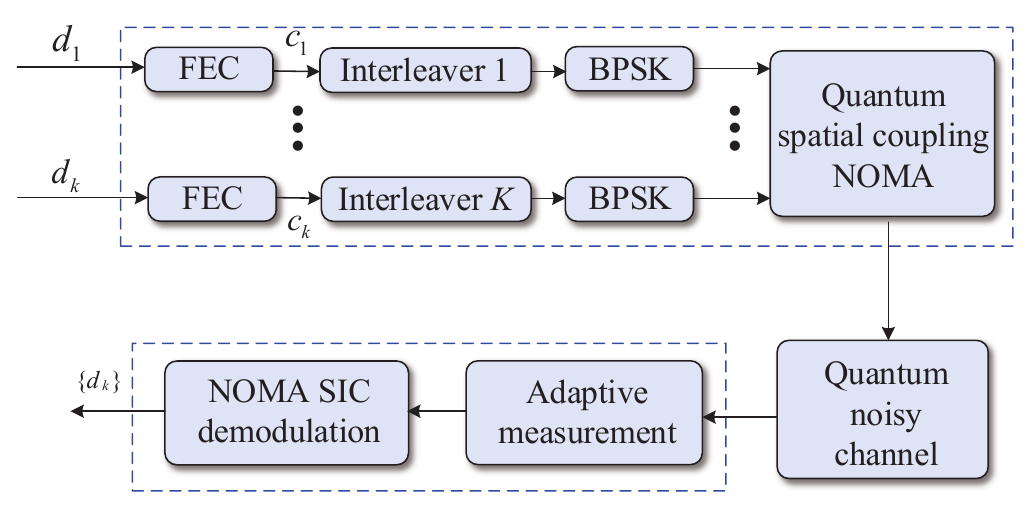}
	\caption{The schematic of the iterative quantum communication system with spatial coupling.}
	\label{fig9}
\end{figure}

\subsubsection{Transmitter Design}

At the transmitter, as shown in the upper part of Fig.~\ref{fig9}, the multi-user information ${{d}_{k}}$ is encoded by the forward error correction (FEC) encoder. The encoded bit sequence ${{c}_{k}}$ is then permuted by the $K$ different interleavers ${{\Pi }_{k}}$ and modulated by binary phase shift keying (BPSK). The multi-user coherent state signal ${{x}_{k}}$ is combined by NOMA superposition block. The coherent-state signals are combined into the transmission channel by the beam splitter network (BSN).

Regarding the spatial coupling transmitting method in q-SC-NOMA, as depicted in Fig.~\ref{fig9}, multi-user interference is reduced. In the proposed scheme, the coupling width is set to 2, which means no matter how many users are in the system, at most two users' signals are superimposed in each time slot. The signal sequence of each user is divided into two parts and transmitted in consecutive time slots. In the superimposed sequence, the first segment is not interfered by other users. The first user's information can be decoded accurately and interference in the next segment is determined. In this way, interference in each segment can be determined and information of every user is decoded sequentially.

\subsubsection{Receiver Design}

In the BPSK scheme, $10$ quantum states can be detected under the interference of quantum noise. Quantum noise is attributed to the evolution of the channel density operator. The model of the quantum multi-user channel illustrates the relationship between channel inputs and channel outputs explicitly.
Considering thermal noise, the received coherent state signal can be represented by the elements of an infinite-dimensional Hilbert space.

The modulated bit sequence is carried by coherent state signals emitted by lasers, which can be represented by a set of orthogonal bases of the Hilbert space.
The receiver first utilizes the beam splitters
to divide the received coherent state signals into $N$ partitions.
The divided state is then combined with the local oscillator field via EOM to obtain a superposed field.
The number of photons of the average intensity is detected through SPD.
After $N$ partition measurements, the measurement output could be generated in the last measurement.
The lower part of Fig.~\ref{fig9} shows the iterative SIC detection process for $K$ users in the q-SC-NOMA receiver, which contains two modules: MUD and DEC. The q-SC-NOMA starts decoding from the information sequence of the first-arrival user.

At the first iteration, the prior information is zero. In the subsequent iterative process, the external information fed back by the DEC module is used as a priori information of the MUD module. Note that for any user, the prior information comes from the sum of the previous iterations and the last decoded user output.

In this system, the Gaussian noise approximation is utilized to model the multi-user interference.
In the q-SC-NOMA system, all external information from the previous time periods are added together. The previous external information is added together and used in the next iteration. Meanwhile, in q-SC-NOMA, except for the segments at both ends,
the other segments are misplaced superpositions of two users' transmitted sequences.
Therefore, the transmitted sequence of each user is divided into two segments and the external information is calculated as two users.
A standard APP process is used to decode input information and output external information in DEC module. DEC outputs the result of hard decision $\tilde{d}$ after several iterations.

Future communication systems will pay more attention to the protection of communication security and user privacy by exploiting quantum communication. In addition, quantum technology can also play an important role in improving the transmission data rate and reliability. NOMA is a promising technology for 6G and future wireless communication, which is expected to be deeply integrated with quantum technology in the future. In this section, we introduce two schemes, the first scheme introduces a combination of QKD and NOMA, which jointly uses QKD and traditional NOMA to enhance security and data rate, respectively. The second scheme is an iterative q-SC-NOMA scheme that addresses the non-orthogonal properties inherent in coherent state multi-user communication, which can reduce computational complexity and provide high robustness to many practical quantum domain defects, such as pattern mismatches and dark counts.

\section{Covert Communication for NOMA}
\label{sec:covert}

Under some circumstances, protecting the content of communications using information-theoretical security techniques is far from sufficient, and the communication itself is often required to hide from being detected by applying covert communication techniques \cite{Shihao_WCM2019}. The target of covert wireless communication is to explore the physical dynamics and non-orthogonal signal processing schemes to hide the amount of wireless information that can be covertly exchanged between legitimate transceivers, subject to a negligible probability of being detected by a warden. Generally, ideas that enable covertness
can be classified into two categories in terms of the effects on the warden. One is to reduce the signal leakage and weaken the signal power at the warden, and the other is to increase interference dynamics and introduce uncertainty at the warden to cover the secret signal. Therefore, different approaches have been proposed to achieve covert communication for NOMA, which are briefly summarized in Table~\ref{table-covert} and reviewed below.

\begin{table*}[t]
\caption{Summary of Covert Communication Techniques for NOMA}
\centering
\small
\begin{tabular}{|l"c"c"c"c|}
\Xhline{1.2pt}
 \makecell[c]{\bf Technique}&\makecell[c]{\bf Effect on Warden}&\makecell[c]{\bf Deployment  \\ \bf Scalability}&\makecell[c]{\bf Power  \\ \bf Consumption}&\makecell[c]{\bf Complexity}\\
\Xhline{1.0pt}
\makecell[c]{Multi-Antenna\\ Beamforming}&\makecell[c]{Increase power dynamics and/or \\nullify/weaken power leakage}&High&High&High\\
\Xhline{0.5pt}
\makecell[c]{Cooperative \\Relaying}&\makecell[c]{Increase power dynamics\\ and/or weaken power leakage}&High&Medium&High\\
\Xhline{0.5pt}
\makecell[c]{Dynamic Power\\Control}&\makecell[c]{Increase power dynamics}&Low&\makecell[c]{Very high under deep fading\\ and medium otherwise}&High\\
\Xhline{0.5pt}
\makecell[c]{Intelligent \\Surface}&\makecell[c]{Increase power dynamics and/or \\nullify/weaken power leakage}&High&Very low&Low\\
\Xhline{0.5pt}
\makecell[c]{Backscatter \\Communication}&\makecell[c]{Increase power dynamics \\and/or weaken power leakage}&Low&Very low&Low\\
\hline
\end{tabular}
\label{table-covert}
\end{table*}

\subsection{Multi-Antenna Beamforming}

By virtual of spatial degrees of freedom, multi-antenna beamforming techniques can be utilized to improve the covertness of NOMA through directional transmission. In particular, the use of beamforming enables the adjustment of the phase and amplitude of the signals to make the superposed beampattern constructive in the direction of legitimate users while destructive in other directions \cite{Tongxing_TWC2019,WangChao_TCOM2021}. In this way, the secret signals can be concentrated to increase the covert rate and simultaneously the useful signal power can be nullified at the warden to realize low probability of detection. With the faster-than-Nyquist (FTN) signaling, a low-complexity linear beamforming can be designed to decouple the MIMO FTN signaling into independent sub-streams, thus benefitting the maximization of the covert rate \cite{Zhang_TVT2024}. As the operating frequency goes higher and the number of transmit antennas becomes larger, additional performance gains for covert communication can be obtained. For instance, by utilizing the channel spatial sparsity in both the angle and power domain, hybrid analog and digital beamforming is applied to achieve covert communication \cite{Bai_TCOM2023}. Theoretical results in \cite{Bai_TCOM2023} show that the covert rate increases considerably when the channels become much more sparser. The near-field beamfocusing with ELAA can make the beampattern more focused towards the legitimate users, and hence almost no signal power can be measured by the warden \cite{Zhang_ICCW2023}. However, the accuracy of beamforming and/or beamfocusing for the covertness improvement depends heavily on the availability of the exact CSI, which may incur excessive signaling overheads of pilot transmissions. To strike a tradeoff between the complexity and performance, beam sweep can be used to obtain the optimal beam pair to transmit the secret signals, instead of directly estimating the high-dimensional MIMO channels \cite{Xing_CL2023}. Furthermore, multi-antenna techniques can make use of the additional jamming signals to cover the secret signal transmissions. Specifically, by separating the antenna array into the secret signal part and the jamming part, the jamming beamforming can be combined with the covert beamforming to mislead the warden \cite{Ci_Globecom2021,Qian_IoT2023}. Also, the jamming signals can be transmitted by distributed jamming helpers and/or full-duplex receivers \cite{WangChao_TWC2022,Jiang_TVT2020,Zheng_TWC2021}.

\subsection{Cooperative Relaying}

The high-power signal should be exploited for the communication scenario with severe path loss or obstacles to offset the weakness channel, which, however, deteriorates the covertness performance. Fortunately, by creating additional transmission links, the cooperative relay can increase the uncertainty of warden's signal detection \cite{covert1,covert2,covert3,Lin_TCOM2023,Gao_IOT2021,Long_TVT2020}. Consider a scenario in which a relay tries to transmit its own information to the destination on top of forwarding the information from the source. Note that the uncertainty is inherently embedded in the forwarding strategies from the relay to the destination, where the positive covert rate can be achieved without any extra uncertainty sources \cite{threenode1,Bai_TIFS2022}.
To further improve the covert performance in the above three-node scenario, the self-sustained relay has been introduced to achieve peak-shaving and valley-filling for the power detection of the source. Specifically, the self-sustained relay can employ energy harvesting schemes to store the energy at the first phase of transmission, then adaptively forward the information from the source and itself to confuse malicious power detection \cite{threenode2}.
By leveraging the randomness of the wireless channels, a multi-hop relaying scheme is employed to improve covert performance in the presence of multiple collaborating Willies over additive white Gaussian noise. The results showed that the designed multi-hop channels create more uncertainty to Willies' detection. When the distance between the source and destination is large, the muti-hop relaying scheme can substantially enhance covert performance compared to the one-hop covert communication \cite{multi_hop_covert}.
Besides, the movable relay can also increase the channel uncertainty by utilizing the dynamic position and Doppler frequency offset. In \cite{UAV_covert}, the authors try to hide the information transmission from a source to its destination while avoiding being detected by a warden. To this end, the work jointly optimized the UAV trajectory and transmit power for achieving the covert capacity maximization \cite{UAV_covert}.

\subsection{Dynamic Power Control}

The key idea of covert communication is to make the received signal power at the warden as random as possible, which indicates dynamic power control plays an important role to guarantee covertness \cite{Yan_TIFS2018}. One simple approach is to use the random transmit power to create energy uncertainty at the warden. In particular, the random transmit power can be either generated in a continuous or a discrete manner. For the continuous case, the transmit power is generated following a uniform distribution or other continuous distributions and is bounded between zero and its maximum value \cite{Jiang_SJ2020,LiQiang_WCL2023,Tao_WCL2021}. For the discrete case, the number of transmit power levels are quantized based on its continuous power distribution, and in each slot, the transmitter randomly chooses one power value from its total number of transmit power levels for sending secret signals \cite{Yan_TIFS2018}. For both random transmit power controls, the signal power received at the warden varies rapidly across different slots, thus making the warden very difficult to judge whether there is the secret signal transmission. Numerical results provided in \cite{Yan_TIFS2018} show that the covertness performance achieved by the discrete random transmit power strategy is improved with the increased number of power levels, i.e., $M$, and the performance approaches to that of the continuous random transmit power strategy when $M$ goes to infinity. Another approach is the channel inversion power control, which is originally proposed in \cite{Weber_TIT2007,Peer_TCOM2005} for achieving near-capacity, and can be also adopted in covert communication. As its name suggests, the channel inversion power control varies the transmit power based on the inversion of the legitimate channels, such that the received signal power remains a constant at the desired users but varies frequently according to channel fading at the warden, thus introducing useful uncertainty \cite{Zlatanov_CL2022,Ma_WCL2021,Wang_WCL2022,Hu_TVT2019}. Nevertheless, both random transmit power control and channel inversion power control have their drawbacks. For example, the increase of $M$ can benefit the covertness in the discrete random power control, but this also increases the system complexity. Moreover, the transmit power changes from one slot to another, and hence a high dynamic range power amplifier is required by the transmitter which may lay a heavy burden on analog/digital circuit designs. In addition, when the wireless channel undergoes deep fading, the use of the channel inversion power control would result in an infinite transmit power, which is impossible in practice.
Consequently, we can conclude that although being a straightforward covert strategy to confuse the warden, the dynamic power control exhibits some limitations in improving the covert communication performance.

\subsection{Intelligent Surface}

Interestingly, the combination of intelligent surface and NOMA is a win-win solution to covert wireless communication. The rationale behind is based on two facts: 1) The superposition NOMA transmission of overt users can be exploited as natural shelters for a covert user, such that the secret signal can be embedded in overt channels to facilitate covert communications \cite{Tao_WCL2020,Hai_CL2023,Duan_CL2023}; 2) The use of intelligent surfaces can control the wireless channels to direct the desired signals to the intended receiver and simultaneously nullify the energy leakage at the warden \cite{LuXiao_Network2020,Li_CL2022,WuYingjie_WCL2022,Si_TCOM2021,Zhou_TWC2022}. Furthermore, an outstanding merit of the intelligent surface enabled NOMA that motivates the use of it for covert communication is its implementation simplicity, such that no additional signal processing, e.g., random transmit power/jamming or channel inversion power control, is needed to introduce uncertainty. While such merit cannot be promised by solely using NOMA or intelligent surfaces \cite{Lu_TWC2022,Lu_ICCC2021}.
In fact, it is rigorously proved in \cite{Lu_TWC2022} that by optimizing the phase shifts of the RIS based on the legitimate NOMA users' CSI, the received signal power over the RIS reflection channels changes randomly at the warden, thus creating uncertainty to the warden's binary test. Moreover, the uncertainty level increases with an increase in the number of RIS elements, and a higher covert rate can be achieved in a green and sustainable manner \cite{Lu_TWC2022}. This motivates the use of other new intelligent surface architectures to improve the covert communication performance. For example, by appropriately generating both random phase shifts and amplitudes at the active RIS, the uncertainty of the warden's received signal power is significantly increased \cite{Chen_WCL2023}. Meanwhile, the active RIS can also increase the uncertainty of the random power from an interference source \cite{Zhu_WCL2023}. Both lead to an enhanced covertness. With the use of aerial RIS carried by UAVs, more sources of uncertainties, such as uncertainties in RIS phase shifts and UAV trajectories, can be introduced to confound the warden's energy detection \cite{Chen_WCL2024, Qian_TVT2023,Wang_TCOM2023}. In \cite{Xiao_TWC2024}, a friendly full-duplex receiver in the refraction space of the STAR-RIS transmits the jamming signals with a varying power
to confuse the warden, who is located in the reflection space of the STAR-RIS. As a consequence, the covert rate is maximized and the full-dimensional covert communication is achieved. Furthermore, it is also expected that the application of movable reflection and/or refraction elements at the intelligent surface can generate significant randomness to the warden and guarantee covert wireless transmissions \cite{Wu_arXiv2024,Sun_arXiv2023,Hu_arXiv2023, Zhu_CL2023}.

\subsection{Backscatter Communication}

To further improve the covertness, the BC technique is employed at the covert NOMA user. Different from the conventional NOMA where the covert NOMA user generates  signals by itself and
superposed it on  the non-covert NOMA user's signal, the BC-based covert NOMA user modulates its information on the non-covert NOMA user's signals to protect its transmission from being detected \cite{10036471,10320343,8761392,9363596,10371390}. The benefits introduced by  BC can be summarized from the following  three  perspectives. First, the reflection coefficient is set by the covert NOMA user and hard to know by the eavesdropper. Thus,  the change of reflection coefficient can be combined with the  uncertainty existing in the physical layer, such as the random power of the ambient signals \cite{8761392}, the randomness of the channel gain \cite{9363596}, and the randomness of information transmission time \cite{10371390}, to enhance the covertness. This is because the optimal detection threshold of the eavesdropper is a function of the exact reflection coefficient. If such  knowledge is absent at the eavesdropper, it has to modify its detection threshold by assuming that the reflection coefficient follows a given distribution. By doing so, the detection threshold used by the eavesdropper deviates from the true one, which will increase the error probability of detection.    Second, the signal backscattered by the convert NOMA user is usually weaker than that of the conventional one due to the multiplicative channel, which makes BC-based covert NOMA user's transmission   more covertness \cite{10036471,10320343,8761392,9363596,10371390}.
Third, due to the simple circuit, the modulation rate of the BC-based covert NOMA user  is much slower than that of an ambient source. In such a case, the communication between the BC-based covert NOMA user and its associated receiver  is similar to the spread spectrum communication \cite{9193946}, which can be used to reduce the power of  covert signals in a given spectrum and to enhance the covertness.

\section{Future Research Directions and Challenges}
\label{sec:future}

As the research on security provisioning of NOMA is still in a very early stage, there are many research opportunities and challenges remained to be explored and addressed. In this section, we identify some initial ideas and promising research trends on secure NOMA communications, as elaborated next.

\subsection{Integrated Sensing and Communication}

With the frequency band of wireless communication moving from conventional sub-6G to millimeter-wave (mmWave) and terahertz (THz), next-generation wireless systems tend to have very high time/angular-domain resolution that enables the dual-function of sensing and communication, namely ISAC, making it possible to support many emerging applications that require both ubiquitous connectivity and accurate sensing capabilities, such as autonomous driving, indoor robot navigation, extended reality, industrial IoT, and etc \cite{DAPAR2016,LiuFan_JSAC2022,LiuFan_SPM2023}. In this novel wireless system paradigm, two fundamental radio technologies, say communication and sensing, can fully share the same radio waveform, frequency band and hardware platform to improve the spectral efficiency as well as reduce the hardware cost.

As a new spectrum sharing fashion, NOMA can benefit ISAC by largely eliminating the interference among communication/sensing signals and providing extra power-domain degrees of freedom for jointly optimizing the performance of sensing and communication \cite{Wang_CL2022,Mu_JSAC2022,Dou_TVT2023}. In fact, the interplay between NOMA and ISAC can also strengthen the security and covertness of wireless systems. In particular, by sensing the surroundings of legitimate links, the ISAC transmitter is able to estimate the distance, angle and even motion information of unknown targets that are potential eavesdroppers/wardens \cite{Wei_CM2022,Su_TWC2021,Su_TWC2023,Xu_TCOM2022,HuaMeng_TWC2024,YangZhu_TCOM2022}. Then, with the aid of sensing results, the ISAC transmitter can better design the beamforming for legitimate signals, whilst generating the highly directional AN beam to confuse the potential eavesdroppers/wardens, which may inspire some interesting future works.

On the other hand, by using NOMA, the ISAC transmitter can send multiple communication signals along with a dedicated sensing signal with fixed waveform. When the waveform of the dedicated sensing signal is foreknown by legitimate receivers, they can measure the reception of the sensing signal to refine their channel estimation results. Since the performance of SIC in NOMA systems relies on the accuracy of channel knowledge at the receiver side, the refined channel estimation results can help the SIC-based detection better eliminate the intrinsic inter-user interference of NOMA, thereby achieving higher secrecy capacity. In addition, when the sensing signal employs continuous waveform and occupies more transmit power than communication signals, the sensing signal can also be utilized to shield the communication activities for covertness enhancement. Therefore, it is interesting to study the power allocation and channel estimation refinement, since those issues not merely determine the sensing-communication tradeoff but the secrecy/covertness enhancement of ISAC systems as well.

\subsection{Satellite Communication}

As an indispensable component of 6G and its beyond, satellite communication can provide anytime ubiquitous coverage to users in anywhere with any type of services. However, the hundreds to tens of thousands of kilometers long distance between the satellites and ground users requires the high-power wireless transmission over the two-way satellite-ground links, in order to combat with path loss, atmospheric attenuation, penetration loss, etc. Such a way not only guarantees the reception quality of satellite communication, but also largely increases the risk of intercepting/eavesdropping the information signal, especially when the eavesdroppers locate close to the legitimate beam. Recently, the integration of NOMA and satellite communication has demonstrated its superiority in providing massive low-latency access by using NOMA to serve additional users, since multiple close-proximity users can be served by just one spot beam \cite{Satellite1,Satellite2,Satellite3}. In fact, the rationale of NOMA can also be exploited to secure satellite communication in terms of covertness and security. This can be interpreted as follows.

Since there are a large number of existing satellite signals with well-known physical-layer standards, e.g., Digital Video Broadcasting (DVB)-S/S2 and its extensions \cite{Satellite4,Satellite5,Satellite6}, those signals can be utilized to improve the covertness of satellite communication links that convey confidential information. In particular, the legitimate satellite transmitter can send the covert signal by mimicking the waveform and the frequency band of existing signals. With this approach, Warden can hardly identify the legitimate transmission activity due to the similar waveform as well as the co-located frequency band to the existing signals. The practical applications of this NOMA-inspired covert transmission should consider the limited onboard processing capability of satellite and develop the low-complexity implementation for waveform reconfiguration and power control. Therefore, these important issues merit further investigation.

Considering the security and privacy issues, when eavesdroppers are located near the legitimate ground user, preventing eavesdroppers from overhearing confidential information is challenging in general. To overcome this challenge, the idea of NOMA can be exploited to secure the legitimate transmission by using the multi-beam capability of the satellite transmitter. In specific, the satellite transmitter generates the legitimate spot beam pointed to the ground user and also interferes the eavesdroppers with multiple AN beams. However, since the eavesdroppers are close to the ground user, AN beams are very likely to be non-orthogonal to the legitimate beam, indicating that the legitimate ground user is also interfered by multiple AN beams. Thus, it is necessary to wisely design the legitimate/AN beampatterns with proper power allocation among those beams, which deserves more future research efforts.

\subsection{New Spectral Resources}

One paradigm shift for the next generation mobile networks is to exploit new spectrum resources, such as the mmWave, THz, and visible light frequency bands, to support ultra-high-speed wireless transmissions and enhance information security \cite{XiaohuYou_WCM2023,Robert_JSTSP2016,Akyildiz_PC2014, Han_WCM2023,Hass_CM2014}. However, as the frequency goes from the microwave band to visible light band, new spectrum features raise unprecedented challenges to the design of next generation mobile networks.

\begin{figure}[t]
  \centering
  \includegraphics[width=3.3in]{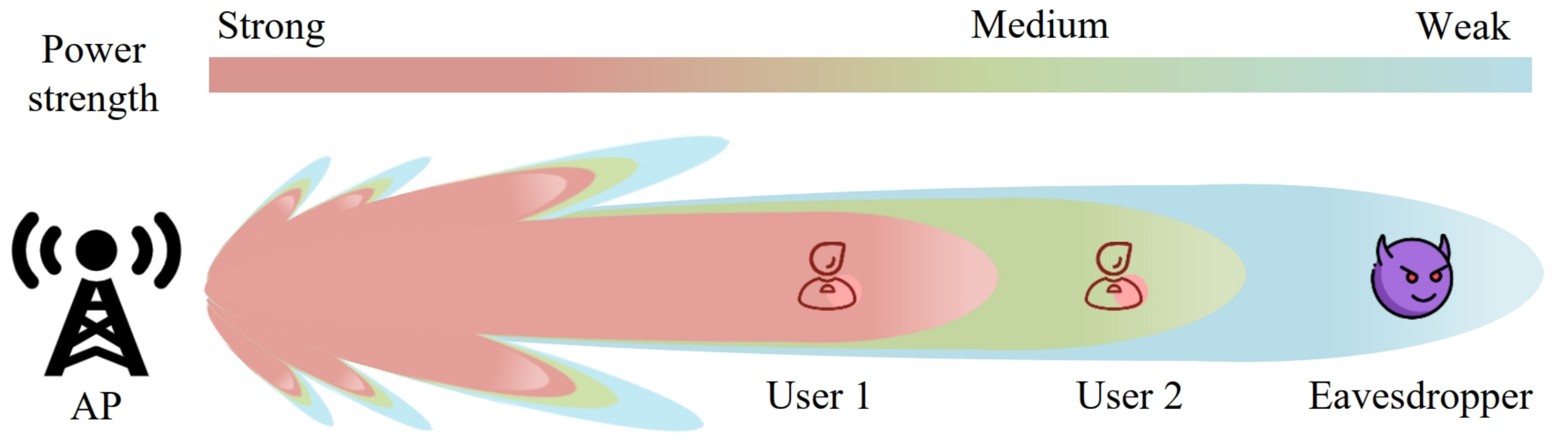}\\
  \caption{Illustration of using the water vapor and molecular absorption for secrecy and covertness enhancement.}\label{fig:mmWave_THz}
\end{figure}

The water vapor and molecular absorption effect is one of the key spectrum properties in mmWave and THz bands, which brings significantly large path loss and noise and degrades the communication links' quality. To this end, various approaches have been proposed to mitigate the harmful effect caused by the water vapor and molecular absorption, such as transmitting signals in spectral windows \cite{Han_TSP2016} or adjusting the frequency band dynamically with the distance to avoid distance-dependent path loss peaks \cite{Jornet_ICC2019}. Despite so, such detrimental effect can be leveraged as a security advantage to benefit secure and covert wireless communications. For example, by an appropriate absorption
peak modulation targeted at minimizing the water vapor and molecular absorption at the legitimate users while magnifying such absorption at the eavesdropper, the received signal quality at the eavesdropper is largely deteriorated and reliable transmissions at the legitimate users are promised \cite{GaoW_TWC2021,Gao_TWC2021,Fang_TTST2022, Zhao_SJ2021}. Fig.~\ref{fig:mmWave_THz} provides a figure of exploiting the water vapor and molecular absorption for secrecy and covertness enhancement. Moreover, recall that visible light communication (VLC) is immune from molecular absorption but is affected by air particles, and thus, it is possible to take advantages of their individual spectrum features and devise a hybrid THz and VLC scheme to enhance secure and covert communication performance, which motivates further research investigations.

Another important feature of these high frequency bands is the beam squint and/or beam split effect, due to the use of ultra-wide band transmissions. This effect  makes the beams disperse to the surrounding directions far away from the targeted users and thus renders the array gain loss as well as considerable communication performance degradation \cite{Tan_JSAC2021,Wang_CM2018}. While the effect of beam squint/split can be good news to security- and covertness-aware NOMA designs in mmWave/THz systems, since the misaligned beams can be used not only for serving additional users to improve spectral efficiency and support massive connectivity \cite{DingZ_CL2022,DingZ_WCL2022,DingZ_TCOM2023}, but also for
sensing malicious eavesdroppers or wardens probably located in different physical directions, which enables predictive beamforming designs to improve the security and covertness performance \cite{GaoF_TCOM2023,LuoH_TWC2024}. In this approach, the misaligned beams are a useful part to the system, and hence it is necessary to optimally allocate the available resources (i.e., power and subcarriers) to balance the sensing and communication performance with secrecy and covertness considerations.

\subsection{Advanced Antenna Architecture}
With the revolutionary innovations of meta-materials and meta-surfaces, holographic MIMO emerges as a disruptive technology to fulfill the visions of ubiquitous demanding services \cite{GCAlex_2020}. Contrary to the phase array antennas in traditional MIMO systems, holographic MIMO utilizes the ultra-thin, extremely large, and nearly continuous holographic-based leaky-wave antennas or the photonic tightly coupled antenna arrays that consist of numerous sub-wavelength metallic/dielectric radiation elements, where these elements can be tuned in software to reshape the incentive electromagnetic waves according to the principle of holography \cite{TGong_2024}. By exploring the intelligent manipulations in the EM domain, holographic MIMO is conceived to simultaneously provide remarkable throughput improvements and security enhancement. Specifically, holographic MIMO enjoys spatially continuous apertures and thus enables signal processing entirely carried out in the electromagnetic (EM) domain without the digital domain. The resulting flexible processing capability facilitates the system to reconfigure the uncontrolled wireless environments into a programmable component, where the signal deterioration can be alleviated and the security-related issues can be incorporated during the system design.
The flexible processing capability in the EM domain enables the holographic MIMO system to alleviate or even eliminate the reflected and refracted components of EM waves. In this context, security and covertness can be significantly improved since eavesdroppers receive few reflection/refraction signal components.
Furthermore, since the involved radiation elements are placed in sub-wavelength distances \cite{NShlezinger_2021}, holographic MIMO can explore the mutual coupling among adjacent elements and achieve ultra-high spatial resolution, where the confidential signals can be transmitted by forming pencil-like beams to prevent potential eavesdropping \cite{TLMarzetta_2019}. Notably, the realized super-directivity becomes more dominant when the number of elements continuously increases.
Last but not the least, due to the extremely large aperture size \cite{MMovahhedi_2019, GBWu_2021, RStevenson_2016}, holographic MIMO may operate in the near-field region (i.e., the Fresnel region). Therefore, near-field holographic MIMO can sense the angle and distance of both legitimate subscribers and eavesdroppers, which paves the way for ISAC-enhanced security designs.

On the other hand, an enhanced spatial degree of freedom (DoF) can be achieved by enabling the phase array antenna with maneuverability, termed the movable antenna (MA) and fluid antenna (FA). Specifically, by flexibly adjusting the positions of array elements within a predetermined region \cite{HuGuojie_SPL2024}, the NOMA network can explore continuous channel variation in the spatial field and enjoy additional design flexibility \cite{WangHonghao_2024}, which breaks the performance limitation resulting from the traditional fixed-position setting of antenna array \cite{HuGuojie_CL2024}. In this context, secure equivalent channels and power allocation can be tailored for confidential information transmission in NOMA networks.
Moreover, the extra spatial DoFs enable a joint desired signal enhancement and interference suppression, it is possible to exploit the inherent interference management in NOMA networks, in which the capacity of the legitimate channel is increased while the capacity of the wiretap channel can be decreased simultaneously \cite{HuGuojie_2024}.

\subsection{Net Zero Communication}

Net zero communication is a key enabling technology to realize green and sustainable networking for the next generation IoT. The core idea of net zero communication is to drive the wireless devices relying on disruptive techniques, such as RF energy harvesting, backscattering, and low power computing, without the need for recharging plugins and replacing batteries \cite{Oppo,Naser_IoTZ2023,LiDong_WCM2023}. There are many reasons for advancing this technology. First,
the integration of net zero communication and NOMA is expected to realize extreme connectivity of low power wireless devices in the next generation mobile networks. For instance, with the energy harvesting capability, the users located next to the transmitter can first gain energy from radio waves harvested in space and then use the harvested energy to serve the remote users \cite{Yuanwei_JSAC2016}. Moreover, the near users in NOMA can also perform backscattering to forward the incident signals to the far users, eliminating the need to generate carrier signals \cite{Chen_TWC2021}. By doing so, a very low cost NOMA communication architecture is achieved.
Second, the low power consumption of net zero communication makes it an ideal technology for secure and covert communication, since the low power transmission helps hide a signal below the noise floor, thus making the eavesdropper very difficult to detect or intercept the signal transmission. In particular, some recent studies suggest that the exploitation of noise modulation (e.g., embedding information in the noise variance and allowing non-coherent
detection at the receiver side) to convey digital information not only consumes extremely low power, but also guarantees covert and unconditionally
secure communications \cite{Mucchi_TVT2022,Basar_TCOM2023,Basar_WCL2024}.

While it is promising to apply net zero communication to enable massive access and secrecy transmissions, there are many challenges remaining to be solved. For example, the computation power and transmission resource of net zero communication are quite limited, it is inappropriate to implement existing complicated security and covertness mechanisms, such as multi-antenna beamforming and active cooperation, which urgently require the development of lightweight security and covertness mechanisms applicable to net zero communication. The waveform designs are crucial in the coding and modulation process of net zero communication, which directly affects the stability of wireless power supply of energy-constrained users as well as the achievable data rate of the network. Hence, more in-depth research is needed to optimize the waveform of net zero communication to satisfy diverse QoS requirements including reliability, security, and energy efficiency. In addition, although the use of extremely low power transmission facilitates information security and covertness, the wireless coverage of net zero communication is tightly circumscribed. The application of intelligent surfaces in net zero communication is expected to benefit coverage extension, energy transmissions, and security enhancement \cite{Lu_Proc2022,Wu_PIEEE2022}, which directs interesting future research works.

\subsection{Generative AI}

The key feature of generative AI is its ability to creatively generate new information from learned patterns, rather than merely imitating or categorizing known information. This technology may have implications for NOMA in communications, especially in optimizing and enhancing the performance of communication systems. Future research should focus on making NOMA technology smarter, more versatile, and more secure.

First, generative models can be used to more accurately model and predict communication channel behavior \cite{X48}. In NOMA systems, CSI plays a crucial role in user pairing and decoding using generative decoders \cite{X49}. Machine learning and game theory algorithms can intelligently predict channel behavior, and generative deep learning models with memories, such as Long Short-Term Memory (LSTM) and Recurrent Neural Networks (RNN), can effectively predict channel behavior. Generative AI can also be used to optimize resource allocation in NOMA systems \cite{X50}, where power allocation to users becomes complex as the number of users increases. Intelligent allocation of power and spectrum resources is achieved by learning and adapting to different channel conditions among users. For example, generative deep learning algorithms and unsupervised learning algorithms can effectively solve the user clustering problem \cite{X51,X52}.

Second, generative AI is expected to play a great role in NOMA security because the presence of mobile environments, increasing heterogeneity, and hyperdensity increase the complexity of wireless networks, posing challenges to traditional NOMA security techniques.
Generative AI can identify potential security threats by learning a large amount of security data and attack samples \cite{X53}. It can analyze patterns and anomalies in communication data to detect possible intrusions, malicious behaviors, or cyber attacks. With the help of generative models, the system can monitor the NOMA communication network in real time to identify and block unauthorized access and malicious activities. This helps in preventing intrusions and taking timely defensive measures accordingly.
In NOMA, multiple user messages are superposed in a single resource block, which can also cause many security issues \cite{X54}. These security issues can be addressed by using encryption and decryption schemes at the transmitter and receiver sides. This process increases the delay and processing requirements.
The research direction in cybersecurity should focus on developing generative AI models with dynamic and adaptive features to cope with increasingly severe attack scenarios and preparing state-of-the-art datasets characterized by balanced classes of real application data frameworks to validate these models as possible solutions to real cyber attacks.

Furthermore, it would be insightful if future works could shed more light on the aspect of ``AI-native'' wireless networks, where the issues of energy efficiency, latency, and privacy are more critical. This is because the green communication will be mandatory, given that the promising generative AI algorithms (e.g., GAN, VAE, DMs) would require excessive energy consumption due to their iterative process for training and inference. The generative AI will be a viable solution in that it can effectively deal with the dynamic and complex distribution because of the combined signaling and propagation characteristics of secure NOMA transmission via PLS. It is of paramount importance to implement the energy-efficient generative AI algorithm with reduced latency and guaranteed privacy, for which it is necessary to address the key issue of scalable generative AI in mobile edge networks with offloaded tasks.

\section{Conclusion}
\label{sec:conclusion}

This tutorial provided a comprehensive overview of the state-of-the-art PLS techniques to safeguard the next generation multiple access networks. In particular, the theoretical fundamentals and real-world deployments of NOMA, information-theoretic security, quantum-safe security, low detection covertness techniques were discussed in detail. Various physical layer technologies for security and covertness provisioning of NOMA were summarized and explained. Finally, the tutorial was concluded by discussing several potential challenges of using PLS techniques to secure NOMA communications and pointing out the promising future research directions.

\section*{Acknowledgement}

The authors would like to thank the current and past members of the research groups at Xidian University and Xi'an Jiaotong University, for their contributions to the work reported in this paper.

\end{document}